%% file: Cosmic-Superstring.tex
\documentclass[12pt]{article}

\usepackage{amssymb,amsmath,mathrsfs,epstopdf,slashed,color}
\usepackage{tikz}
\usetikzlibrary{matrix}
\usetikzlibrary{positioning}
\usepackage{longtable}
\usepackage{booktabs}

\usepackage{graphicx}
\usepackage{placeins}
\usepackage{hyperref}
\usepackage{cite}

\allowdisplaybreaks[1]

\makeatletter

\@addtoreset{equation}{section}
\makeatother

\newcommand{\ba}{\begin{eqnarray}}
\newcommand{\ea}{\end{eqnarray}}
\def\be{\begin{equation}}
\def\ee{\end{equation}}

\newcommand{\msun}{{\rm M}_\odot}

\newcommand{\simless}[0]{\mathbin{\lower 3pt\hbox
   {$\rlap{\raise 5pt\hbox{$\char'074$}}\mathchar"7218$}}}
\newcommand{\simgreat}[0]{\mathbin{\lower 3pt\hbox
   {$\rlap{\raise 5pt\hbox{$\char'076$}}\mathchar"7218$}}}
\newcommand{\gta}[0]{\simgreat}
\newcommand{\lta}[0]{\simless}

\title{\bf Detection of Low Tension Cosmic Superstrings
}
\author{David F. Chernoff and S.-H. Henry Tye}

\begin{document}

\begin{titlepage}

\setcounter{page}{0}
  
\begin{flushright}
 \small
 \normalsize
\end{flushright}

\vskip 2.2cm
\begin{center}

{\Large \bf Detection of Low Tension Cosmic Superstrings
}

\vskip 1.2cm
  
{\large David F. Chernoff${}^1$ and S.-H. Henry Tye${}^{2,3}$}
 
 \vskip 0.6cm

 ${}^1$ Department of Astronomy, Cornell University, Ithaca, NY 14853, USA \\
 ${}^2$ Jockey Club Institute for Advanced Study and Department of Physics \\ Hong Kong University of Science and Technology, Hong Kong\\
 ${}^3$ Department of Physics, Cornell University, Ithaca, NY 14853, USA

 \vskip 0.4cm

Email: \href{mailto: chernoff@astro.cornell.edu, iastye@ust.hk}{chernoff at astro.cornell.edu, iastye at ust.hk}

\vskip 0.8cm
  
\abstract{\normalsize Cosmic superstrings of string theory differ from conventional cosmic
strings of field theory. We review how the physical and cosmological
properties of the macroscopic string loops influence experimental
searches for these relics from the epoch of inflation. The universe's
average density of cosmic superstrings can easily exceed that of
conventional cosmic strings having the same tension by two or more
orders of magnitude. The cosmological behavior of the remnant
superstring loops is qualitatively distinct because the string tension
is exponentially smaller than the string scale in flux
compactifications in string theory. Low tension superstring loops live
longer, experience less recoil (rocket effect from the emission of
gravitational radiation) and tend to cluster like dark matter in
galaxies. Clustering enhances the string loop density with respect to
the cosmological average in collapsed structures in the universe.  The
enhancement at the Sun's position is $\sim 10^5$. We develop a model
encapsulating the leading order string theory effects, the current
understanding of the string network loop production and the influence
of cosmological structure formation suitable for forecasting the
detection of superstring loops via optical microlensing, gravitational
wave bursts and fast radio bursts. We evaluate the detection rate of
bursts from cusps and kinks by LIGO- and LISA-like experiments.
Clustering dominates rates for $G \mu < 10^{-11.9}$ (LIGO cusp),
$G \mu<10^{-11.2}$ (LISA cusp), $G \mu < 10^{-10.6}$ (LISA kink); we
forecast experimentally accessible gravitational wave bursts for $G
\mu>10^{-14.2}$ (LIGO cusp), $G \mu>10^{-15}$ (LISA cusp) and $G \mu>10^{-
14.1}$ (LISA kink). 
}

\vspace{0.3cm}
\begin{flushleft}
 \today
\end{flushleft}
 
\end{center}
\end{titlepage}

\setcounter{page}{1}
\setcounter{footnote}{0}

\tableofcontents

\parskip=5pt

\section{Introduction}

Cosmic strings are one-dimensional topological defects formed by spontaneous symmetry breaking in the early universe. They were first proposed in the 1970s \cite{Kibble:1976sj} and have evoked ongoing cosmological interest (see Ref. \cite{Vilenkin:2000} for review). The initially formed defects (of order one per horizon via the Kibble mechanism) quickly evolve into a scaling network of horizon-size long strings and loops of different sizes, with properties dictated largely by the string tension $\mu$.  The hypothesis that a cosmic string network might actively source the density fluctuations for structure formation in our universe was extensively studied in the 1980s and 1990s and found to require tension $G\mu \simeq 10^{-6}$, where $G$ is the Newton constant (taking $c=1$). In that scenario the fraction of cosmic string energy content in the universe is roughly
\be
\label{Ocs}
 \Omega_{string} \simeq \Gamma G \mu 
\ee
where, numerically,  $\Gamma \simeq 50$. If cosmic strings were responsible for the fluctuations their contribution to the energy content in the universe would be negligible. However, in models with actively generated fluctuations the power spectrum of the cosmic microwave background radiation (CMBR) has {\it no} acoustic peaks. The discovery of the acoustic peaks in the CMBR power spectrum in late 1990s rules out cosmic strings as the primary source of fluctuation and strongly supports the inflationary universe scenario. Observational bounds on the cosmic string tension and energy content continue to improve.
  In the last year, pulsar timing limits on the stochastic
  background from strings improved from $G \mu < 10^{-9}$ \cite{Sousa:2016ggw} to $G \mu \lta 1.5 \times 10^{-11}$
\cite{Blanco-Pillado:2017oxo,Blanco-Pillado:2017rnf}.
 Very recently (after this paper was substantially finished) limits on cosmic strings for LIGO observing run O1 were reported \cite{Abbott:2017mem}:
 $G \mu < 10^{-10}$ (stochastic background) and
 $\lta \,3 \times 10^{-7}$ (bursts)
  for model choices closest to our own. We have added brief comments and will elaborate at a future time.\footnote{The quoted limits are for model $M=2$ with intercommutation probability $p=10^{-2}$ in Fig. 5 \cite{Abbott:2017mem}}

A long standing goal of theoretical physics is finding a consistent framework for quantum gravity and nature's known force fields and matter content. String theory is the leading candidate. Because of its rich structure and dynamics, an explicit string theory realization of the standard model of strong and electroweak interactions remains elusive. As a result, the search for evidence of string theory in nature turns out to be very challenging. One promising avenue is hunting for the progeny of strings of string theory stretched to horizon sizes. The study of inflation in string theory (see Ref. \cite{Baumann:2014nda} for a review) has revealed routes to the formation of macroscopic strings at the conclusion of the inflationary epoch. Such strings behave very much like the original cosmic strings \cite{Jones:2002cv,Sarangi:2002yt,Jones:2003da,Copeland:2003bj} but with tensions that easily satisfy the present observational bounds. Although similar in many respects, they differ in a number of significant ways. To distinguish them from the traditional cosmic strings, we refer to them as cosmic superstrings. In this paper, we show superstrings can have properties consistent with today's observational bounds and still be detectable in the near future. Encouraged by the recent spectacular success of LIGO \cite{LIGO}, we shall present our estimate of the detectability of cosmic superstrings (for $G \mu > 10^{-15}$) via gravitational wave bursts from cusps and kinks \cite{Damour:2001bk,Damour:2004kw}. Gravitational wave searches combined with microlensing searches \cite{Chernoff:2007pd,Chernoff:2014cba} can teach us a lot about what types of strings might be present. If cosmic superstrings are discovered then measurements will provide valuable information about how our universe is realized within string theory and go a long way towards addressing the question ``is string theory the theory that describes nature?'' Any positive detections, of course, will provide the most direct possible insights.

Since superstring theory has 9 spatial dimensions, common experience suggests 6 of them are compactified. Turning on quantized fluxes \cite{Giddings:2001yu,Kachru:2003aw} in the presence of $D$-branes \cite{Polchinski:1995mt} yields a warped geometry having throat regions connected to a bulk space. A typical flux compactificaton in Type IIB string theory can have dozens to hundreds of throats. In one possibility, the brane world scenario, visible matter is described by open strings living inside a stack of 3 spatial dimensional D3-branes sitting at the bottom of one of the throats. The D3-branes span the normal dimensions of our universe. The mass scale at the bottom of a throat is decreased (warped) by orders of magnitude compared to the bulk scale, which is simply the string scale $M_S$, taken here to be a few orders of magnitude below the Planck scale $M_P = G^{-1/2} \simeq 10^{19}$ GeV. Superstrings sitting at a bottom have tensions decreased by the same factor so tension $\mu$ is orders of magnitude below that implied by the string scale $M_S^2$. 

Reheating at the end of inflation excites the light string modes that constitute the standard model particles and marks the beginning of the hot big bang. The production of cosmic superstrings after inflation has been studied mostly in the simplest scenario in string theory, namely the $D$3-${\bar D}$3-brane inflation in a warped geometry in flux compactification \cite{Dvali:1998pa,Dvali:2001fw,Burgess:2001fx,Kachru:2003sx}.  The energy source for reheating is the brane-anti-brane annihilation at the end of inflation. In a flux compactification, this energy release happens in a warped throat, namely the inflationary throat. The energy released can also go to light string modes and strings with horizon-scale sizes. The annihilation of the $D$3-${\bar D}$3-brane pair easily produces both $F$-strings and $D$1-strings in that throat.
In a simple brane inflationary scenario, using the PLANCK data\cite{Planck:2013}, one finds that $G \mu < 10^{-9}$\cite{Firouzjahi:2005dh,Bean:2007hc}. There may be numerous throats so the standard model throat may differ from the inflationary throat and also a host of other spectator throats.
Energy released in the inflationary throat spreads to other, more warped throats. Reheating is expected to generate standard model particles in the standard model throat and cosmic superstrings in throats warped at least as much as the inflationary throat\cite{Kofman:2005yz,Chen:2006ni,Chialva:2005zy}. 

Brane-flux inflationary scenarios generally yield similar outcomes. For other inflationary scenarios in string theory, the picture is less clear, though even a very small production of $F$-strings and $D$1-strings will eventually evolve to the scaling solution, so it may not be unreasonable to assume that such strings are produced irrespective of the details of the particular inflationary realization.
The cosmic superstrings and the particles tend to sit at the throat bottoms due to energetic considerations. The string network in each throat is expected to evolve independently of the other throats though all cosmic strings are visible to us via their gravitational interactions. Each network reaches a scaling solution that is insensitive to initial conditions and largely set by string tensions appropriate to the throat.

We shall start with ordinary cosmic strings, which have been extensively studied \cite{Vilenkin:2000}, and list how properties of superstrings in string theory differ and how each difference enhances or suppresses the prospects for detectability. To describe order of magnitude changes to the probability of detection, we introduce a single parameter ${\cal G}$ to summarize why cosmic superstrings offer much better chances than ordinary cosmic strings. In this over-simplified picture, we compare the fraction of cosmic superstring energy content in the universe to that of the conventional $\Omega_{string}$ (\ref{Ocs}),
$$\Omega_{superstring} \sim {\cal G} \Omega_{string}\simeq \left(\frac{N_sN_T}{p}\right) \Omega_{string}$$
where $N_s$ is the effective number of species of strings within a single warped throat (e.g., $N_s \sim 1$ to $4$). Here $p \le 1$ is the effective intercommutation probability. For usual cosmic strings, $p \simeq 1$ while $p \le 1$ for superstrings, and can be as small as $p \sim 10^{-3}$ \cite{Jackson:2004zg}. It is pointed out that it may go like $p^{2/3}$ (instead of $p$) in $\Omega_{superstring}$ \cite{Avgoustidis:2005nv}. $N_T$ is the effective number of throats in the flux compactification, throats with cosmic superstrings sitting at its bottom; actually, we should only count those with string tensions above the eventual observational limit. Here we have in mind $G \mu > 10^{-18}$.
Overall, we expect $1 \ll {\cal G} < 10^4$.
Combining this $\cal G$ factor enhancement with the enhancement coming from the clustering of low tension cosmic superstrings (following dark matter) in our galaxy (a density enhancement factor ${\cal F} \sim 10^5$ \cite{Chernoff:2007pd,Chernoff:2009tp}) gives hope for detecting microlensing of stars with optical surveys and gravitational wave bursts at advanced LIGO.

Cosmic superstrings differ from ordinary cosmic strings in a number of fundamental ways:  

(1) There are 2 types of strings, namely fundamental strings, or $F$-strings, and $D$1-branes, i.e., $D$-strings \cite{Copeland:2003bj}. The intercommutation (reconnection) probability $p$, which is $p \simeq 1$ for vortices, can be $p \ll 1$ for superstrings \cite{Jackson:2004zg}. This property has already been incorporated in a number of cosmic superstring network studies \cite{Sakellariadou:2004wq,Avgoustidis:2004zt}.

(2) A $F$-string sitting at the bottom of throat $i$ has tension $G\mu_i \sim GM_S^2 h_i^2 \ll GM_S^2$, where $h_i \ll 1$ is the warp factor at the bottom. An empty throat without branes will have its own strings with a spectrum of tension \cite{Copeland:2003bj,Firouzjahi:2006vp}. The string networks may contain junctions and beads \cite{Gubser:2004qj,Siemens:2000ty,Leblond:2009fq}. A throat with $D$3-branes (or $\bar D$3-branes) at its bottom will have only $D$-strings there \cite{Leblond:2004uc}.  This is because branes allow open $F$-strings inside them so the closed $F$-strings inside branes tend to break into tiny open strings.
Interactions between strings from different throats are expected to be very weak. We introduce an effective number $N_s$ of types of strings to reflect the presence of the tension spectra present.

(3) Depending on the Calabi-Yau manifold chosen by nature, we expect dozens or hundreds of throats in a typical flux
compactification. Throats with different warped geometries result in
different types of cosmic superstring tension spectra with the
fundamental tensions substantially lower than the string scale, since strings 
tend to sit at the bottoms of the throats. We
introduce an effective number $N_T$ of throats with string tensions
$G\mu > 10^{-18}$, the lower limit of detectability in the forseeable
future for both gravitational wave stochastic backgrounds
\cite{Blanco-Pillado:2017rnf} and optical microlensing.\footnote{For
  microlensing the lower limit of detectability may be crudely
  estimated as follows.  The angular size of a typical star at a
  typical distance in the galaxy is comparable to the deficit angle
  for a string with $G \mu \sim 10^{-13}$. The state of the art for
  measuring relative flux variations of bright nearby stars
  in exoplanet searches is about
  $10^{-5}$. A string with $G \mu \sim 10^{-18}$ would lens
  approximately $10^{-5}$ of the stellar disk and create a
  hypothetical relative flux variations of this size.}

(4) All strings in string theory should be ``charged'' under a two-form field, so cosmic superstrings will emit axions (i.e., two-form fields in 3+1 dimensions) in addition to gravitational waves. Because of this additional decay mode, the density of some types of cosmic string loops may be significantly decreased. Although the emission rate of axions has been generally studied in Ref. \cite{Firouzjahi:2007dp,Gwyn:2011tf}, the emission rate of a particular axion by a specific string depends strongly on axionic properties such as mass, coupling ``charge'' to strings and decay rate to two photons. 

(5) Cosmic strings that move (oscillate) in a throat will have a tension varying in time and from point to point along its length\cite{Avgoustidis:2007ju,Avgoustidis:2012vc}. 

For low tension cosmic strings ($G \mu <10^{-9}$), clustering of string loops in our galaxy can enhance the cosmic string density by many orders of magnitude similar to the clustering of dark matter. Five orders of magnitude are expected at the solar position and more at the center of the Galaxy. Clustering substantially increases the potential of detection\cite{Chernoff:2007pd,Chernoff:2009tp}. This property applies to ordinary low tension cosmic strings as well.

Increasingly comprehensive studies of gravitational wave bursts from cosmic string network have appeared \cite{DePies:2007bm,Leblond:2009fq,Kuroyanagi:2012wm,Kuroyanagi:2012jf,Blanco-Pillado:2013qja}. Some have already included (1), the low $p$ effect \cite{Jackson:2004zg,Sakellariadou:2004wq,Avgoustidis:2004zt}. We will highlight the other effects, in particular (2) and (3), which can dramatically raise the prospects for detection. Following Ref. \cite{Chernoff:2007pd}, Ref. \cite{DePies:2009mf} has included the clustering effect. Here we provide a more detailed analysis following a better understanding of the clustering effect \cite{Chernoff:2009tp}.
We shall describe a simple cosmic string model (with a string tension $\mu$ so $G \mu < 10^{-7}$ and loop size relative to the horizon size $\alpha \sim 0.1$) and discuss how each of the above effect may modify the properties and detectability in microlensing and gravitational wave search/observation. In general, (1)-(3) tend to enhance while (4) tends to decrease the detectability via gravitational wave.  The main analysis in this paper focuses on the clustering of low tension strings like dark matter in galaxies and its effect on their detectability via microlensing and gravitational wave bursts. In microlensing, caustics are also possible if the string segment is not straight when compared to the star behind it \cite{Uzan:2000xv}. 

We do not know the precise compactification geometry so there are quite a number of uncertainties in determining the intrinsic string properties and the string network evolution dynamics. Given this state of current understanding this modeling though precise should be considered as no better than an order of magnitude estimate. Nonetheless, we find with this analysis that a wide range of superstring tensions are potentially detectable and often by several different types of experiments. We attempt to provide enough details to illustrate how the predictions/estimates may vary with respect to the input assumptions/physics as our understanding/knowledge continues to improve.

Following the discussion of the properties of the cosmic superstrings in Sec. 2,
we outline in Sec. 3 three separate methods by which loops may
be detected: cusp emission of axions followed by conversion to
photons, microlensing of stellar sources of photons and emission of
gravitational waves. We then provide a detailed astrophysical model that
summarizes the properties (number density, lengths, velocities, etc.)
of string loops relevant to forecasting experimental outcomes.
Clustering of low tension loops is a significant effect that enhances
the ability of experiments to detect loops. Here we
concentrate on estimating the gravitational burst rate for LIGO/VIRGO
and LISA taking account of the enhancements from the local source population.
For example, we find that cusp bursts from loops in the halo of our
Galaxy dominate the contribution from the rest of the homogeneous universe for
LIGO for $10^{-15}< G \mu < 10^{-13}$; likewise, cusp bursts for LISA
for $G \mu < 10^{-11}$ are halo-dominated.  Elsewhere, we will employ
the model to forecast the detection rates for microlensing and
axion-mediated photon bursts.

\section{Properties of Cosmic Superstrings}

The first suggestion that string theory's strings might manifest as cosmic superstrings was contemplated in the heterotic string theory\cite{Witten:1985fp}.  However, among other issues the tension of superstrings in that description is far too high to be compatible with data. With the discovery of $D$-branes \cite{Polchinski:1995mt}, the introduction of warped geometries in flux compactification \cite{Giddings:2001yu,Kachru:2003aw} and the development of specific, string theory based inflationary scenarios, the prospect has improved dramatically. %
In the brane world scenario, the cosmic superstrings are produced after the inflationary epoch and evolve to a scaling network. The network includes long, horizon-crosssing strings and sub-horizon scaled loops. These are the objects of interest for experimental searches.

Of the 9 spatial dimensions in Type IIB string theory, 6 dimensions (i.e., $y^m$) are compactified into a Calabi-Yau like manifold,
\be 
\label{bmetric}
ds^2= h^2(y^m)dx^{\mu}dx_{\mu} + g_{mn}(y) dy^mdy^n
\ee
where $x^{\mu}$ span the usual 4-dimensional Minkowski spacetime, so
 $$M_P^2 \simeq M_S^8 \int d^6y \sqrt{g_6(y)} h(y)^2$$
where $g_6$ is the determinant of $g_{mn}$. The manifold consists of the bulk, where $h(y^m) \simeq 1$, and smoothly connected throats.  At the bottoms of the throats (i.e., tips of deformed cones), we expect $h(y^m) \ll 1$. A typical compactification can have dozens or hundreds of throats, each with its own warp factor $h_j$. In a simple brane world scenario, one throat, namely the standard (strong and electroweak) model (S) throat, has a stack of $D$3-branes sitting at the bottom, with warp factor $h_S \ll 1$. This stack spans our 3-dimensional observable universe. All standard model particles are open string modes inside the branes.  The Higgs Boson mass $m_H$ is considered natural if it satisfies $m_H \sim M_Sh_S$. Since Type IIB string theory has only odd-dimensional branes, i.e., $D$(2n+1)-branes, it does not have $D$2- or $D$0-branes but has $D$1-branes, so there are $D$1-strings but no membrane-like or point-like defects.  Both $D$1-strings and fundamental $F$-strings can form cosmic superstrings. Closed strings may be born and move in space outside the $D$3-branes. 

The ends of an open $F$-string must end on a brane. Both closed $D$-strings and $F$-strings will be present in a throat if it has neither $D$3-branes nor $\bar D$3-branes. If a closed $F$-string comes in contact with the brane it will fragment into open $F$-strings with ends inside the brane. It will not survive as a cosmic superstring. However, a $D$-string may swell inside a $D$3-brane and persist, behaving like a vortex instead of a strictly one-dimensional object\cite{Leblond:2004uc}. Likewise, if we live inside $D$7-branes wrapping a 4-cycle, the same phenomenon happens: only $D$-strings survive as cosmic superstrings in the S throat and other throats with branes.

\subsection{Tension Spectrum}

Typically, strings of all sizes and types will be produced towards the end of inflation, e.g., during the collision and annihilation of the $D$3-$\bar D$3 brane pair as energy stored in the brane tensions is released\cite{Jones:2002cv,Sarangi:2002yt,Jones:2003da,Sarangi:2003sg}. The lower string modes are effectively particles but some of the highly excited modes are macroscopic, extended objects.
Large fundamental strings (or $F$-strings) and/or $D$1-branes (or $D$-strings) that survive the cosmological evolution become cosmic superstrings \cite{Copeland:2003bj}. 

In 10 flat dimensions, or in the bulk in a flux compactification, supersymmetry dictates that the tension of the bound state of $p$ $F$-strings and $q$ $D$-strings is \cite{Schwarz:1995dk},
\ba
	\label{flat}
	T_{{p,q}} = T_{F1} \sqrt{p^2  +\frac{q^2}{g_s^2}}\, .
	\label{pqtension10}
\ea
Coprime combinations of $(p,q)$ can form strings with junctions \cite{Copeland:2003bj}, so their zipping and unzipping will be part of the string evolution dynamics \cite{Avgoustidis:2014rqa}. For $(p,q)$ not coprime, simpler states of fewer $F$ and $D$-strings exist having equivalent energy per component. Recent network studies of this idealized spectra strongly suggest that cosmic superstrings evolve dynamically to a scaling solution with a stable relative distribution of strings with different quantum numbers \cite{Tye:2005fn}, very much like ordinary cosmic strings of either Abelian Higgs or Nambu-Goto type \cite{Vilenkin:2000}. The strings' scaling density decreases roughly $\propto T_{p,q}^{-N}$, where $N \sim 8$, a rapid falloff for higher $(p,q)$. We shall consider scenarios where at least some of the lower $(p,q)$ strings of more realistic spectra are stable enough to realize the scaling solution. Generally if the $F$-strings are stable we expect more $F$-strings than $D$-strings since $g_s<1$. In that case the total number density of all cosmic strings will be comparable to that of $F$-strings with $(p,q)=(1,0)$, enhanced by a factor $1/g_s^N$ relative to D-strings with $(p,q)=(0,1)$.

In a more realistic scenario the compactified manifold is not flat but contains warped throats. Since reheating after inflation (e.g., the $D$3-$\bar D$3-brane annihilation) is expected to take place at the bottom of a throat, some of the cosmic superstrings will be produced in that part of the manifold. If $D$3-branes are left in the bottom of the throat, the $F$-strings will fragment while the $D$-strings will be metastable, presumably surviving as cosmic strings \cite{Leblond:2004uc}. In an empty throat new $F$- and $D$-strings will survive and form bound states, resulting in a spectrum of string tensions with junctions and probably beads. The particulars depend on the geometry of the throat but it is illustrative to consider the tension spectrum in the well-studied Klebanov-Strassler (KS) throat \cite{Klebanov:2000hb}. This is a warped deformed conifold with an $S^3$ fibered over $S^2$. Let $r$ be the distance from the bottom of a throat on the manifold and $R$ be the characteristic length scale.  
The bulk is connected to the edge of the throat at $r=R$, where
\be
\label{KMR}
R^4=\frac{27 \pi g_sN}{16 M_S^4}, \quad \quad N=KM
\ee
where $N=KM$ is the number of $D$3-charges and integers $K$ and $M$ are the NS-NS and RR fluxes respectively. These integers are expected to be relatively large. The tip of a conifold sits at $r=0$. Here, the $S^3$ has a finite size if the conifold is deformed (without breaking supersymmetry), while the $S^2$ has a finite size if the conifold is resolved (breaking  supersymmetry), so $r =r_i \gtrsim 0$ at the bottom of the throat. At the top ($r \simeq R$) the warp factor is
$h(r=R) \simeq 1$ and at intermediate locations $h(r) \simeq r/R$. In terms of the fluxes the warp factor at the bottom of the $i$th throat is
\be
\label{warpi}
h_i= h_i(r_i \simeq 0) = e^{-2\pi K_i/g_sM_i} \ll 1 
\ee
and a $(p,q)$ bound string near that point has tension \cite{Firouzjahi:2006vp}
\ba
\label{Tanswer}
T_{p,q} \simeq  \frac{M_S^2h_{i}^{2}}{2 \pi} \sqrt{  \frac{q^2}{g_s^2} + 
\left(\frac{b M_i}{\pi}\right)^2 \sin^2\left(\frac{\pi (p-qC_0)}{M_i}\right)},
\ea  
where $b=0.93$ is a number numerically close to one, $C_0$ is the RR-zero form scalar expectation value there, and the integer $M_i \gg 1$ is the number of fractional D3-branes, that is, the units of 3-form RR flux $F_3$ through the $S^{3}$ in the KS throat. For integer $K_i$, the infrared field theory at the bottom of the $i$th throat is a pure $N=1$ supersymmetric Yang-Mills theory, and the warp factor $h_i$ is expected to be small. The mass of the bead at the junction is \cite{Gubser:2004qj}
\be
m_b= \frac{h_iM_S}{3}\sqrt{\frac{g_s}{4 \pi}} \left(\frac{b M_i}{\pi}\right)^{3/2} .
\ee
Ref. \cite{Leblond:2009fq} argues that the cosmic string network will evolve to a scaling limit for modest integers $M_i >10$. The string and bead properties of other geometric throats is an interesting, open question.

\subsection{Production of Cosmic Superstrings in Brane Inflation}

The production of cosmic superstrings in the early universe depends on the inflationary scenario in string theory. 
The simplest is probably brane inflation\cite{Dvali:1998pa,Dvali:2001fw,Burgess:2001fx,Kachru:2003sx}, in which brane-anti-brane annihilation releases energy towards the end of the inflationary epoch that generates closed strings. $D$1-strings can be viewed as topological defects in the $D$3-$\bar D$3-brane annihilation so they are produced via the Kibble mechanism. $F$1-strings may be viewed as topological defects in a S-dual description produced in a similar way, since the Kibble mechanism depends only on causality, irrespective of the size of the coupling. As a result, horizon size strings are produced.

In the simplest brane inflationary scenario, we focus on two of the many throats in the compactified manifold, namely the inflationary throat $A$ and the standard model throat $S$. Because of the warped geometry a mass $M$ in the bulk becomes $h_{A}M$ at the bottom of throat $A$, where $h_{A} \ll 1$  (\ref{warpi}) is the warp factor there. Since $\bar D$3-branes are attracted towards the bottoms of throats, let us suppose there is a $\bar D$3-brane sitting at the bottom of the $A$ throat. A $D$3-brane in the bulk will be attracted towards the $\bar D$3-brane and inflation (driven by the potential energy from the brane-anti-brane tensions)   happens as it moves down the throat. The inflaton $\phi$ is proportional the brane-anti-brane separation in the throat. The attractive potential is dominated by the lightest closed string modes, namely, the graviton and the RR field, yielding a Coulomb-like $r^{-4}$ potential where $r$ is distance from the $\bar D$3-brane at the tip. 
The warped geometry dramatically flatten the inflaton potential $V(\phi)$
so the attraction is rendered exponentially weak in the throat. 
For a canonical kinetic term we have $\phi = \sqrt{T_3}r$. The simplest inflaton potential takes the form \cite{Kachru:2003sx}
\be
\label{infpot}
\begin{split}
V(\phi) =   V_A + V_{D \bar D} &=2T_3h_A^4(1-\frac{1}{N_A}\frac{\phi_A^4}{ \phi^4}) \\
&= \frac{64\pi^2\phi_A^4}{27N_A} \left( 1 - \frac{\phi_A^4}{N_A \phi^4} \right)
\end{split}
\ee
where the $D$3-brane tension $T_3=M_S^4/(2\pi g_s)$ is warped to $T_{3}h_A^4$. Note that this inflaton potential has only a single parameter, namely $\phi_A^4/N_A$. Crudely, $h(\phi) \sim \phi/\phi_{edge}$, where $\phi=\phi_{edge}$ when the $D$3-brane is at the edge of the throat at $r=R$. Likewise, at the bottom $\phi=\phi_{A}$, the warp factor is $h_{A} = h(\phi_{A})= \phi_{A}/\phi_{edge}$. The inflaton $\phi$ is an open string mode and the attractive tree-level gravitational plus RR potential can also be obtained via the one-loop open string contribution.
The scale of the potential is reduced because $N_{A} \gg 1$ is the $D$3 charge of the throat.
We consider an ordering
$$ 0 \le \phi_A \lesssim  \phi_f \le \phi \le \phi_i < \phi_{edge}$$
where inflation begins at $\phi=\phi_i$ and ends at $\phi_f$, when a tachyon appears signaling the annihilation of the brane-anti-brane pair. At least 55 e-folds of inflation must take place inside the throat to achieve consistency with observations. 

The combination $\phi^4_A/N_A$ in the inflaton potential $V(\phi)$ (\ref{infpot}) is constrained by the magnitude of the power spectrum in the Cosmic Microwave Background Radiation and one finds \cite{Firouzjahi:2005dh,Shandera:2003gx,Shandera:2006ax}
$$n_s=0.967, \quad \quad  r \simeq 10^{-9}$$
and the tension of $D$1-strings is
\be
\label{Gtension}
G \mu \simeq \frac{4 \times 10^{-10}}{\sqrt{g_s}}  .
\ee
and $h_{A} \sim 10^{-2}$, with a value dependent on details of the throat. 
The $F$-string tension is smaller by a factor of the string coupling $g_s$: i.e., $\mu_F = g_s \mu$ where $g_s <1$.
This is consistent with the present observational bound \cite{Sousa:2016ggw}.

Towards the end of inflation (near $\phi_f$), as the $D$3-$\bar D$3-brane separation $r$ decreases, an open string (complex) tachyonic mode appears at
\be
\label{tachyon}
\frac{m_{tachyon}^2}{M_S^2} = M_S^2r^2 - {\pi}
\ee
which triggers an instability due to tachyon rolling. As $\phi$ decreases the $\phi^{-4}$ Coulomb-like form of the potential is chopped off, leaving $V(\phi)$ with a relatively flat form and possessing an imaginary component \cite{Sarangi:2003sg}.  In the closed string picture, this happens precisely when the weakening Yukawa suppression of the massive closed string modes' contribution to the potential is overtaken by the rapidly increasing degeneracy of excited closed string modes
$A(n) \rightarrow (2n)^{-11/4} \exp (\sqrt{8\pi^2n})$. Here,
$n$ is the excitation level (a string with center of mass
$m$ has level $n=m^2/{8\pi M_S^2}$). The contribution to $V(r)$, in the large $n$ approximation, is
$$V(r) \propto -r^{-4} \sum_n n^{-11/4} \exp \left(\sqrt{2 \pi n}\left[\sqrt{\pi} - M_Sr\right]\right)$$
where the $\sqrt{\pi}$ term comes from the degeneracy while the $-M_Sr$ term comes from the Yukawa suppression factor $\exp(-mr)$.
Comparison to Eq(\ref{tachyon}) reveals that the exponential growth of degeneracy leads to a divergent $V(r)$ precisely at the point where the tachyon appears.
 Regularization introduces an imaginary part for $V(\phi)$, which may be interpreted, via the optical theorem, as the width per unit world volume for a $D$3-$\bar D$3-brane pair decaying to $F$ strings \cite{Sarangi:2003sg},
\be
\Gamma = {\rm Im} [V(\phi)] \simeq \frac{\pi}{2} h_A^4 \left(\frac{ |m_{tachyon}^2|}{4 \pi} \right)^2 .
\ee
The appearance of this imaginary part of $V(\phi)$ is due to the large Hagedorn degeneracy of the massive modes and the implication is that
$D$3-${\bar D}$3-brane annihilation leads to very massive closed string modes. 
The energy released first goes to on-shell closed strings. For large mass $m$, the transverse momenta of these strings are relatively small,
$$\frac{<k_{\perp}^2>}{m^2} \sim \frac{6}{\sqrt{\pi}} \frac{M_S}{m} $$
so a substantial fraction of the annihilation energy
goes to form massive non-relativistic closed strings. 
Although the above discussion is for the $D$3-${\bar D}$3-brane annihilation channel to $F$ strings, we expect production of $D$1-strings as well, since one may view a $D$3-brane as a di-electric collection of $D$1-strings \cite{Myers:1999ps}. The process of $D$3-${\bar D}$3-brane annihilation producing vortex-like $D$1-strings has been studied in the boundary string field theory framework \cite{Jones:2002sia}. The detailed, quantitative mass distribution of the strings is not critically important as long as evolution proceeds to a scaling cosmic superstring network independent of the initial distribution \cite{Tye:2005fn}. No monopole-like or domain-like defects are produced since there are no $D$0-branes or $D$2-branes present in the Type IIB string theory framework adopted here.
 
Some of the $D$3-${\bar D}$3-brane energy goes to closed $D$1-strings and $F$1-strings in the $A$ throat; the rest is dumped into other throats including the $S$ throat, which initiates the hot big bang.  Energetics favor heat transfer to any throat with a larger warp factor than that of the $A$ throat, creating cosmic superstrings of lower tension than those in the $A$ throat.

It is interesting to note that all the energy released by the $D$3-$\bar D$3-brane annihilation goes to closed strings first \cite{Chen:2006ni}. In the absence of other branes, this is clear, since open strings end on branes and, after the brane annihilation, no branes exist to anchor endpoints. To understand the fate of an open string in the $D$3-brane consider the $U(1)$ flux tube between its two ends. After annihilation, the flux tube together with the open string now forms a closed string. If an open string stretches between a spectator brane and the $D$3-brane to be annihilated, there is a flux tube linking it to another end of a similar string (its conjugate). After annihilation, the flux tube plus the connected open strings form an open string attached to the spectator brane.

Related brane (or brane-flux) inflationary scenarios, share many relevant properties, leading to cosmic superstring production in the manner described. Other stringy inflationary scenarios may also generate cosmic superstrings and classical strings towards the end of the inflationary epoch. This is an important problem to investigate. Our general viewpoint is that since reheating must be present at the end of the inflationary epoch to start the hot big bang, and all particles produced are light string modes, some excited strings should be produced and the Kibble mechanism should be applicable to these.
Schematically, cosmic strings contribute to the Hubble parameter $H$,
$$H^2 = \frac{8 \pi G}{3} \left( \Lambda + \frac{\rho_{strings,0}}{a^2}+ \frac{\rho_{matter,0}}{a^3} + \frac{\rho_{radiation,0}}{a^4} \right)$$
where $\rho_{strings,0}$ is the initial energy density of cosmic strings at the end of inflation. Even if $\rho_{strings,0}$ is exponentially small, its role (relative to matter and radiation densities) will grow substantially because $a$ increases many orders of magnitude; string inter-commutation and gravitational decay will jointly drive the system to its attractor solution,
the scaling cosmic string network. A set of diverse inflationary scenarios in string theory may lead to the scaling networks of interest.

\subsection{Low Inter-commutation Probability}

Cosmic superstrings have different properties than vortices in the
Abelian Higgs model. The inter-commutation probability of vortices in
three dimensions approaches $p \simeq 1$. The string density
in the scaling solution is often estimated from numerical simulations
with an assumed or effective value $p=1$.
The situation is more complicated for superstrings in many respects.

First, $p \simeq 0$ for a pair of interacting strings from different warped throats. 
A string network in each throat
evolves and contributes separately to the total density. We will discuss
the number of throats in the following section.

Second, within a single throat $p<1$ because the physics of collisions
is more complicated than it is
for the Abelian case. It depends on the relative speed and angle of
the 2 interacting string segments among other things. From
calculations \cite{Jackson:2004zg} we estimate $p \sim g_{s}^{2}$ and
take string coupling $g_{s} \sim 1/10$ as not unreasonable.
When $p<1$ the chopping of long strings into loops is less efficient.
This is the
superstring case. The overall string density must increase to compensate and to
realize the scaling solution but the precise variation
is not well-understood. The one scale model suggests density
$\rho \propto 1/p^2$ but small scale structure
on the string raises the effective intercommutation probability when
two long segments collide. Simulations \cite{Avgoustidis:2005nv}
suggest that the density $\rho \propto {1}/{p^{2/3}}$.

Third, cosmic superstrings in a single throat will be present with a
variety of tensions and charges \cite{Tye:2005fn}. The effective
number of independent types per throat $N_s$ is not well understood in this
context. It is unclear how the presence of
beads (i.e., baryons) in the tension spectrum %
will impact the evolution of the string network.  The network may contain
multiple beads, so-called necklaces \cite{Siemens:2000ty,Shlaer:2005ry}.

Let us write the scaling from the density of Nambu-Goto strings to superstrings
in a single throat as
$$\Omega_{string} \rightarrow \Omega_{superstring} \simeq
\frac{N_s}{p} \Omega_{s} \sim \frac{N_s}{g_{s}^{2}} \Gamma G \mu$$
where $\mu$ is the $F$-string tension, $N_s$ is the effective number
of non-interacting types of strings and bound states in a throat,
e.g. $N_s \sim 1$ to $4$.  There are significant uncertainties in evaluating
the enhancement in terms of $p$ and $N_s$.

\subsection{Multi-throats}

As discussed earlier, a typical 6-dimensional manifold has multiple throats.  Assuming there are 2 throats along each dimension, we have $2^6=64$ throats while 3 along each dimension yields $3^6=729$ throats, so it is not hard to imagine that a typical manifold has many throats. For example, one of the best studied manifold ${\bf CP}^4_{11169}$ has, in the absence of any specific symmetry imposed, as many as 272 throats \cite{Candelas:1994hw}. Denote the number of throats by $N_T$.

The annihilation in the inflationary throat heats the entire manifold.
The heating may drive the birth of scaling string networks in the
subset of throats which possess greater degrees of warping. (The last
epoch of inflation will have diluted away all networks sourced by
previous annihilation events.) In general, each throat has its own
geometry, warp factor and set of string tensions. For example, since
only $D$-strings survive in the S throat, and Eq.(\ref{Tanswer}) shows
there is no binding energy for multiple $D$-strings, we expect only
one tension in the S throat, the minimal number.  The tension spectra
of other throats will be at least as complicated.  The multiplicity of
throats, the range of warping and the possible complexity of the
spectra in each throat is the source of the generic expectation that there
exist a wide range of string tensions for future experiments to
target.

If there are more $\bar D$3-branes than $D$3-branes in a throat, then some number of $\bar D$3-branes will be left behind there after all pairs have annihilated. Let us consider the dynamics of $p$ $\bar D$3-branes inside a KS geometry, the deformed conifold with $M$ units of RR 3-form flux around the 3-sphere.
If the number $p$ of $\bar D$3-branes left is not too small compared to $M$, then the system will roll to a nearby supersymmetric vacuum with $M-p$ number of $D$3-branes sitting at the bottom of the throat. This happens via the nucleation of an NS 5-brane bubble wall \cite{Kachru:2002gs}.  This decreases $K$ by one unit, so the warp factor  goes from $h= e^{-2\pi K/g_sM}$ to  $h= e^{-2\pi (K-1)/g_sM}$, that is, it is less warped.
 If $p \ll M$, then the system is classically stable, but it may decay later via quantum tunneling again via the 
brane-flux annihilation.  If this has happened already, a new cosmic superstring network might have been produced relatively late.

\subsection{Cosmic Strings in an Orientifold}

$F$1-strings are charged under the Neveu-Schwarz (NS) $B_{\mu \nu}$ field ($B_2$, with same strength as gravity) while the $D$1-strings are charged under the Ramond-Ramond (RR) field $C_{\mu \nu}$ ($C_2$) with a definite $D$1-charge. Since $C_{\mu \nu}$ (or $B_{\mu \nu}$) is a massless anti-symmetric tensor field, we can introduce an axion field $a$ related to it via the field strength $F_3$,
$F_{\alpha \mu \nu} = \partial_{ [ \alpha} C_{\mu \nu ]} = \epsilon_{\alpha \mu \nu \beta} \partial^{\beta} a$. 
The massless tensor field has only one degree of freedom in 4-dimensional spacetime. Since $F_{\alpha \mu \nu}$ is invariant under a gauge transformation $C_{\mu \nu} \rightarrow C_{\mu \nu}  + \partial_{[\mu} A_{\nu]}$, we infer that the massless $a$ has a shift symmetry, $a \rightarrow a+ {\rm constant}$. 

So cosmic superstring loops can emit axions as well as gravitons \cite{Firouzjahi:2007dp}. However, 
in a more realistic orientifold construction, both $C_2$ and $B_2$ are projected out \cite{Copeland:2003bj,Gwyn:2011tf}. Pictorially, the orientifold projection reverses the orientation of a $D$1-string, i.e., turns it to a $\bar D$1-string, so the $D$1-string effectively becomes a $D$1-$\bar D$1 bound state, which is unstable. However, the $D$1-string inside a warped throat is far separated from the $\bar D$1-string in the image throat, so the decay time is expected to be much longer than the age of the universe. That is, they are expected to be cosmologically stable. 

Furthermore, in any flux compactification of orientifolds, there are multiple complex structure moduli as well as K\"ahler moduli. As a result, we expect multiple axions to be present. Since a 2-form field is dual to an axion, one expects there are strings charged under each axion. What are these strings? Are they additional strings beyond the $D$1- and $F$1-strings? Since at least one axion is associated with each throat, one is led to entertain the possibility that the $D$1- and $F$1-strings inside a throat are charged under the corresponding axions associated with that throat.   

So we expect the radiation of light axions as well as gravitons by any string in any throat.
For a cosmic $D$1-string with an observable tension $\mu_j$ at the bottom of the $j$th throat, we expect the coupling interaction takes the form 
$$ S \sim  \int \left[ \frac{\mu_j}{g_s} g_{\mu \nu} + b_j\mu_j C_{\mu \nu} \right] d \sigma^{\mu \nu}$$
where $g_{\mu \nu}$ is the 4-dimensional metric, $b_j$ an order unity parameter and $C_{\mu \nu}$ is now the dual of the relevant axion while the string is described by
$d\sigma^{\mu \nu} = ({\dot x}^{\mu} x'^{\nu}-{\dot x}^{\nu} x'^{\mu})d\tau d\sigma$,
where the dot and the prime indicate derivatives with the world sheet variables. We shall define $N_T$ to be the number of throats in which the strings decaying via axions do not overwhelm its gravitational wave emission.

\subsection{Domain Walls Bounded by Closed Strings}

In a more realistic scenario, an axion will have a mass. There are 2 ways it can pick up a mass:

(1) If we identify the above $A_{\mu}$ in the gauge transformation of
$C_{\mu \nu}$ as a massless gauge field, we see that $C_{\mu\nu}$ can
become massive by absorbing $A_\mu$. 
This is like the standard Higgs mechanism in which a
gauge field $A_{\mu}$ becomes massive by absorbing the massless
``axion'' in spontaneous symmetry breaking.

(2) Non-perturbative (instanton) effects typically generate a
potential term of the form $V(a) \simeq -M_S^4e^{-S_{inst}} \cos a$,
which breaks the shift symmetry of $a$ to a discrete symmetry. The effect
is typically exponentially small. It is more convenient to rewrite as
$$V(\phi) \simeq m_a^2(f/M)^2 \left(1 - \cos \left(\frac{M \phi}{f}
\right) \right) \rightarrow \frac{m_a^2}{2}\phi^2+ . . .$$ where
$\phi$ is the axion with a canonical kinetic term,
$f$ is the axion decay constant or its coupling
parameter and $M$ is the integer related to the $Z_M$ symmetry for the
$F$-string (\ref{KMR}).

For a potential of the above form with $M>1$, a closed string loop can become the
boundary of a domain wall, or membrane. The tension of the membrane is of order
$$ \sigma \sim m_a f^2 .$$
In general, an axion mass is hardly restricted; it can be as heavy as some standard model particles or as light as  $10^{-33}$ eV.
One intriguing possibility is that this axion can contribute substantially to the
dark matter of the universe as fuzzy dark matter \cite{Hu:2000ke,Hui:2016ltb}.
If so, its contribution to the energy
density is roughly given by $\rho_{a} = m_a^2f^2$ while its mass is estimated 
to be $m\simeq 10^{-22}$ eV $\simeq 10^{-33} M_P$.
Hence, $\rho_{a}=m_a^2f^2 \simeq 10^{-118} M_{Pl}^4$ 
$f \simeq 10^{-10} M_{Pl}$ and $\sigma \simeq 10^{-69}
M^3_{Pl} \simeq 10^{-14}$ GeV$^3$.

On simple energetic grounds the membrane tension dominates the cosmic string
tension for large loops, i.e. when loop of size $r$ satisfies $r>2
\mu/\sigma$. Write $\mu = (\Lambda/M_{Pl})^2$ for string energy scale
$\Lambda$, adopt and fix the membrane parameters above and take the
loop size equal to the size of the universe today $r \sim 4.2$
Gpc. The membrane energy dominates if the string tension is less than
a critical size: $\Lambda < \Lambda_c$ with $\Lambda_c/M_{Pl} = 2.5 \times
10^{-5}$, or string energy scale $\Lambda_c<3.1 \times 10^{14}$ GeV.
Observationally, however, the string loops of greatest interest today
are much smaller than the horizon scale today.  Their size is
set by the condition they can just evaporate in the age of the
universe. Assuming gravitational radiation determines the rate of
evaporation the loop size today is $\ell = \Gamma G \mu t_0$ and the
condition $\ell > 2 \mu/\sigma$ is independent of $\mu$. For such loops
the string tension dominates over membrane tension at any epoch such
that $t < 2/(G \Gamma \sigma) \sim 2 \times 10^8 t_0$.

\subsection{Varying Tension}

So far, we have been assuming that cosmic superstrings sit at the bottoms of the throats. In general, they can move around the bottoms. Because of the deformation of a throat (from a conifold), the bottom of a throat is at $r=r_i$, which is small but not zero. For a Klebanov-Strassler (deformed) throat \cite{Klebanov:2000hb}, the bottom is $S^3$, so a cosmic superstring at the bottom of a throat can move around. In fact, it may oscillate \cite{Avgoustidis:2007ju,Avgoustidis:2012vc}and at times move to $r > r_i$. Observationally, the tension of an upward displaced piece of the string would appear to be larger, since the local warp factor $h(r)=r/R$ is bigger (i.e., closer to the bulk). 
Tension varying along a string and/or in time is a direct consequence of the extra dimensions and warped geometry. Observation of such a behavior can be very informative.

\vspace{3mm}

\subsection{Comparing Cosmic Superstring Density to Cosmic String Density}

Suppose the typical mass scale of our standard model throat (S throat) is of order of the electroweak (or supersymmetry breaking ) scale, i.e., TeV scale. The CMBR observations (see Eq.(\ref{Gtension}) implies the inflation throat (A throat) has a much higher scale $\sqrt{\mu} \simeq 10^{14}$ GeV. The energy released from the $D$3-${\bar D}$3-bane annihilation will be able to heat up our universe (i.e., our branes) \cite{Chen:2006ni,Chialva:2005zy}. In addition to the S and A throats, consider another throat C with a warped factor $h_{C}$. Let the reheating (RH) temperature at the beginning of the hot big bang be $T_{RH} < \sqrt{\mu}$. We have argued that strings in the $C$-throat will be produced if $T_{RH}>h_{C}M_{s}$. Hence, in addition to cosmic strings in the $A$-throat, we expect small tension cosmic strings will appear in throats with large warping. These light cosmic strings interact very weakly with cosmic strings in the $A$-throat. On the other hand, if $T_{RH}  < h_{C}M_{s}$ string production will be suppressed by a Boltzmann factor. When the number of cosmic strings produced is less than one per horizon it may still be possible to reach the scaling solution if the string loop decay rate is much smaller than the expansion rate and if there are sufficient long (superhorizon) strings present. The onset of the scaling of the cosmic string network is delayed.

Beads (or baryons) on cosmic superstrings typically move at similar speeds as strings themselves, since they are being dragged along by the motions of the strings. One expects that the beads may merge or annihilate each other along the strings while junctions are being created and removed. Numerical
investigations for necklaces \cite{Siemens:2000ty} indicate that string loops with many beads tend
to have periodic self-intersecting solutions, so string loops may quickly chop themselves up into
smaller and smaller loops, some of which will be free of beads/baryons.
As a result, the superstring network may end up with smaller loops and hence the pulsar timing bounds on string tension should be relaxed somewhat \cite{Hindmarsh:2016dha}. Since cosmic superstrings with junctions and baryons are more involved than simple necklaces, a detailed study is important to pin down the loops sizes and the effective bound from the pulsar timing data. 

Let us summarize here. The number density of cosmic strings in the universe, when compared to the Nambu-Goto model or the Abelian Higgs model, is enhanced by 3 factors: the decreased intercommutation probability $p$, the effective number $N_s$ of string species in each throat, and the number of throats $N_T$ each of which has an independent scaling string network with fundamental tension $\mu_j$  ($j=1,2,...,N_T$) and subdominant axion emission. The overall enhancement is
\be
{\cal G}=\frac{N_s}{p \mu} \sum_j \mu_J \simeq N_sN_T/p \gg 1, \quad \quad \Omega_{superstring} \sim {\cal G}\Gamma G \mu
\ee
where $\mu$ is some average tension.
Based on the above contributions ${\cal G}$ can easily be as big as ${\cal G} \sim 10^4$, with a distribution in tensions that are roughly bounded by $G\mu \lesssim 2 \times 10^{-10}$.  The tension in our S throat might be as small as $G\mu \sim 10^{-30}$ (i.e., TeV scale). On theoretical grounds
it might be as high as GUT scale but observationally the highest tension in any throat should not exceed $G \mu \simeq 2 \times 10^{-10}$.
If strings are created at energy scales below $T_{RH}$ it is easy to imagine scenarios where there are dozens of throats with separate scaling superstring networks. 

In estimating the probability of detectability, and for the sake of simplicity, we gather all differences of cosmic superstrings from ordinary cosmic strings into a single scaling parameter ${\cal G}$. At times, we take ${\cal G} = 10^2$ as the canonical value for a fixed $F$-string tension.
It is clear that further studies, the properties of cosmic string spectrum 
(including baryons), their productions, stabilities and interactions, and the cosmic evolution of the network as well as their possible detections will be most interesting. It is reasonable to be optimistic about the detectability of cosmic superstrings, but this is far from guaranteed. 

There are other inflationary scenarios in string theory, mostly with the inflaton as a closed string mode, in contrast to brane inflation, in which the inflaton is an open string mode. Although the reheating process has not yet been carefully studied, energy released towards the end of inflation is expected to go to closed strings directly, so the production of some cosmic superstrings may be expected. 

If any throat still contains a few $\bar D$3-branes today, the system would have relaxed to a non-supersymmetric NS 5-brane ``giant graviton'' configuration; that is, these $\bar D$3-branes can provide the uplift of our universe from a supersymmetric Anti-deSitter space to a non-supersymmetric deSitter space with a small positive cosmological constant. If so, our universe today is classically stable but not fully stable and will decay at some point in the future.

\section{Possible Detections of Cosmic Superstrings}

In the braneworld scenario, there are many warped throats in the
Calabi-Yau manifold, one of which must contain the standard model
branes but all of which may contain cosmic superstrings. Strings
in the standard model throat are limited to D-strings
thickened in a stack of D3 branes (or D7 branes wrapping 4-cycles) \cite{Leblond:2004uc}.
Generically the throats have different warp factors, so the string tensions 
span a range of values.
It is noteworthy that the strings in
these other throats may dominate the string content of the
universe. Throats without D3 branes may
harbor a spectra of bound states of F- and D-strings. Each
throat contains its own scaling string network. We have subsumed
all these effects in the detectability parameter ${\cal G} \sim 10^2$.

How can all these strings be detected?  Let us mention 3 possibilities, starting 
with the least promising one first.

{\bf Fast radio bursts:} Fast radio bursts have been observed at cosmological distances with some repetitions but no evidence for periodicity thus far \cite{Lorimer:2007qn,Thornton:2013iua}. Among other astrophysical possibilities, cosmic strings have been suggested as a possible source of such bursts.

Strings may carry charges and
interact with fields present within the throats they occupy. For
example, superconducting cosmic strings can carry currents and interact with
electromagnetic fields
\cite{Vachaspati:2008su,Cai:2011bi,Zadorozhna:2009zza,Cai:2012zd,Yu:2014gea,Ye:2017lqn}. To
be able to emit standard model photons, such strings must sit in the
same throat as the standard model particles. If string segments annihilate
in the standard model throat, they can generate particles/fields
belonging to the standard model. Historically, there have been many
proposals to explain cosmic rays, neutrinos and gamma ray bursts in
this manner \cite{Berezinsky:2001cp,Gruzinov:2016hqs}.  
In general, cosmic superstrings are not superconducting but these
considerations are important.

A string in string theory is charged under a specific 2-form field,
i.e., an axion field in 4 spatial dimensions. Strings are universally
coupled to gravity and specifically to the axions under which they are
charged. Cusps on strings have long been identified as sources for
gravitational wave bursts.  A cusp is a bit of string that momentarily
approaches the speed of light. In doing so a small region of the string
doubles back on itself for a short period of time.  Since it is
charged under an axionic field, it behaves like a string-anti-string
pair, completely unstable to annihilation and decay via
axionic and gravitational wave bursts. In essence, gravitational and axionic beams emerge from the tip
simultaneously when it is moving close to the speed of light. The
production of axions is similar to that of gravitational waves (in
terms of beaming, periodicity, etc.).  Both would appear as bursts
with the same characteristic time-dependence.

Assuming generic mixing a light axion (a closed string mode) produced
in any throat may decay in the standard model throat to give two
photons. In fact, no other standard model particle products are
possible. An observer in the standard model throat may hope to detect
not only gravitational waves but also photons from the cusp. Although the
gravitational waves bursts are expected to be beam-like, the photons
that result from the decay of the axion bursts will have larger
angular spread, giving rise to diffuse radiation (the photons still
suffer the relativistic headlight effect).  However, when the axion
beam passes through a magnetic field the Primakoff effect can take
place, due to the coupling $ \propto a{\bf E \cdot B}$, converting axion $a$
to a photon. Since the inter-galactic magnetic field
carries little momentum, the momentum of the axion is largely carried
by the photon produced, so this stimulated axion decay yields a
beam of photons, in roughly the same direction as the axionic beam. 
This might be the origin of some of the fast radio
bursts observed. 

To test the idea we suggest a study of the
correlations of fast radio bursts with gravitational wave
observations. Such a study is practical because the angular direction
to certain fast radio burst sources are precisely known. If strings are
responsible the radio bursts and gravitational wave emission will be
correlated in both space and time.  Ref. \cite{Brandenberger:2017uwo} considered a
somewhat different physical picture in which the cosmic string cusps
decay directly to produce radio signals. Such a string must sit in the
same throat as the standard model just like the superconducting cosmic
strings that emit standard model photons. Both the axion-mediated
and direct, standard model bursts from cusps would give similar
space and time correlations of radio and gravitational wave
bursts. Strings that can decay directly in the standard model throat
will have intrinsically shorter lifetimes and are likely to represent 
a small subset of all the superstrings in the Calabi-Yau manifold.

{\bf Microlensing:}
The usual pictorial description of cosmic string lensing in 3+1 begins
with a deficit angle in the geometry of the disk perpendicular to the
string. When source, string and observer are all nearly aligned there
exist two straight line paths from source to observer that
circumnavigate the string in opposite senses, clockwise and
counterclockwise. This leads to double images, i.e., cosmic string lensing.
When the string tension is high enough ($G \mu \sim 10^{-7}$), lensing 
of galaxies is possible, i.e., double images of the galaxy. For low tensions, 
the deficit angle is small, so point-like lensing is possible only for objects 
of small angular size like stars. For a typical distant star, we cannot resolve 
the double images, but only observe a doubling of flux, i.e., microlensing.
Microlensing of stars have been discussed in Ref. \cite{Chernoff:2007pd,Chernoff:2014cba}.

If the string lies in another throat no standard
model photons can ``circumnavigate'' the string in the sense that the above 
simple picture implies. Nonetheless we show in Appendix \ref{stringlensing} how
the geometry bends the photon path from a source in the S throat to reach the observer in
the S throat when the string lies in another throat.  This issue is
important because the most sensitive tool for direct detection of low
tension strings may turn out to be microlensing, a variant of normal
lensing in which the observer measures flux changes without resolving
the lensed images. String microlensing has been studied in some detail
\cite{Bloomfield:2013jka}.

{\bf Gravitational wave bursts:}
Gravitational interactions are the traditional means of detecting the
presence of minimally coupled cosmic strings. String-sourced plasma
perturbations may be imprinted on the CMB power spectrum.  Strings may
create resolved images of background galaxies by lensing. These are
examples of the direct consequences of a gravitational interaction. A
somewhat less direct method of probing the string content is to
measure the light element yield of big bang nucleosynthesis since
extra mass energy alters the cosmological expansion rate. Finally, one
can hope to measure the gravitational emission in the form of bursts
and a stochastic background with pulsar timing arrays and LIGO.  For a
recent review see \cite{Chernoff:2014cba}. In this paper, we present a more 
detailed analysis of the rate of gravitational wave bursts expected from cosmic superstrings 
for  LIGO/VIRGO and LISA.

In summary, the hunt for superstrings will be based on both
gravitational and axionic degrees of freedom because they are
sensitive to the string content of all the throats in a Calabi-Yau
manifold.

\section{Models for cosmologically-generated loops}

To calculate the expected rate of gravitational wave bursts,
stellar microlensing occurrences or two photon decays of emitted
axions requires
a model for the string loop sources. We begin by describing
the number of strings in the universe as a whole including the
distribution of loop sizes. We utilize results for the dynamical
motion of strings in growing matter perturbations to estimate the
concentration of strings within our own Galaxy and on larger scales.

Building a model makes explicit the dependence of
the demography of the string population on microscopic parameters like
string tension, number of throats, number of effective string species
and probability of intercommutation even if they are not precisely
known. We include as appendices a detailed description of the
model and describe the context and most important consequences below.

\subsection{Two Loop Sizes}

In Kibble's description of the network \cite{Kibble:1984hp} long
strings are stretched by the universe's expansion and
intercommutations chop out loops which ultimately evaporate. With
stretching, chopping and evaporation the scaling solution is an
attractor and all the macroscopic cosmic strings properties (length of
string per volume, correlation length, etc.) appear to
scale \cite{Kibble:1976sj}. Virtually all analytic descriptions of network evolution for
traditional cosmic strings begin with these processes. Cross sections,
rate coefficients and efficiencies have generally been derived from
simulations in which a realization of the network is followed in a
large enough spacetime volume to infer its statistical properties.
Luckily, such simulations rapidly enter the scaling regime so that the
macroscopic properties can be established. However, important
differences amongst various simulations have been observed, especially
regarding the small structures on the network's long strings and the
size of the loops formed from such strings. Since the loops provide
the main observational diagnostic for superstrings, this is an
important point and a consensus has only recently emerged.

Some early simulations generated only tiny loops at the smallest
available grid scale
\cite{Bennett:1987vf,Bennett:1989ak,Bennett:1990uza,Allen:1990tv}
while others \cite{Albrecht:1984xv,Albrecht:1989mk} found the network
created predominantly large loops with sizes within a few orders of
magnitude of the scale of horizon. Small sized features on long strings
are expected to damp by gravitational radiation so that there is a
natural physical cutoff but all simulations omit the direct
calculation of the gravitational backreaction. It may be reasonable to
imagine the grid scale cutoff plays a similar, dissipative role.  Why
small loops should predominate in some simulations and not others was
unclear.  Grid-based and discrete numerical simulations of
cosmological string dynamics are generally expected to treat large
scales easily and accurately so why only some simulations led to large
loops was also perplexing.

In fact, intercommutations generate substructure on the long strings
so the dynamics at the horizon scale cannot be separated
from that on small scales \cite{Austin:1993rg}. Recent work
finds the string substructure has a fractal character which
influences dissipation and small loop formation
\cite{Polchinski:2006ee,Dubath:2007mf,Polchinski:2007rg,Polchinski:2007qc}.  The
current understanding is that string loops of two characteristic scales
are generated in a scaling cosmological network during epochs of
powerlaw expansion. Roughly 5-20\% of the string invariant length
(invariant length equals the total energy per spatial increment divided by
tension) that is chopped out of the expanding, horizon-crossing
strings finds its way into large loops where ``large'' means the size
at time $t$ is roughly $\ell/t \sim 10^{-4}-10^{-1}$
\cite{Martins:2005es,Ringeval:2005kr,Vanchurin:2005pa,Olum:2006ix,BlancoPillado:2010sy,BlancoPillado:2011dq,Blanco-Pillado:2013qja}.  In other words,
large loop are comparable to the size of the horizon at formation. The remaining
part of the excised invariant string length yields very small loops
which move relativistically and evaporate in less than a
characteristic expansion time ($\ell \propto (G \mu)^{1.2}$ and $(G
\mu)^{1.5}$ or $H \tau \sim (G \mu)^{0.3}$ and $(G \mu)^{0.13}$ for
radiation era and matter era, respectively) \cite{Polchinski:2007qc}.

The mechanism for the production of the small loops today is
intimately tied to the small scale structure introduced by
intercommutation on horizon scales at earlier epochs. When a loop is
chopped out of smooth string (with continuous paths on the tangent
sphere for right and left moving modes) the loop configuration
typically contains a transient cusp in which two nearly parallel
string segments approach each other and then recede (this occurs each
period of loop oscillation). If the loop forms from horizon crossing
string that is not smooth then the first approach to the cusp
configuration results in intercommutation, explosive sub-loop
formation and excision of the cusp from the
loop. Ref. \cite{Polchinski:2006ee} showed on analytic grounds that
after many e-folds of expansion the large horizon crossing loops are
replete with small scale structure. This mechanism and
ultimately gravitational backreaction dispose of the bulk of
the string invariant length. The important question is how much of the
network escapes this fate and what are the properties of those loops?
Numerical simulations have not yet been able to follow the buildup and
scaling of the small scale structure on long strings because its
generation takes much longer than the macroscopic measures usually
used to judge whether a simulation has reached the scaling regime but
the properties of the larger loops have now converged and the
uncertainties that remain due to the small loop effects are
subdominant\cite{Blanco-Pillado:2013qja}.

To summarize the essentials: the model, that used in our previous
analyses \cite{Chernoff:2009tp}, assumes the fraction $f \sim 0.05 -
0.2$ of the invariant length goes into large loops of size $\ell =
\alpha t$ at time $t$ for $\alpha \sim 0.1$. For comparison,
simulations for the radiative era \cite{Blanco-Pillado:2013qja} imply
$\alpha \sim 0.1$ (their $\alpha$ is written in terms of the horizon
scale $2t$) and $f \sim 0.1$. The remaining fraction $1-f$ goes into
small loops of invariant size $\ell \sim (G \mu)^{1.2}$ moving
relativistically (radiation era).  The results are similar to model
$M=2$ used for constraining cosmic strings from the first Advanced
LIGO/VIRGO observing run \cite{Abbott:2017mem}.  The large loop
distribution is effectively established on horizon scales utilizing a
small fraction $f$ of the excised network.  In our model after a large
loop forms it evolves independently, shrinking in size; it is the
primary object of interest.  The rest of the excised network forms
small, short-lived loops (suddenly, as soon as a cusp appears) at roughly the
same epoch as fraction $f$ makes large loops.
More detailed descriptions of the evolution of
small loops (e.g.  $M=2$ based on \cite{Blanco-Pillado:2013qja} and
$M=3$ based on \cite{Lorenz:2010sm,Ringeval:2017eww})
combine simulation results and
theoretical models. These models give different small loop distributions;
one extrapolates simulation results ($M=2$) and the other fits a
theoretical model for ongoing (not explosive) loop formation ($M=3$).

\subsection{Velocity One Scale Model and Loop Density}

Let $V$ be physical volume and $E$ be the energy of a network of long
(horizon-scale) strings of tension $\mu$. Let $L$ be the length such
that there is 1 string of invariant length $L$ in volume $V=L^3$
(loops are not included in $L$). The physical energy density is
$\rho_\infty = E/V = \mu L/V = \mu/L^2$. From the encounter rate
of strings with other strings and intercommutation probability $p$
we deduce the expected energy
transformed from network to loops (loop formation) and vice-verse
(reconnection). The newly formed loop size distribution is assumed to
scale with the horizon size. We account for how the energy in the
network's long strings is increased by stretching, lost by formation
of loops and gained by reconnection and conversely how the energy in
the loop population is altered (loops are assumed small compared to
the horizon and have negligible stretching). This is just Kibble's
original network model
\cite{Kibble:1984hp,Bennett:1985qt,Bennett:1986zn}
with the addition of $p \ne 1$ and omitting the reconnection
terms which may be shown to be small.
The Velocity One Scale (VOS) model
\cite{Martins:1995tg,Martins:2000cs,Battye:1997hu,Pogosian:1999np}
is supplemented with simulation-determined fits to describe chopping and
velocity.

Write $L = \gamma t$. In exact scaling $\gamma$ is constant but
we regard it and other quantities like it
as slowly varying. The summary of the model is
\ba
\rho_\infty & = & \frac{\mu}{ \gamma^2 t^2} \\
\frac{t}{\gamma}\frac{d\gamma}{dt} & = & -1 + Ht \left( 1 + v^2 \right) + \frac{C(t) p v}{2 \gamma}
\ea
where $H$ is the Hubble constant and $C(t)$ is chopping efficiency
parameter fit from numerical simulations. The equation for the velocity $v$
\cite{Martins:1995tg} is
\ba
\frac{dv}{dt} & = & \left( 1 - v^2 \right) H \left( \frac{k(v)}{Ht \gamma} - 2 v \right) \\
k(v) & = & \frac{2 \sqrt{2}}{\pi} \frac{1 - 8 v^6}{1 + 8 v^6}
\ea
where $k(v)$ is a fit. When $Ht$, $C(t)$ and $v(t)$ are constant in time
this gives the exact scaling solution. We treat these
as two coupled ODEs for $\gamma$ and $v$ to be solved numerically.
We begin at large $z$ when $Ht=1/2$ setting the left hand sides to
zero ($d/dt \to 0$) and find an equilibrium point for $\gamma$ and
$v$.  As $z$ decreases the system begins to evolve because $C$ and $H t$ vary.
We infer $H t$ from the multicomponent $\Lambda$-CDM
cosmology.

The rate at which energy is transformed to loops ${\dot E}_{\ell}$ is
\be
\frac{\dot E_{\ell}}{a^3} = C(t) \rho_\infty \frac{pv}{L} = C(t) \mu \frac{p v}{\gamma^3 t^3}
\ee
where $a$ is the scale factor. We integrate this using
the solutions for $\gamma$ and $v$ from the VOS expressed
as a single modestly varying function ${\cal A}$ so that
\be
\frac{\dot E_{\ell}}{a^3} = \frac{\mu}{p^2 t^3} {\cal A} .
\ee

\subsection{Loop Size Distribution Born of Large Loops}
We assume that the large loops formed at a given time $t$ have size $\alpha t$ and consume a fraction $f$ of the
invariant length (energy) being chopped out of the network. Large
loops are non-relativistic with comparable geometric and invariant lengths. The birth rate density for loops of size $\ell$ at time $t$ is
\be
\label{dndtdl}
\frac{d n_\ell}{dt d\ell} =
\frac{f {\cal A}}{\alpha p^2 t^4} \delta(\ell - \alpha t) .
\ee
The loops evaporate by gravitational and axionic emission at
constant rate. The length at $t$ for a loop born at $t_b$ with size $\ell_b$ is
\be
\label{looplengthatt}
\ell[\ell_b, t_b, t] = \ell_b -  \Gamma G \mu (t - t_b)
\ee
Integrating over birth times and sizes gives the
differential loop size density 
\ba
\frac{dn}{d\ell} & = & \frac{ {\cal A}_b f \alpha^2 }{p^2} \left( \frac{a_b}{a} \right)^3
\frac{ \Phi^3 }{(\ell +  \Gamma G \mu t)^4} \\
t_b & = & \frac{ \ell +  \Gamma G \mu t}{ \alpha \Phi } \\
\Phi & = & 1 + \frac{ \Gamma G \mu}{ \alpha }
\ea
for $t_b < t$ and $\ell < \alpha t$. The quantities
${\cal A}_b$ and $a_b$ are evaluated at $t=t_b$. The form for
$dn/d\ell$ peaks at $\ell=0$ but the quantities of greatest
observational interest are weighted by $\ell$ or higher powers.
The characteristic dissipative scale for the large loops is
$\ell_d =  \Gamma G \mu t$.

If $G \mu < 7 \times 10^{-9} (\alpha/0.1)(50/\Gamma)$ the loops
near $\ell_d$ today were born before equipartition, $t_{eq}$. For a simple numerical estimate today
near the dominant small end of size spectrum
in the radiative era $t<t_{eq}$ we write
\ba
\frac{a_b}{a} & = & \left( \frac{a_b}{a_{eq}}\right) \left( \frac{a_{eq}}{a} \right) \\
& \simeq & \left( \frac{t_b}{t_{eq}}\right)^{1/2}
\left( \frac{t_{eq}}{t} \right)^{2/3}
\ea
and using $x \equiv \ell/\ell_d$ we have
\ba
\ell \frac{dn}{d\ell} & = & \frac{x}{(1+x)^{5/2}}
\left( \frac{ {\cal A} {\it f} } {p^2} \right)
\left( {\Gamma G \mu} \right)^{-3/2}
\left( \frac{\alpha t_{eq}}{t_0} \right)^{1/2}
\left( \frac{1}{t_0} \right)^3
\label{approxdndl}
\ea
In the radiative era
${\cal A}_b \sim 7.68$ is close to constant;
write the other string-related parameters
$p=1$, $f=0.2 f_{0.2}$, $\alpha=0.1 \alpha_{0.1}$ and $\Gamma=50 \Gamma_{50}$;
and from $\Lambda$CDM $t_{eq}=4.7 \times 10^4$ yr and
$t_0=4.25 \times 10^{17}$ s. These give a baseline
result that applies to cosmic strings
\ba
\ell \frac{dn}{d\ell} & = &1.15 \times 10^{-6} \ \ \frac{x}{(1+x)^{5/2}}
\left(\Gamma_{50} \mu_{-13}\right)^{-3/2} f_{0.2} \alpha_{0.1}^{1/2} \ \ {\rm kpc}^{-3}\\
x & = & \frac{\ell}{\ell_d} \\
\mu_{-13} & = & \left( \frac{ G \mu}{10^{-13}} \right)
\ea

The characteristic length and mass of the loops just evaporating
are
\ba
\ell_d & = & 0.0206 \ \Gamma_{50}  \mu_{-13} \ \ {\rm pc} \\
M_{\ell_d} & = & 0.043 \ \Gamma_{50} \mu_{-13}^2 \ \ \msun 
\ea
Numerical results for mass densities based on the approximate forms are
\ba
\frac{d \rho}{d\ell} & = & \mu \ell \frac{dn}{d\ell} \\
& = & 
2.41 \times 10^{-6} \ \ \frac{x}{(1+x)^{5/2}} \left(\Gamma_{50} \right)^{-3/2} \mu_{-13}^{-1/2} f_{0.2} \alpha_{0.1}^{1/2} \ \ \msun {\rm kpc}^{-3} {\rm pc}^{-1} \\
\frac{d \rho}{d\log M_\ell} & = & \ell \frac{d\rho }{d\ell} = \mu \ell^2 \frac{dn}{d\ell} \\
& = & 
4.98 \times 10^{-8} \ \ \frac{x^2}{(1+x)^{5/2}} \left(\Gamma_{50} \right)^{-1/2} \mu_{-13}^{1/2}  f_{0.2} \alpha_{0.1}^{1/2}\ \ \msun {\rm kpc}^{-3} \\
\frac{d \Omega_{\ell,0}}{d \log M_\ell} & = & \frac{1}{\rho_{cr,0}} \frac{d \rho}{d\log M_\ell} \\
& = & 
3.66 \times 10^{-10} \ \ \frac{x^2}{(1+x)^{5/2}} \left(\Gamma_{50} \right)^{-1/2} \mu_{-13}^{1/2} f_{0.2} \alpha_{0.1}^{1/2}
\ea

\begin{figure}[ht]
\centering
\includegraphics{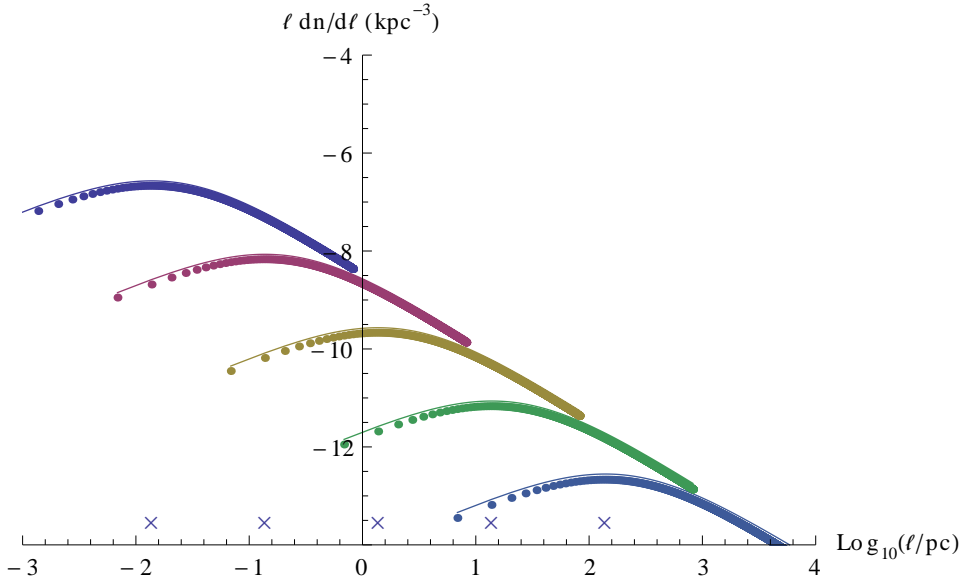}
\caption{\label{dndlogl} The size distribution of loops today for a range of string tensions $ G\mu = 10^{-13}$ to $10^{-9}$ at the current epoch $t=t_0$. Solid lines are exact; dotted lines are approximate; ``x'' marks the expected peak $x=2/3$. 
The density is the average, homogeneous density in the universe without
clustering; intercommutation $p=1$, fraction of large loops formed $f=0.2$ and scale of large loop size $\alpha=0.1$.}
\end{figure}

In Fig. \ref{dndlogl} the solid lines show
$\ell dn/d\ell$ today ($t=t_0$) for $G \mu = 10^{-13}$ to
$10^{-9}$ in powers of 10; each peak is near $\ell = (2/3) \ell_d$.
The approximate results are denoted with the ``dots'' which are quite
close to the exact evaluation given by smooth lines.
Note that both sets of lines have peaks near $x=2/3$.

We will denote all these results as ``baseline'' -- they
apply to one species of normal cosmic strings with
intercommutation probability $p=1$, spatially
averaged throughout the universe.

Modifications to the baseline that originate in the differences
between superstring and field theory are lumped into
a common factor ${\cal G}$ including the
reduced intercommutation probability of superstrings as follows
\be
\left( \frac{dn}{d\ell} \right)_{homog} = {\cal G} \left( \frac{dn}{d\ell} \right)_{baseline} .
\ee
The homogeneous, cosmologically averaged, superstrings have loop densities
that exceed the baseline densities by the factor $1 < {\cal G} < 10^4$.

\subsection{String loop clustering}
If a loop is formed at time $t$ with length $\ell =
\alpha t$ then its evaporation time $\tau =
\ell/\Gamma G \mu$. For Hubble constant $H$ at $t$ the
dimensionless combination $H \tau = \alpha/(\Gamma G \mu)$
is a measure of lifetime in terms of the universe's age.
Superstring loops with moderate $\alpha$ and very small $\Gamma G \mu$ live
many characteristic Hubble times.

New large loops are born with mildly
relativistic velocity.  The peculiar
center of mass motion is damped by the universe's expansion.
A detailed study \cite{Chernoff:2009tp} of the competing effects (formation time,
velocity damping, evaporation, efficacy of anisotropic emission of
gravitational radiation) in the context of a simple formation
model for the galaxy shows that loops accrete when $\mu$ is
small. The degree of loop clustering relative to dark matter
clustering is a function of $\mu$ and approximately independent
of $\ell$. Smaller $\mu$ means older, more slowly moving
loops and more effective clustering.

The spatially dependent dark matter enhancement in a collapsed
object is
\be
{\cal E} = \frac{\rho_{DM}}{\Omega_{DM} \rho_c}
\ee
where $\rho_{DM}$ is the dark matter density and $\rho_c$ is
the critical density. The dark matter enhancement is
very substantial throughout the Galaxy. At the
local position ${\cal E} \sim 10^{5.5}-10^6$.

The formation of the Galaxy by cold dark matter infall
inevitably is accompanied by loops with low center of mass
motions. The tension dependent
enhancement to the homogeneous distribution of loops is
\ba
   {\cal F} & = &{\cal E} \beta(\mu)
\ea
where $0 < \beta(\mu) \lta 0.4$. For a fixed tension
there is only weak $\ell$ dependence
of ${\cal  F}$, i.e. the enhancement is roughly independent
of the individual loop length. The specific form for $\beta$
derived for the Galaxy 
is given in the appendix. Lower
tension strings behave more and more like cold dark matter,
i.e. $\beta$ increases as $\mu$ decreases. 
In fact, $\beta$ does not reach $1$ partially because loops do not
survive in the Galaxy forever, each is eventually
accelerated by the rocket effect and ejected before complete evaporation
occurs. The tension dependent enhancement
saturates ($\beta \to 0.4$) near $\mu = 10^{-15}$.

The {\it local} string loop population is enhanced by the factor ${\cal F}$ with
respect to the homogeneous distribution.
Since dark matter is
strongly clustered it follows that string loops with small
$\mu$ are strongly clustered.

We summarize the enhancement of the local Galactic population by
\be
\left( \frac{dn}{d\ell} \right)_{local} = {\cal F} \left( \frac{dn}{d\ell} \right)_{homog} = {\cal F} {\cal G} \left( \frac{dn}{d\ell} \right)_{baseline} .
\ee
This is the basis for rate calculations of microlensing and of gravitational wave bursts. Large ${\cal F}$ and large ${\cal G}$ make microlensing and
gravitational wave detections of nearby loops feasible.

\section{Estimate of the Rate of Gravitational Wave Bursts}

Loops and long horizon-crossing strings will generate gravitational
radiation. Strong emission is expected when a string element
accelerates rapidly, notably at kinks and cusps.  Here, we concentrate
on the bursts expected from large loops and on the determination of
the confusion limit which delimits the stochastic gravitational wave
background of unresolved bursts from the same sources. We do not
address the emission from the small loops or from the long strings or
from any other sources.

There are many current and future experiments with the potential
to make direct detections of gravitational wave emission from string loops
and/or set upper limits on
it. These include Earth-based laser interferometers (LIGO\cite{TheLIGOScientific:2014jea}, VIRGO\cite{Accadia:2011zzc}, KAGRA\cite{Aso:2013eba}; for overview
\cite{Evans:2014cwa}),
space-based inteferometers (LISA\cite{Audley:2017drz}, DECIGO\cite{Kawamura:2011zz}) and pulsar timing arrays
(NANOGRAV\cite{Arzoumanian:2015liz}, European timing array\cite{Lentati:2015qwp},Parkes\cite{Hobbs:2013aka}).

The calculation methodology was formulated in\cite{Damour:2004kw,
  Damour:2001bk,Damour:2000wa,Siemens:2006yp} and we schematically
follow the treatments \cite{Kuroyanagi:2012wm,Kuroyanagi:2012jf} with
several modifications that play an important role in Earth-based
experiments \cite{Chernoff:2007pd,DePies:2007bm,Chernoff:2009tp,Chernoff:2014cba}). The major changes with respect to previous treatments are:
\begin{itemize}
\item Tension-dependent clustering of string loops in the Galaxy
halo (${\cal F}$).
\item Only a small portion of the long
  strings' invariant length transformed into large loops (fraction $f=0.1$
  to make large loops of parameterized size $\alpha=0.1$). The rest is
  lost for the purpose of direct detection of gravitational
  wave emission.
\item Enhancements of the string density with respect to field theory strings on account of multiple species of strings in each throat,
diminished intercommutation probability and multiple throat (${\cal G}$).
\end{itemize}
There is no clustering in \cite{Abbott:2017mem}; their
      model $M=2$ has a similar fraction of large loops and they explore
      a range of $1/p$ comparable to the range ${\cal G}$ we have discussed.

We outline the methodology and provide examples of the results.

\subsection{Homogeneous Methodology}

We follow \cite{Kuroyanagi:2012wm,Kuroyanagi:2012jf} for calculating
event rates in a homogeneous universe with ${\cal F}={\cal G}=1$.  The
birth rate density for loops (eq. \ref{dndtdl}) is expressed in terms of
$\gamma=L/t$. The function $\gamma(t)$ is numerically derived for a
$\Lambda$CDM cosmology (eqs. \ref{modelstart}-\ref{modelend}). The
cosmological treatment is essentially exact although the description
of the network presumes that it is close to scaling at all times and
this will fail (1) when the network first forms, (2) at the
radiation-matter transition and (3) on large scales at late times as $\Lambda$ comes to dominate
expansion. For the loops of interest none of these are consequential.
After a loop is formed its length shrinks at a constant
rate until complete evaporation (eq. \ref{looplengthatt}).

The Fourier transform of the gravitational wave
amplitude $h(f)$ observed at Earth with frequency $f$ from a single
passage of a cusp or kink on a loop at red-shift $z$ and having loop
length $\ell$ at the time of emission has
asymptotic form\cite{Damour:2001bk}
\ba
h_{cusp} f & = & \frac{A_{cusp} G \mu \ell^{2/3}}{f_{em}^{1/3} r(z)} \\
h_{kink} f & = & \frac{A_{kink} G \mu \ell^{1/3}}{f_{em}^{2/3} r(z)} \\
r(z) & = & \int_0^z dz' \frac{1}{H(z')} \\
f_{em} & = & f(1+z)
\ea
where $r(z)$ is comoving distance, $H(z)$ is the Hubble constant and
$f_{em}$ is the emission frequency. The transform $h(f_{em})$ vanishes for
frequencies (approximately) less than the loop fundamental $f_{em} < 2/\ell$.
The numerical quantities $A_{cusp}$ and $A_{kink}$ are order unity
coefficients \cite{Damour:2001bk}. We conservatively fix $A_{cusp} = A_{kink} = 1$
(cf. $A_{cusp} \sim 2.68$ in \cite{Damour:2001bk,Kuroyanagi:2012wm,Kuroyanagi:2012jf}).
Each cusp or kink on a loop emits beamed radiation with angular scale $\Theta$
and solid angle $\Omega$ and is observed (at Earth) with
repetition frequency $f_{rep}$:
\ba
\Theta & = & \left( \frac{ f_{em} \ell }{2} \right)^{-1/3} \\
\Omega_{cusp} & = & \pi \Theta^2 \\
\Omega_{kink} & = & 2 \pi \Theta \\
f_{rep} & = & \frac{2}{\ell (1+z)} .
\ea
Assume that there are $n$ active cusps or kinks per loop per
fundamental period (and note separate ``cusp'' and ``kink'' labels are
omitted when the same form applies to both types). Typically, a loop has an even number of cusps and an integer number of kinks. For numerical
examples we take $n_{cusp}=2$ and $n_{kink}=4$. The solid angle for
the kink presumes the beam pattern traces a great circle on the sky.

Let $R$ be the rate of reception of signals of frequency $f$
at Earth and let $dR/dzdt_b$ be the differential rate
with respect to the emission redshift $z$ of loops born at time $t_b$.
Distinguishing the emission rate density and birth rate density for
clarity gives
\ba
\frac{dR}{dz dt_b} & = & \left(\frac{dn}{dt}\right)_{em,z} \frac{dV}{dz} \times
\frac{\Omega \  n \ f_{rep}}{4 \pi}  \\
\left(\frac{dn}{dt}\right)_{em,z} & = &
\left(\frac{dn}{dt}\right)_{b,z_b} \left( \frac{ a(z_b) }{ a(z) } \right)^3
\ea
because the loop density scales like cold matter.

It is straightforward to transform from time of birth to Fourier
amplitude $d/dt_b \to d/dh$ by first writing $\ell$ in terms of
$h$, $f$, $z$ and $G \mu$ for a given cosmology
\ba
\ell_{cusp} = \left( \frac{ hf r f_{em}^{1/3} }{A_{cusp} G \mu} \right)^{3/2} \\
\ell_{kink} = \left( \frac{ hf r f_{em}^{2/3} }{A_{kink} G \mu} \right)^3
\ea
and then substituting these expressions into the birth time $t_b$ and
differentiating to find $dt_b/dh$
\ba
t_b & = & \frac{ \ell +  \Gamma G \mu t}{\alpha + \Gamma G \mu} \\
\left(\frac{dt_b}{dh}\right)_{cusp} & = & \frac{3 \ell_{cusp}}{2 h (\alpha + \Gamma G \mu)} \\
\left(\frac{dt_b}{dh}\right)_{kink} & = & \frac{3 \ell_{kink}}{h (\alpha +  \Gamma G \mu)}
\ea

With the change of variables
\be
\frac{dR}{dz dh}  =  \frac{dR}{dz dt_b} \frac{dt_b}{dh}
\ee
the rate for cusps and kinks is evaluated as an integral over $z$ for any
given $h$
\be
\frac{dR}{dh} = \int dz \frac{dR}{dz dh} .
\ee
Single strong signals can be separated from the background of
overlapping signals if they occur less frequently than $f$,
the frequency at which observations are made. We will define
the amplitude $h_*$ at the frequency of highest sensitivity $f$
by
\be
\int_{h_*}^\infty \frac{dR}{dh} dh = f .
\ee
The rate per log amplitude is $h dR/dh$. Generalizing to
a frequency-dependent amplitude $h_*(f)$ we write
the stochastic background as
\be
\Omega_{GW}(f) = \frac{2 \pi^2 f^3}{3 H_0^2} \int_0^{h_*} dh h^2 \frac{dR}{dh} .
\ee

The homogeneous changes scale $\frac{dn}{dt_b} \to f {\cal G}
\frac{dn}{dt_b}$ where $f$ is the fraction of the long string length
that enters large loops and and ${\cal G}$ accounts for the number of
effectively independent species of strings.

\subsection{Clustering}

We treat clustering inhomogeneity in a spherically symmetric
fashion as if the Earth were at the center of the Galaxy,
ignoring the CDM density variation at radii less than $R_{gc}=8.5$ kpc,
the Sun's distance from the center. We retain the density variation for
$R_{gc}<r<R_{TA}$ where $R_{TA}$ is the turn around radius of the halo
of the Galaxy in a spherically symmetric infall model. The measured
rotation curve of the galaxy and the age of the universe imply
$R_{TA}=1.1$ Mpc \cite{Chernoff:2009tp}. This power law description
of the CDM halo density is accurate beyond the
central regions where baryons concentrate out
to the scale where infall times become comparable to the
age of the universe. Quantitatively,
\ba
\frac{\rho_{DM}(r)}{\rho_c \Omega_{DM}} =
10^{3.2} \left( \frac{r}{100 \ {\rm kpc}} \right)^{-2.25} .
\ea
The density scaling of the infall model is close to that of an
ideal, flat rotation curve for which $\rho \propto r^{-2}$.

Although very simple, the spherically symmetric description is
suitable starting a few kpc from the Galactic center and reaching
half way to Andromeda, i.e. $\sim 400$ kpc. Beyond the midpoint a monolithic
collapse can't be inaccurate, of course. The bulk of the matter in the
halo lies at large distance but the falloff of burst
signals with distance emphasizes nearby sources. The detailed density
profile at the Galactic center (core or cusp) is less important than
the total mass within $r<R_{gc}$ which is well-fixed by the rotation
curve. Likewise, the burst rate will not be overly sensitive to the
cutoff at the turn around radius because nearby sources are easier to
detect.

Consider a sphere of radius $R_{gc}$ with
inner flat overdensity which joins smoothly onto a
density profile like that of the infall model for $r>R_{gc}$.
The CDM overdensity and the tension-dependent enhancement factor
give the loop overdensity:
\ba
    {\cal E}(r) & = &
    \max( 1, \frac{\rho_{DM}(\max(r,R_{gc}))}{\rho_c \Omega_{DM}} ) \\
        {\cal F}(r) & = & \max(1, \beta(\mu) {\cal E}(r) ) .
\ea
A schematic picture shows the relationship between the power law
density profile in CDM and the adopted ${\cal F}(r)$.
\begin{figure}[ht]
  \centering
\includegraphics{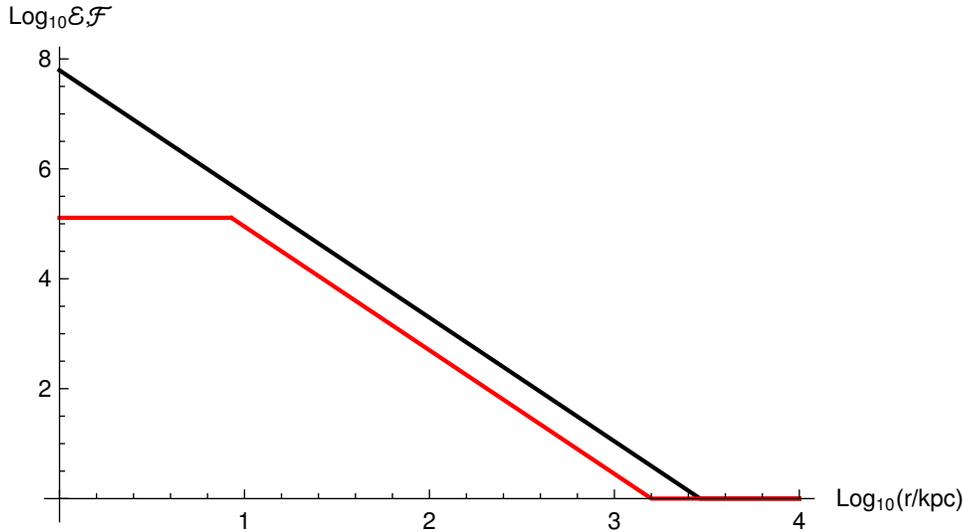}
\caption{The black line is the power law CDM profile that forms in
  spherical infall. The red line is the adopted string enhancement for $G \mu = 10^{-13}$. It is based on the truncated CDM profile at $r < R_{gc}$,
  the position of the Sun. This
  string loop distribution has been used to estimate the local contribution
  to gravitational wave bursts that emanate from the
  Galaxy and from larger distances.}
\end{figure}
The outer edge is close to the point where $\rho = \rho_c \Omega_{DM}$.
The enhancement factor is included in the calculation by the replacement
\be
\frac{dn}{dt} \frac{dV}{dz} \to {\cal F}(r=z/H) \frac{dn}{dt} \frac{dV}{dz}
\ee
which alters the integration at small $z$.
This is a conservative means of estimating the effect
of clustering on detection because it discounts the
most concentrated inner regions.

\subsection{Noise}

For a given $h(f)$ the signal to noise over a band $[f_{lo},f_{hi}]$
is 
\be
\rho = 2 \sqrt{ \int_{f_{lo}}^{f_{hi}} \frac{ |h|^2 } {S(f)} df }
\ee
where $S(f)$ is the one-sided power spectral density 
(or, equivalently $S(f)=A(f)^2$ where $A(f)$ is amplitude
spectral density). 

The magnitude of the Fourier transform (in
the continuum limit) of an observed cusp or kink signal
follows an approximate power law form in $f$:
$h(f) \propto f^{-n}$ with $n=4/3$ for a cusp and $n=5/3$ for a kink.
The cusp's envelope is smooth and the kink's somewhat more variable.
The non-zero range of $h(f)$
extends from the frequency of the loop fundamental to a value
set by how the direction of observation relates to
the beam's emission axis.  If the observational
band $[f_{lo},f_{hi}]$ falls within the intrinsic range of the
power law form then it suffices to calculate the
signal to noise using a single frequency $f_{char}$.
Write $(h f)_{char} = h(f_{char}) f_{char}$. 
For a given instrument with fixed $f_{lo}$, $f_{hi}$ and $S(f)$
the quantity of interest is
\be
\frac{1}{(hf)_{inst}} = 2 \sqrt{ \int_{f_{lo}}^{f_{hi}} \left( \frac{f_{char}}{f} \right)^{2n}
    \frac{ df } {f_{char}^2 S(f)} }  
\ee
so that the signal to noise is simply written
\be
\rho = \frac{(h f)_{char}}{ (hf)_{inst} } .
\ee
Calculating $h dR/dh$ at $f=f_{char}$ for independent
$h$ is equivalent to calculating $dR/d \log \rho$
as function of signal to noise $\rho$. We will present
results in terms of $dR/d \log \rho$ where $R$ is measured in events per year.

The selection of the frequency band to describe a given instrument
must balance several considerations: ideally the range of $[f_{lo},f_{hi}]$ should
be small so that the power law approximation has maximal validity and
large to encompass as much of the signal as possible. Specifically,
$f_{lo}$ should be small because large loops have higher amplitudes at
smaller frequencies and $f_{hi}$ should be large enough to reach the most
sensitive part of the instrument's noise curve. Our choices
for $[f_{lo},f_{hi}]$ are given in the Table \ref{tabnoise}. For $S(f)$
we use an analytic noise model for LISA and numerical values of a
plotted noise curve for LIGO.
\footnote{Taken from ``LISA Unveiling a hidden Universe'', an ESA study chaired by Danzmann and Prince, Feb. 2011, section 3.4 available at
  \url{sci.esa.int/science-e/www/object/doc.cfm?fobjectid=48363} and
AdvLIGO noise curve in document LIGO-T0900288-v3 available at
\url{dcc.ligo.org/public/0002/T0900288/003/AdvLIGO\%20noise\%20curves.pdf}
} The calculations take $f_{char}=f_{hi}$ and the sensitivities
to cusp and kink bursts are given in Table \ref{tabnoise}.
\begin{table}
  \begin{center}
    \begin{tabular}{ccccc}
      \hline
  Instrument & $f_{lo}$ & $f_{hi}$ & $(hf)_{inst}$(cusp) & $(hf)_{inst}$(kink) \\
  LISA & $3 \times 10^{-5}$ & $5 \times 10^{-3}$ & $3.45 \times 10^{-22}$ & $2.70 \times 10^{-22}$ \\
  LIGO & $10$ & $220$ & $8.97 \times 10^{-24}$ & $5.32 \times 10^{-24}$
\end{tabular}
    \caption{\label{tabnoise}All frequencies in Hz. $f_{char}=f_{hi}$. }
    \end{center}
\end{table}

\subsection{Results}

Calculations with ${\cal G}=1$, $\alpha=0.1$,
$f=0.1$ and $\Gamma=50$ describe loops formed by traditional field theory
(FT) strings based on the current understanding of network
evolution. Superstring (SS) calculations assume ${\cal G}=10^2$. We
find that for both LISA and LIGO the locally clustered strings can
dominate the statistics of detected bursts over specific ranges of
string tension. This statement is true for cusps and kinks in both FT
and SS calculations.

\begin{figure}[ht]
  \centering
\includegraphics{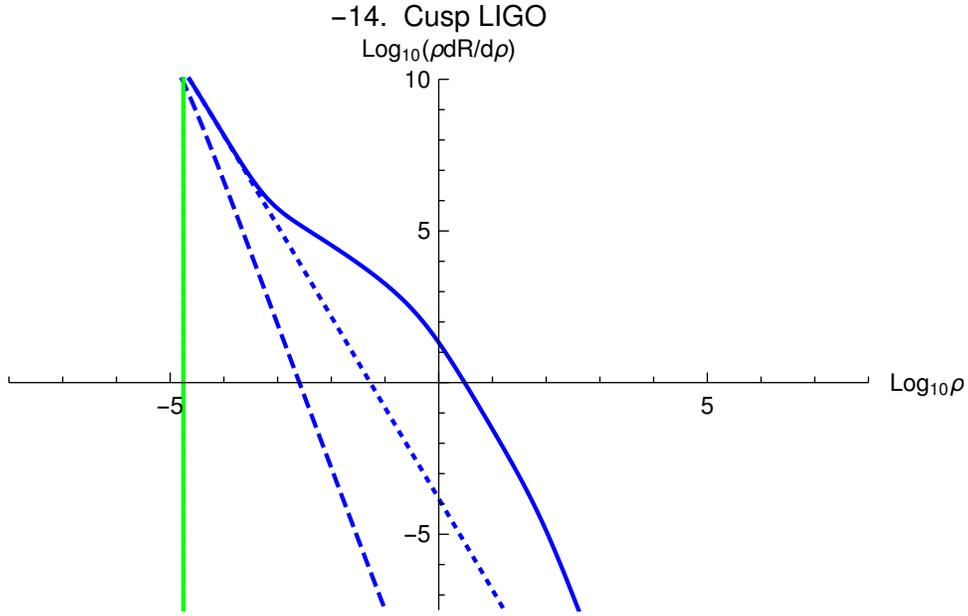}
\caption{\label{LIGO_Cusp_-14} Advanced LIGO detects cusp bursts for
  field theory (FT) strings with string tension
  $G \mu = 10^{-14}$.  Blue lines illustrate
  differential rates as a function of log signal to noise ratio. The
  solid blue line is the total rate from all sources, the dotted blue
  line is the rate from homogeneous cosmology (no local clustering) and
  the dashed blue line is the rate for sources with $z>0.68$.}
\end{figure}

Fig. \ref{LIGO_Cusp_-14} shows a very wide view of
LIGO cusp detections for FT strings with $G \mu =
10^{-14}$. The abscissa is $\log_{10} \rho$ and the ordinate is the
$\log_{10}$ of the rate per year (per log interval of $\rho$). Roughly speaking, in
similar graphs in this section we are observationally
most interested if/when the lines enter the upper right
hand quadrant: here the typical rates exceed one per year
for non-trivial signal to noise ratios.

The solid blue line is the total forecast of detections
for FT strings. It includes strings clustered within the halo of the Galaxy plus
those throughout a $\Lambda$CDM homogeneous universe. The dotted
blue line excludes the Galactic halo's local contribution -- it shows
the contribution of the
homogeneous cosmological distribution. The dashed blue line
displays separately high redshift contributions (defined by $z >
0.68$). For a wide range of $\rho$ the detections are dominated by
strings in the local halo. The vertical green line is the confusion
limit for the total (solid) and for the homogeneous components (dotted)
which overlap in this case.

The greatest impact of string loop
clustering on LIGO detections of cusps for field theory strings
occurs near $G \mu \sim
10^{-13.3}$. Over the range $10^{-14.8} <G \mu < 10^{-12.5}$ the solid
blue curves cross into the upper right hand quadrant and the rate of
burst detections is dominated by loops in the halo.

\begin{figure}[ht]
  \centering
\includegraphics{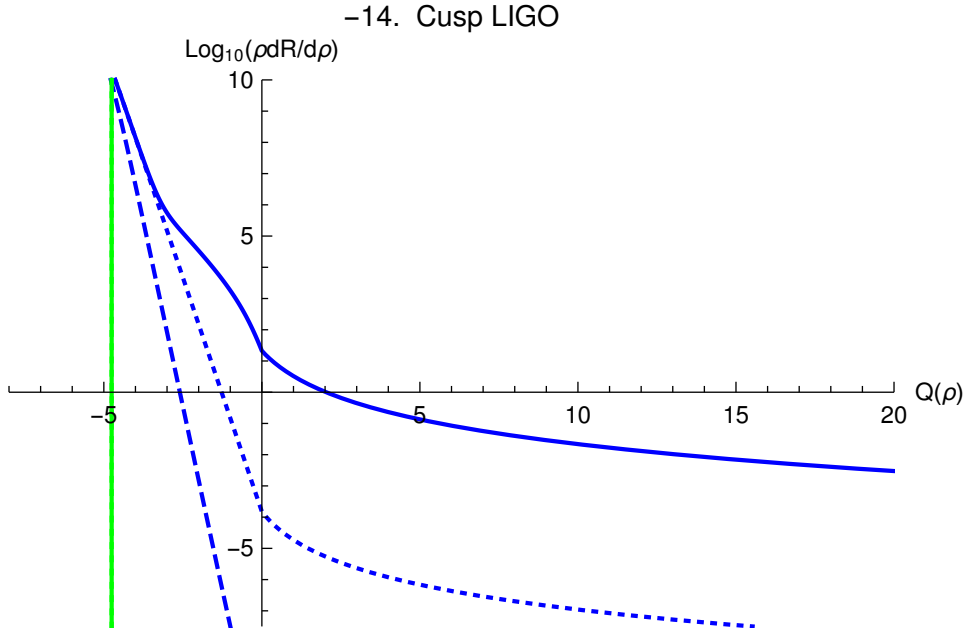}
\caption{\label{LIGO_Cusp_-14.alt2} Advanced LIGO detects cusp bursts
  for $G \mu = 10^{-14}$ for field theory (FT) cosmic strings using
  same line types as Fig. \ref{LIGO_Cusp_-14} with split log-linear
  abscissa. To the left of the y-axis $\log_{10}\rho$, to the
  right $\rho - 1$ where $\rho$ is signal to noise.}
\end{figure}

A more useful way to display the same results is shown in
Fig. \ref{LIGO_Cusp_-14.alt2} where a split log-linear
abscissa using the quantity
\be
Q(\rho) = \left\{
  \begin{tabular}{cc}
    $\rho - 1$ & $\rho > 1$ \\
    $\log_{10} \rho$ & $\rho < 1$ 
  \end{tabular}
  \right.
\ee
which has the effect of spreading out the interesting signal to noise
ratios on the right hand side and compressing the very small ones on
the left.  The right hand side of the plot shows that the situation
with many bursts per year only occurs for $\rho < 3$ ($Q<2$), a
weak signal to noise for an experiment like LIGO.
Signals of this sort are likely to
fall below the threshold for many LIGO based searches. The linear
scale on the right hand side makes this important
fact more obvious than the previous rendition.

\begin{figure}[ht]
  \centering
\includegraphics{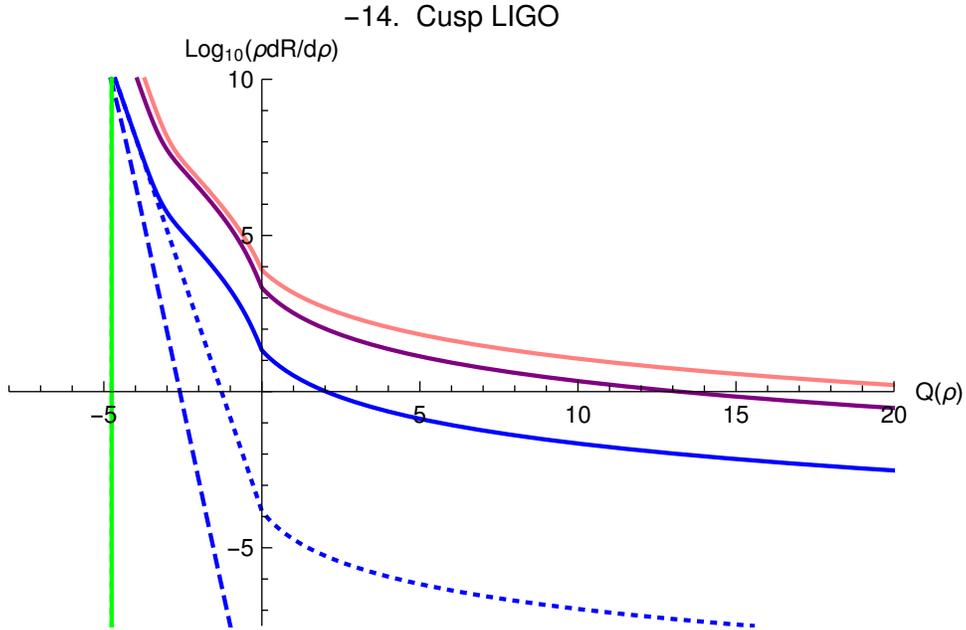}
\caption{\label{LIGO_Cusp_-14.alt3} LIGO detects cusp bursts
  for $G \mu = 10^{-14}$ for both field theory (FT) strings and superstrings (SS).
  The blue and green
  lines are the same as Fig. \ref{LIGO_Cusp_-14.alt2}.
  The purple line shows the superstrings' total detection rate for Advanced LIGO
  the pink line supplements that with a hypothetical factor of 3 decrease in
  the noise spectral density. Only totals are shown for the superstring cases (no dotted or dashed lines and no confusion limits).
}
\end{figure}

Now consider the impact of the move to superstrings shown
in Fig. \ref{LIGO_Cusp_-14.alt3}. We have
argued that several factors suggest 
${\cal G}=10^2$ is a reasonable summary of the enhancement effects
of superstrings over field theory strings. This choice
shifts all rates upward by the same factor and yields the purple
line for the total LIGO burst rate for superstrings of this
tension. The high signal to noise ($\rho > 10$) and large total rate (many
per year) implies such strings are detectable.
As a final consideration we included a hypothetical improvement to
Advanced LIGO that decreases $S(f)$ the power spectral density (PSD)
by a factor of 3. Improved PSD might arise from
  sensitivity upgrades to the
  LIGO/VIRGO detectors and/or from extending the network of detectors
  to include
  KAGRA\cite{Aso:2013eba} and
  India-LIGO\cite{Unnikrishnan:2013qwa}.
This change has the
effect of shifting the purple curve to the right and yields the
pink line. These superstring cusp burst rates are likely to be
realistically detectable. We will refer to this scenario as SS$^*$.
(Below we will also consider a hypothetical improvement by a factor of 3
for LISA as well.)

For each source and experiment we will define the minimally-interesting
minimum tension (MIMT). The MIMT is the minimum
characteristic tension for the curve to enter the upper right quadrant
of figures like these, specifically, $dR/d\log\rho>1$ yr$^{-1}$ at
$Q>0$. Tensions greater than the MIMT {\it might} be seen but tensions
less than the MIMT are {\it very unlikely}. We will also define the
probably-detectable minimum tension (PDMT).  The PDMT is the
minimum characteristic tension for one significant (arbitrarily taken
to be $Q>10$) event per year, or $dR/d\log\rho>1$ yr$^{-1}$ at
$Q>10$. Tensions greater than the PDMT {\it can} realistically be
expected to generate detectable events for the assumed source and
experiment.

\begin{figure}[ht]
\centering
\includegraphics{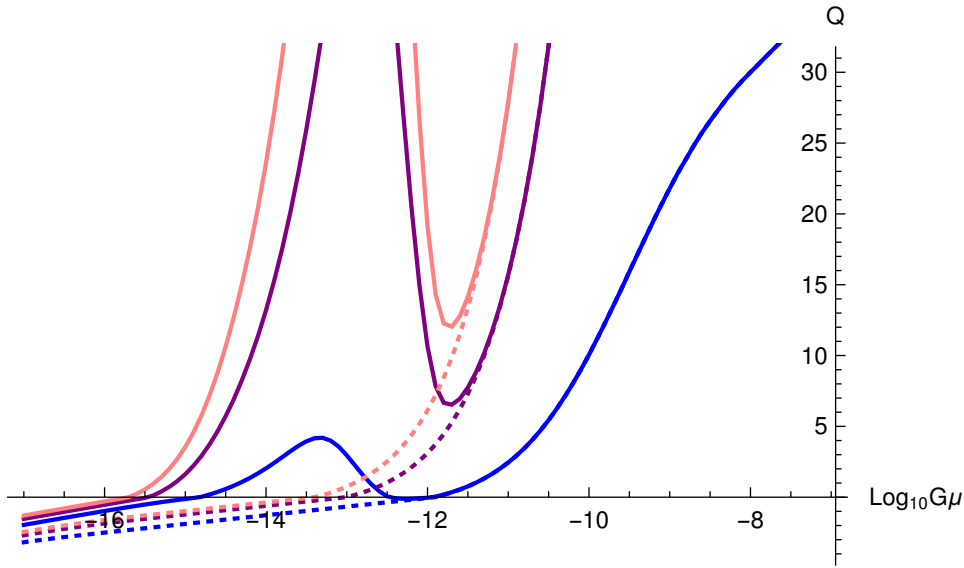}
\caption{\label{fig:yaxLIGOCusp} LIGO cusp detection:
  $Q$ as a function of string tension for $dR/d\log\rho = 1$ yr$^{-1}$.
  The blue line is for field theory (FT) strings,
  the purple line is for superstrings (SS),
  the pink line is for superstrings with reduced noise detector
  (SS$^{*}$).
  (These colors
  are the same as Fig. \ref{LIGO_Cusp_-14.alt3}.) The 
  minimally-interesting minimum tensions (MIMTs)
  correspond to the left-most intersection of the solid line with the
  $Q=0$ axis; the probably-detectable minimum
  tensions (PDMTs) correspond to the left-most intersection of
  the solid lines with the horizontal $Q=10$ lines. The dashed lines
  are the results without clustering. The separation between
  dashed and solid lines of the same color at low $\mu$
  is the enhancement due to clustering.
}
\end{figure}

\begin{figure}[ht]
\centering
\includegraphics{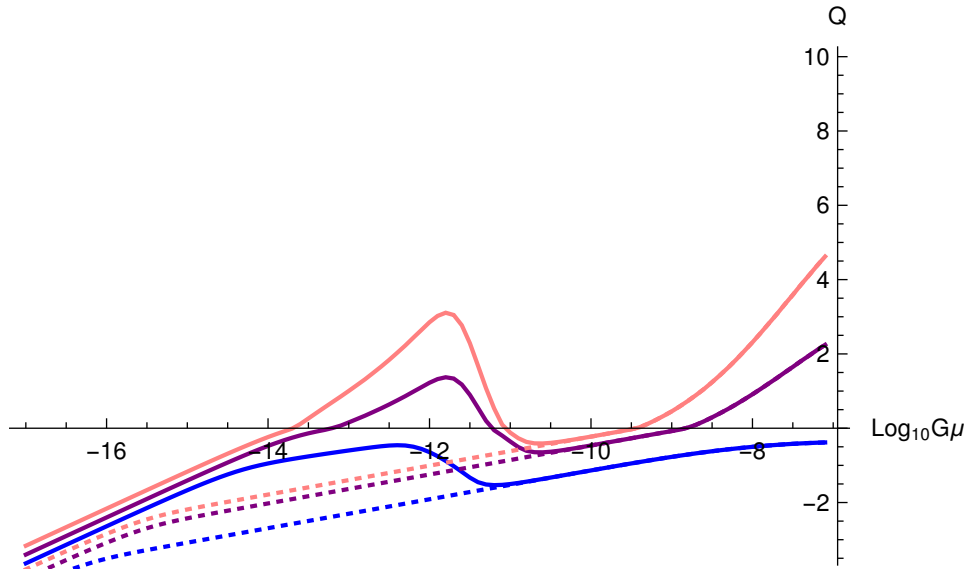}
\caption{\label{fig:yaxLIGOKink}LIGO kink detection: Same as Fig. \ref{fig:yaxLIGOCusp}
but with a reduced vertical scale.
The solid lines never reach $Q=10$ even though there is
an enhancement from clustering.}
\end{figure}

Figure \ref{fig:yaxLIGOCusp} plots $Q$ as a function of
string tension for $dR/d\log\rho = 1$ yr$^{-1}$ for LIGO cusp bursts.
This information is extracted by creating
figures like Fig. \ref{LIGO_Cusp_-14.alt3} for many different tensions
and locating where lines intersect the abscissa.
The dashed lines are calculations without clustering while
the solid lines include that effect. The most immediate
implication is that clustering enhances $Q$ at $G \mu \lta 10^{-12}$.
Large signal to noise detections are expected for a tension
range about two orders of magnitude in width, a range where unclustered
loops primarily give weak signals. The solid lines
cross the horizontal $Q=0$ axis (leftmost) at the MIMT; they cross the 
$Q=10$ ordinate (leftmost) at the PDMT. We will always quote the MIMT
and PDMT taking account of clustering. 
Analogous figures for
LIGO kink detection and LISA cusp and kink detection are
given by Figs. \ref{fig:yaxLIGOKink}, \ref{fig:yaxLISACusp} and
\ref{fig:yaxLISAKink}.
See Tables \ref{tab:LIGOCusp}, \ref{tab:LIGOKink}, \ref{tab:LISACusp} and
\ref{tab:LISAKink} for details.

Now consider LIGO detection rates for kink bursts, summarized in Fig.
\ref{fig:yaxLIGOKink}. The FT strings (blue lines) never cross the $Q=0$
axis (no MIMT, PDMT). The SS (purple lines) and SS$^*$ (pink lines)
do not ever reach $Q=10$ (no PDMT). We do not anticipate
frequent, strong kink events in LIGO.
Generally speaking, the fundamental frequency of these loops
is small compared to the range of frequencies at which LIGO is
sensitive. The power radiated in the higher harmonics falls off more rapidly
for kinks than cusps and so kinks prove to be harder to detect at
the characteristic frequency at which LIGO is sensitive.

This fundamental frequency mismatch between source and detector
strongly motivates consideration of space-based
detectors like LISA that are designed for lower frequencies. For cusp
bursts, LISA can detect FT strings, SS and SS$^*$ with
MIMT $G \mu = 10^{-15.6}$, $10^{-16}$ and $10^{-16.3}$, respectively.
The clustering dramatically enhances the sensitivity at $G \mu \lta 10^{-11}$,
a number somewhat dependent upon FT, SS or $SS^*$.
Conversely, the cusp burst rates for tensions in the range $G \mu \gta 10^{-11}$
do not bear a strong imprint (greater than factor 2 enhancement)
from the local halo population.

\begin{figure}[ht]
  \centering
\includegraphics{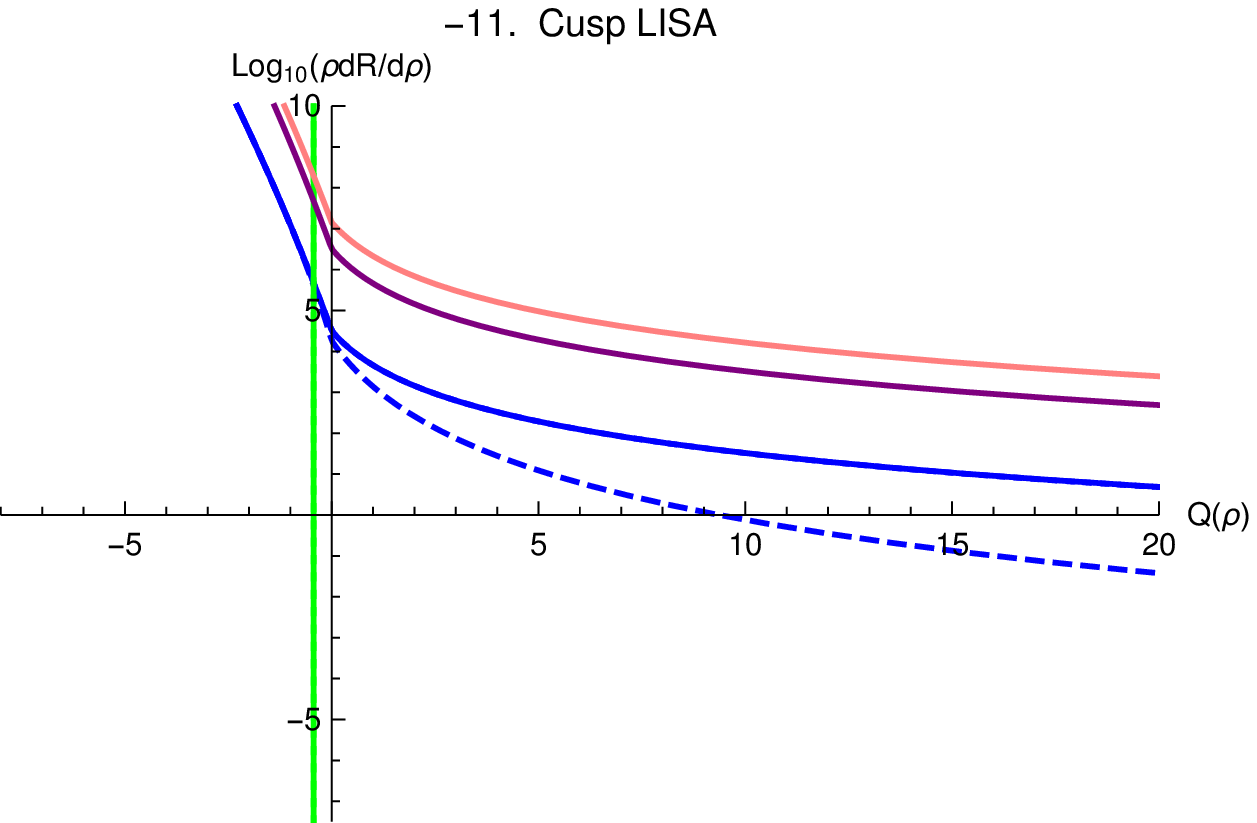}
\caption{\label{LISA_Cusp_-11._alt3} LISA detects cusp bursts
  for $G \mu = 10^{-11}$ for field theory (FT) strings, superstrings
  (SS) and superstrings with less noise (SS$^*$). The line
  types are the same as Fig. \ref{LIGO_Cusp_-14.alt3}. Clustering
  is irrelevant -- the blue solid line for total and blue
  dotted line for homogeneous cosmology give essentially identical
  results.
}
\end{figure}

To illustrate,
Fig. \ref{LISA_Cusp_-11._alt3} shows the situation for $G \mu = 10^{-11}$.
Note that the FT clustered and unclustered results lie on top of
each other.
The universe as a whole provides the dominant source of loops, in part
because high tension strings are less clustered (they do not track the
dark matter profile as closely as low tension ones) and in part
because higher tension strings emit signals of larger intrinsic amplitude
that are detectable at larger distances.

\begin{figure}[ht]
  \centering
\includegraphics{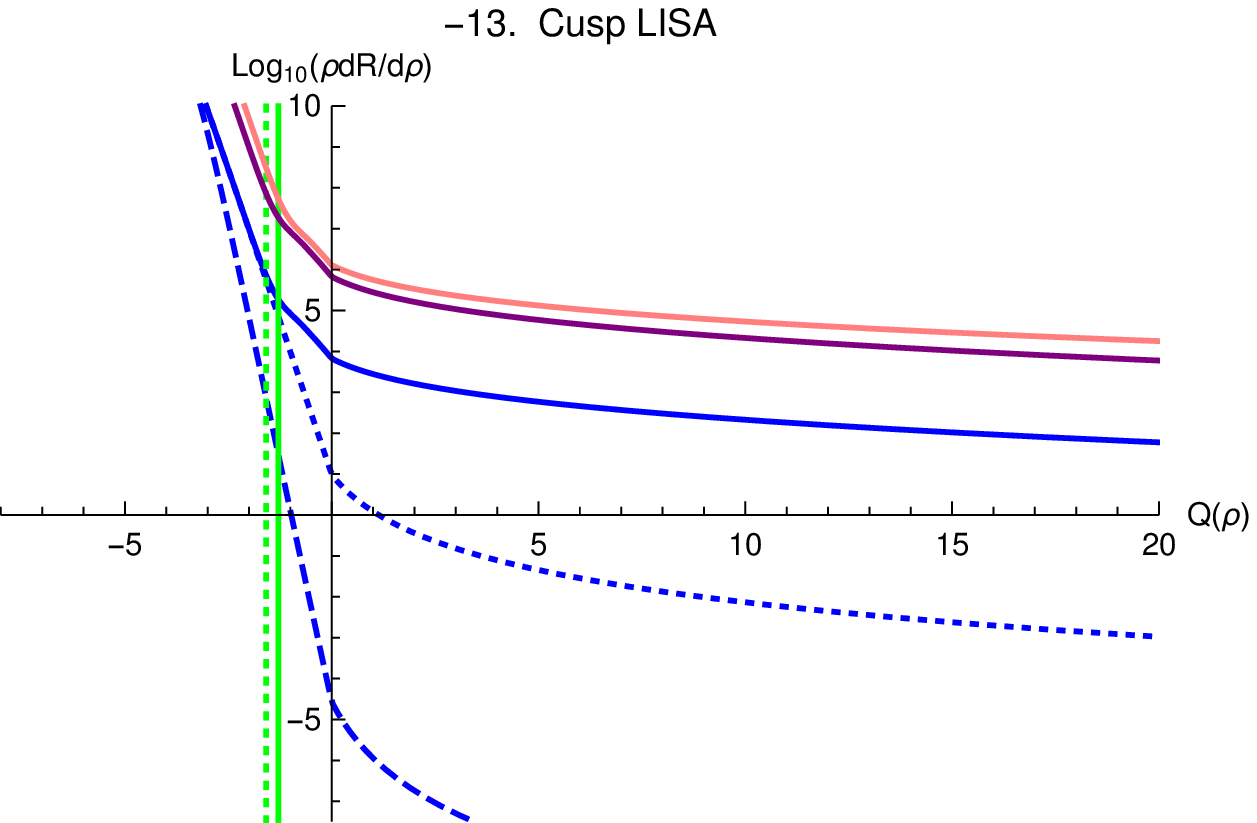}
\caption{\label{LISA_Cusp_-13._alt3} LISA detects cusp bursts
  for $G \mu = 10^{-13}$ from field theory (FT) strings, superstrings (SS) and superstrings with less noise (SS$^*$).  The line
  types are the same as Fig. \ref{LIGO_Cusp_-14.alt3}. The
  highly significant detections $\rho \sim 20$ are
  mostly sourced by the halo -- the blue solid line for total rate is
  much larger than the blue dotted line for homogeneous cosmology.
}
\end{figure}

\begin{figure}[ht]
\centering
\includegraphics{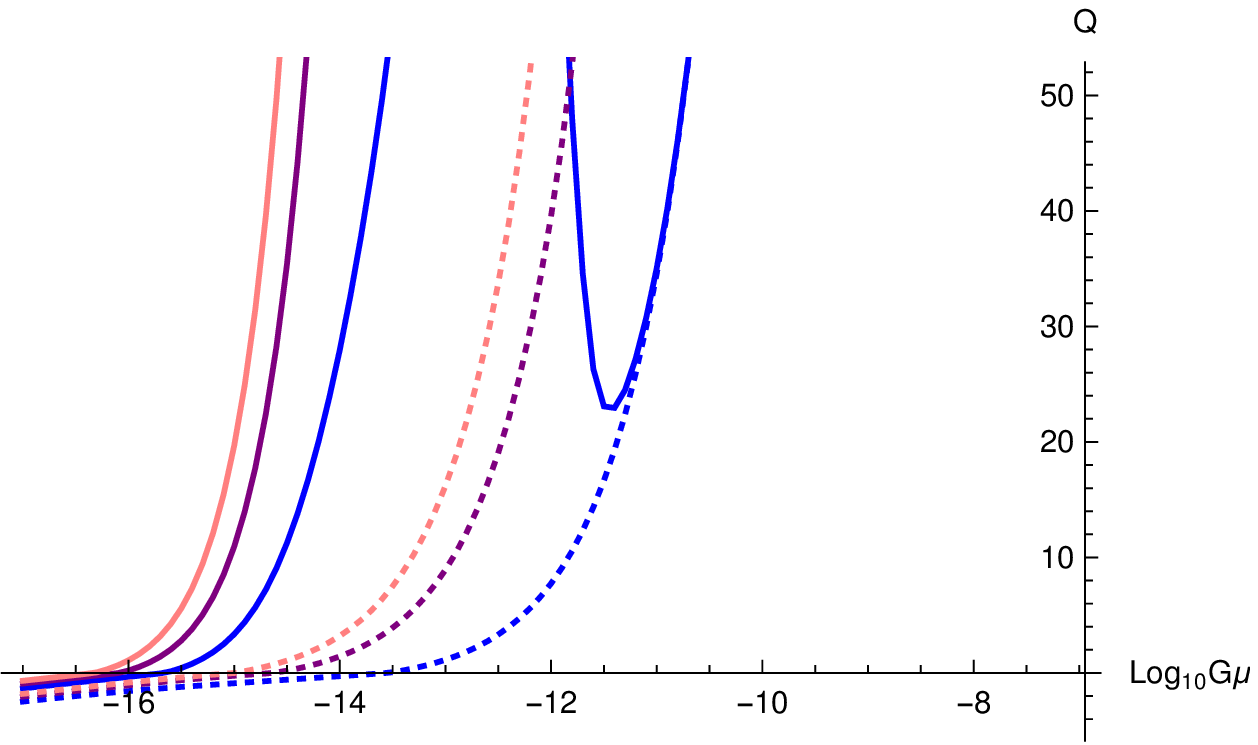}
\caption{\label{fig:yaxLISACusp}LISA cusp detection: Same as Fig. \ref{fig:yaxLIGOCusp}
with a larger vertical scale. The clustering greatly enhances
the sensitivity at small tension.}
\end{figure}

\begin{figure}[ht]
\centering
\includegraphics{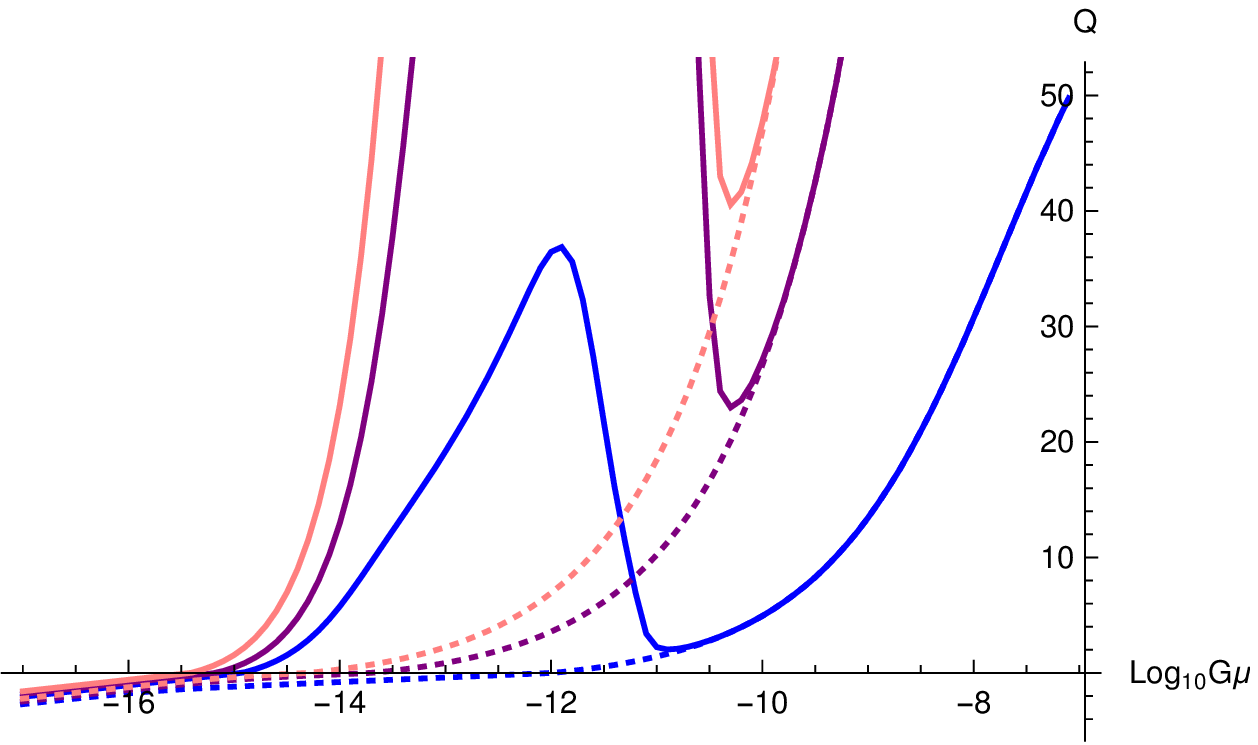}
\caption{\label{fig:yaxLISAKink}LISA kink detection: Same as Fig. \ref{fig:yaxLIGOCusp}
with a larger vertical scale. The clustering greatly enhances
the sensitivity at small tension.}
\end{figure}

Detected bursts from strings with $G \mu < 10^{-11}$ are largely
sourced by the halo population and
all rates are significantly enhanced by clustering.
Fig. \ref{LISA_Cusp_-13._alt3} illustrates
the case for $G \mu = 10^{-13}$.
Note that the clustered and unclustered FT strings are now quite
different.
Of course, the rate of SS detection is enhanced with respect to FT strings
by the factor ${\cal G}=10^2$. On the other hand, the
detectable range in $G \mu$ (see Fig. \ref{fig:yaxLISACusp})
is not significantly widened because
$dR/d\log \rho$ is a steep function of $G \mu$.
In terms of the MIMT the changes are rather small: $G \mu = 10^{-15.6}$ for FT,
$10^{-16}$ for SS and $10^{-16.3}$ for SS$^*$. 

\begin{figure}[ht]
  \centering
\includegraphics{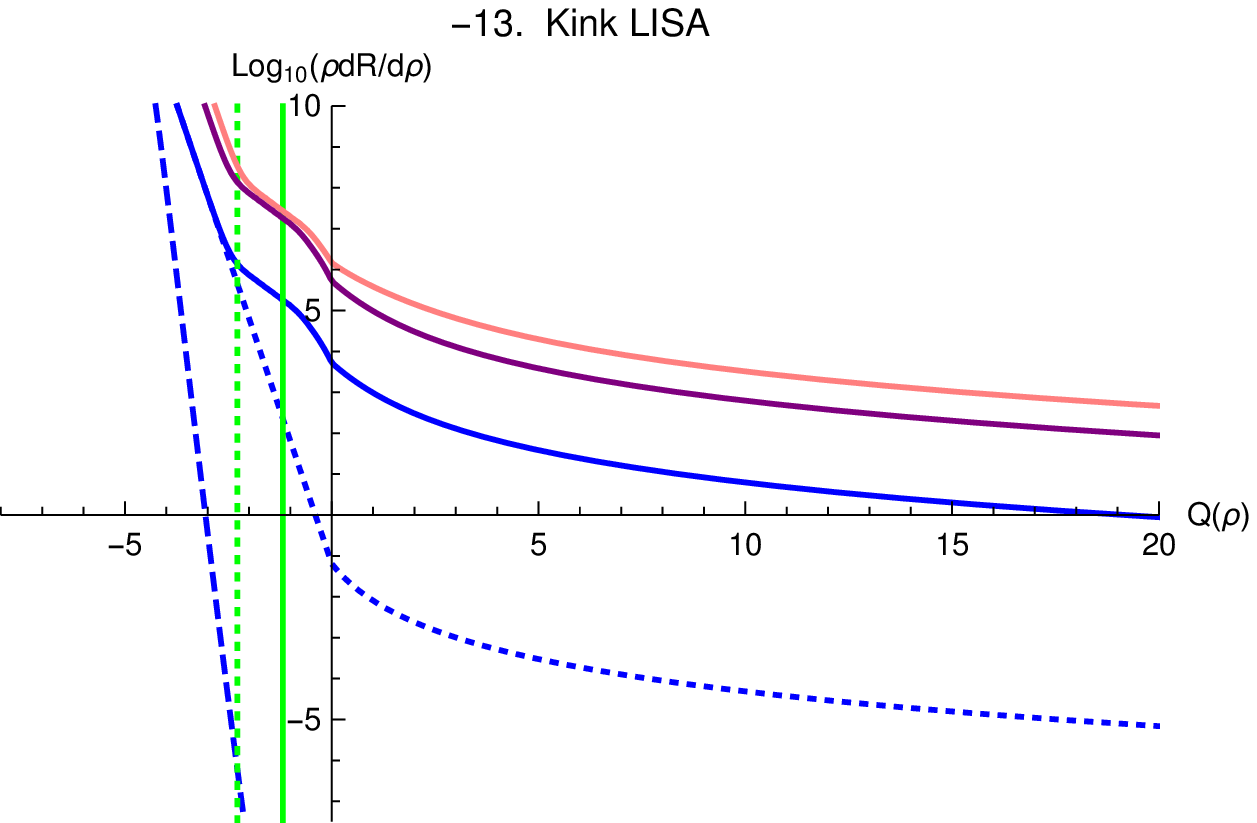}
\caption{\label{LISA_Kink_-13._alt3} LISA detects kink bursts
  for $G \mu = 10^{-13}$ for field theory strings and superstrings.
  The line
  types are the same as Fig. \ref{LIGO_Cusp_-14.alt3}. 
  Most detections
  are sourced by the halo -- the blue solid line for total is
  much larger than the blue dotted line for homogeneous cosmology.
}
\end{figure}

The situation for LISA kinks is summarized in Fig. \ref{fig:yaxLISAKink}.
FT strings with $G \mu \gta 10^{-11}$ traverse the upper
right hand quadrant and those with $G \mu \gta 10^{-9}$
have sufficient numbers and rates to be seen. These
results are not impacted by the clustering. Interestingly,
for $G \mu < 10^{-11}$ the clustering turns on and allows FT
strings to be detected in a lower
range of tensions $10^{-13.6} < G \mu < 10^{-11.3}$. A more extreme
version of this situation holds for SS and SS$^*$.
Kink bursts $G \mu \gta 10^{-11}$
are not sourced by the halo; those below are.
The tension range $G \mu \gta 10^{-14}$ should allow reliable detections.
Fig. \ref{LISA_Kink_-13._alt3} illustrates the situation when clustering
is important and strong signals are produced.
The MIMT for LISA kink bursts
is $G \mu \simeq$ few times $10^{-16}$. The detectable rates for tensions $G \mu <
10^{-11}$ are significantly enhanced by
clustering.

We can condense and summarize the outcomes in terms of the MIMT and PDMT.
Table \ref{tab:MIMTPDMT} lists
LIGO and LISA experiments, for FT, SS and SS$^*$ (all with the
effects of clustering included).
\begin{table}
  \begin{center}
    \begin{tabular}{cccc}\\
      \hline
      \multicolumn{4}{c}{LIGO Cusp}\\
      $Q$ & FT & SS & SS$^*$ \\
      0 & -14.8 & -15.4 & -15.7 \\
      10 & -10.0 & -14.2 & -14.5 \\
      \hline
      \multicolumn{4}{c}{LIGO Kink}\\
      $Q$ & FT & SS & SS$^*$ \\
      0 &  & -13.2 & -13.6 \\
      10 &  &  &  \\
      \hline
      \multicolumn{4}{c}{LISA Cusp}\\
      $Q$ & FT & SS & SS$^*$ \\
      0 & -15.6 & -16.0 & -16.3 \\
      10 & -14.6 & -15.0 & -15.5 \\
      \hline
      \multicolumn{4}{c}{LISA Kink}\\
      $Q$ & FT & SS & SS$^*$ \\
      0 & -14.5 & -15.1 & -15.4 \\
      10 & -13.6 & -14.1 & -14.4
    \end{tabular}
    \caption{\label{tab:MIMTPDMT} $log_{10} G \mu$ for MIMT ($Q=0$) and PDMT ($Q=10$) for
      LIGO cusps, kinks and LISA cusps, kinks. Blank entries mean
      there is no value in the range $10^{-17} < G \mu < 10^{-7}$. These estimates include the effect of clustering. For a characteristic burst rate
      $dR/d\log\rho = 1$ yr$^{-1}$ the line labeled $Q=0$ is the
      tension below which detection is very unlikely. The
      line labeled $Q=10$ is the tension for a strong,
      probably-detectable signal (signal to noise $\rho=11$).
      Greater tensions generally yield stronger signals but
      see Figs. \ref{fig:yaxLIGOCusp}, \ref{fig:yaxLIGOKink}, \ref{fig:yaxLISACusp} and \ref{fig:yaxLISAKink} for the non-monotonic impact of clustering
      on the rate forecasts.
}
  \end{center}
\end{table}
Summarizing the situation for SS (superstring loops with ${\cal
  G}=10^2$, no PSD enhancement), clustering within the Galaxy has a
favorable impact on the forecast for experimentally accessible
gravitational wave bursts (frequency of occurrence $> 1$ yr$^{-1}$ and
S/N $>10$) from cusps on strings with tensions $G \mu < 10^{-11.9}$
for LIGO/Virgo, $G \mu < 10^{-11.2}$ for LISA and for bursts from
kinks for tensions $G \mu < 10^{-10.6}$ for LISA. Frequent, high S/N
detections of cusps are expected for $G \mu \gta 10^{-14.2}$
(LIGO) and $G \mu \gta 10^{-15}$ (LISA) and of kinks for $G \mu \gta
10^{-14.1}$ (LISA). The table provides similar information for FT
(field theory strings) and SS$^*$ (superstrings with enhanced PSD).

More detailed information is available by inspection of
\ref{fig:yaxLIGOCusp}, \ref{fig:yaxLIGOKink}, \ref{fig:yaxLISACusp} and
\ref{fig:yaxLISAKink} and the Tables
\ref{tab:LIGOCusp}, \ref{tab:LIGOKink}, \ref{tab:LISACusp} and
\ref{tab:LISAKink} for numbers.

\newpage
\section{Summary}

We have reviewed the physical basis in string theory for the
occurrence of cosmic superstrings produced during the epoch of
inflation with particular attention paid to the braneworld
scenario. The Calabi-Yau manifold likely hosts many different
varieties of cosmic superstrings (in terms of tension, charge, etc.),
each with its own scaling network, uncoupled except for mutual
expansion of the large dimensions. The main signals of cosmic
superstrings in other throats will be carried by gravitational and
axionic degrees of freedom. We have reviewed the cosmology of
superstrings contrasting it with field theory strings. The warping of
the throats of the Calabi-Yau manifold lowers the string tension. The
loops formed by the scaling network dominate the total superstring
contribution to the critical density. The expansion of the universe
allows low tension strings to slow down enough to cluster. We have
presented a simple model that quantitatively encapsulates these
understandings and allows straightforward evaluation of microlensing
rates for stars in the galaxy, cusp and kink gravitational radiation
and two photon decays from axions in the standard model throat. Here, we
present forecasts for bursts for LIGO and LISA and note that clustering of loop
sources within the Galaxy raises the rates of detection and signal
strengths for low tension strings in these experiments.
Conversely, these results imply that stricter upper limits are
  achievable for bursts from strings in tension ranges where 
  local clustering dominates the signal.
Elsewhere we will discuss the implications for the stochastic background,
microlensing and
two photon production from axions.

\section*{Acknowledgment}

We thank Tom Broadhurst,  Eanna Flanagan, Ariel Goobar, Craig Hogan, Liam
Mcallister, Xavier Siemens, Masahiro Takada and Barry Wardell
 for valuable discussions. DFC acknowledges that this
material is based upon work supported by the National Science
Foundation under Grant No. 1417132.  SHHT is supported by the CRF
Grant HKUST4/CRF/13G and the GRF 16305414 issued by the Research
Grants Council (RGC) of the Government of the Hong Kong SAR.

\newpage

\begin{longtable}[c]{ccccccc}
  \label{tab:LIGOCusp}\\
  \toprule
  $\log_{10} G \mu$ & \multicolumn{6}{c}{Q}\\
                   & \multicolumn{2}{c}{Field Thy} & \multicolumn{2}{c}{Superstrings} & \multicolumn{2}{c}{Enhanced S/N} \\
  & Hmg & Cl & Hmg & Cl & Hmg & Cl \\
  \midrule
  \endfirsthead
  \toprule
  $\log_{10} G \mu$ & \multicolumn{6}{c}{Q}\\
                   & \multicolumn{2}{c}{Field Thy} & \multicolumn{2}{c}{Superstrings} & \multicolumn{2}{c}{Enhanced S/N} \\
  & Hmg & Cl & Hmg & Cl & Hmg & Cl \\
  \midrule
  \endhead
  \input{muQ-LIGO-CuspShort}
  \caption{Q as a function of string tension for LIGO detection
    rate $dR/d\log\rho=1$ yr$^{-1}$ of cusps
    for 3 cases: FT (field theory strings ${\cal G}=1$),
    SS (superstrings ${\cal G}=10^2$ and SS$^*$ (superstrings with
    improved PSD). Separate unclustered (Hmg) and clustered (Cl) calculations
    are reported for each type of source/experiment.
    Entries with $Q<0$ are suppressed.
  }
\end{longtable}

\begin{longtable}[c]{ccccccc}
  \caption{LIGO Kink}
  \label{tab:LIGOKink}\\
  \toprule
  $\log_{10} G \mu$ & \multicolumn{6}{c}{Q}\\
                   & \multicolumn{2}{c}{Field Thy} & \multicolumn{2}{c}{Superstrings} & \multicolumn{2}{c}{Enhanced S/N} \\
  & Hmg & Cl & Hmg & Cl & Hmg & Cl \\
  \midrule
  \endfirsthead
  \toprule
  $\log_{10} G \mu$ & \multicolumn{6}{c}{Q}\\
                   & \multicolumn{2}{c}{Field Thy} & \multicolumn{2}{c}{Superstrings} & \multicolumn{2}{c}{Enhanced S/N} \\
  & Hmg & Cl & Hmg & Cl & Hmg & Cl \\
  \midrule
  \endhead
  \input{muQ-LIGO-KinkShort}
  \caption{Q as a function of string tension for LIGO detection
    rate $dR/d\log\rho=1$ yr$^{-1}$ of kinks
    for 3 cases: FT (field theory strings ${\cal G}=1$),
    SS (superstrings ${\cal G}=10^2$ and SS$^*$ (superstrings with
    improved PSD). Separate unclustered (Hmg) and clustered (Cl) calculations
    are reported for each type of source/experiment.
    Entries with $Q<0$ are suppressed.
  }
  \end{longtable}

\begin{longtable}[c]{ccccccc}
  \caption{LISA Cusp}
  \label{tab:LISACusp}\\
  \toprule
  $\log_{10} G \mu$ & \multicolumn{6}{c}{Q}\\
                   & \multicolumn{2}{c}{Field Thy} & \multicolumn{2}{c}{Superstrings} & \multicolumn{2}{c}{Enhanced S/N} \\
  & Hmg & Cl & Hmg & Cl & Hmg & Cl \\
  \midrule
  \endfirsthead
  \toprule
  $\log_{10} G \mu$ & \multicolumn{6}{c}{Q}\\
                   & \multicolumn{2}{c}{Field Thy} & \multicolumn{2}{c}{Superstrings} & \multicolumn{2}{c}{Enhanced S/N} \\
  & Hmg & Cl & Hmg & Cl & Hmg & Cl \\
  \midrule
  \endhead
  \input{muQ-LISA-CuspShort}
  \caption{Q as a function of string tension for LISA detection
    rate $dR/d\log\rho=1$ yr$^{-1}$ of cusps
    for 3 cases: FT (field theory strings ${\cal G}=1$),
    SS (superstrings ${\cal G}=10^2$ and SS$^*$ (superstrings with
    improved PSD). Separate unclustered (Hmg) and clustered (Cl) calculations
    are reported for each type of source/experiment.
    Entries with $Q<0$ are suppressed.
  }
  
  \end{longtable}

\begin{longtable}[c]{ccccccc}
  \caption{LISA Kink}
  \label{tab:LISAKink}\\
  \toprule
  $\log_{10} G \mu$ & \multicolumn{6}{c}{Q}\\
                   & \multicolumn{2}{c}{Field Thy} & \multicolumn{2}{c}{Superstrings} & \multicolumn{2}{c}{Enhanced S/N} \\
  & Hmg & Cl & Hmg & Cl & Hmg & Cl \\
  \midrule
  \endfirsthead
  \toprule
  $\log_{10} G \mu$ & \multicolumn{6}{c}{Q}\\
                   & \multicolumn{2}{c}{Field Thy} & \multicolumn{2}{c}{Superstrings} & \multicolumn{2}{c}{Enhanced S/N} \\
  & Hmg & Cl & Hmg & Cl & Hmg & Cl \\
  \midrule
  \endhead
  \input{muQ-LISA-KinkShort}
  \caption{Q as a function of string tension for LISA detection
    rate $dR/d\log\rho=1$ yr$^{-1}$ of kinks
    for 3 cases: FT (field theory strings ${\cal G}=1$),
    SS (superstrings ${\cal G}=10^2$ and SS$^*$ (superstrings with
    improved PSD). Separate unclustered (Hmg) and clustered (Cl) calculations
    are reported for each type of source/experiment.
    Entries with $Q<0$ are suppressed.
  }
  \end{longtable}

\newpage

\appendix

In these appendices we provide a full set of details for the calculation
of the loop population for low tension strings formed predominantly
in the radiative era.

\section{Nambu Goto dynamics}

\label{NGderived}
We begin by summarizing the dynamical description of a idealized
Nambu Goto string. Start with the spacetime metric $g_{\mu\nu}$ in 3+1 dimensions. The
spacetime location of a two dimensional worldsheet is
$x^\mu = x^\mu(\sigma,\tau) = x^\mu(\zeta^a)$ where we take $a=0$ (timelike)
and $1$ (spacelike).  The induced
metric is $\gamma_{ab}=g_{\mu\nu} x^{\mu}_{,a} x^{\nu}_b$. The worldsheet action is
\be
S = \int {\cal L} \sqrt{-\gamma} d^2\zeta
\ee
where ${\cal L}$ is a Lagrangian density and $\gamma$ is
$\det(\gamma_{ab})$. The Nambu action is the simplest possibility
${\cal L}= -\mu$, a constant.

Varying the action gives the equations of motion
\be
\gamma^{ab}x^\rho_{,a}x^\sigma_{,b} \Gamma^\kappa_{\sigma\rho} +
\frac{1}{\sqrt{-\gamma}} \frac{\partial}{\partial \zeta^a}
\left( \sqrt{-\gamma} \gamma^{ab} x^\kappa_{,b} \right)
= 0 .
\ee
The stress energy tensor at spacetime point $x^\alpha$,
$T^{\mu\nu}(x^\alpha)$, is
\ba
T^{\mu\nu}\sqrt{-g}|_{x^\alpha} 
& = & 2 \frac{\delta S}{\delta g_{\mu\nu}(x^\alpha)} \\
& = & -\mu \int d^2\zeta \sqrt{-\gamma} \gamma^{ab}x^\mu_{,a}x^\nu_{,b}
\delta^{(4)}\left( x^\alpha - x^\alpha(\zeta) \right) .
\ea

\section{Flat, expanding universe}
\label{flatexpandinguniverse}
Consider the metric of the form 
$ds^2=a(\eta)^2 \left( -d\eta^2 + d{\bf x}^2 \right)$
where $\eta$ is conformal time $d\eta=dt/a(t)$ and ${\bf x}$ is
comoving coordinate. 
Abbreviate ${\dot x}=\partial x^{\mu}/\partial \zeta^0$ and
${x'} = \partial x^{\mu}/\partial \zeta^1$,
make the gauge choices ${\dot x} \cdot {x'}= 0$ and $\zeta^0=\eta$,
and we have
\ba
x^\mu_{,\eta} & = & (1, {\dot {\bf x}}) \\
x^\mu_{,\sigma} & = & (0, {{\bf x}'}) \\
{\dot {\bf x}} \cdot {\bf x'} & = & 0
\ea
Now define
\be
\epsilon \equiv \sqrt{\frac{ {{\bf x}'}^2}{1 - {\dot {\bf x}}^2}}
\ee
and the equations of motion are
\ba
\frac{2 {\dot a}}{a} {\dot {\bf x}}^2 + \frac{\partial \log \epsilon}{\partial \eta} & = & 0 \\
-\frac{ \partial (a^2 \epsilon {\dot x}^i)}{\partial \eta} +
\frac{ \partial (a^2 {x'}^i \epsilon^{-1} )}{\partial \sigma} & = & 0
\ea
with $\dot x^i = \partial x^i/\partial \eta$, conformal time $d\eta=dt/a$
and comoving ${\bf x}$.

The energy or momentum in a comoving volume is
\ba
\int T^{0\alpha} \sqrt{-g} d^3x & = & \mu a \int d\sigma \epsilon \left( 1, \frac{d{\bf x}}{d\eta} \right)^\alpha \\
\epsilon & = & 
\sqrt{\frac{|{\bf x}'|^2}{1-|{\frac{d{\bf x}}{d\eta}}|^2}}
\ea
If we switch from conformal to physical time we have
\ba
\int T^{0\alpha} \sqrt{-g} d^3x & = & \mu \int d\sigma \epsilon \left( 1, {\frac {d{\bf x}}{dt}} \right)^\alpha\\
\epsilon & = & 
\sqrt{\frac{|{\bf x}'|^2}{1-|{\frac{d{\bf x}}{dt}}|^2}} .
\ea

The energy and momentum for normal cosmological time and comoving
coordinates is
\ba
\left( E, P^i \right) & = & \int T^{0\alpha} g_{00} \sqrt{-g} d^3y \\
  & = & a \int d\sigma \epsilon \left( 1, {\dot x}^i \right) .
\ea

\subsection{Kinks and Cusps}

In flat space, the evolution of a closed string ${\bf x}(\sigma, t)$ is given by, in the gauge
$${\dot {\bf x}} \cdot {\bf x'}  =  0, \quad \quad {\dot {\bf x}^2} + {\bf x'}^2  =  0$$
\be
{\bf x}(\sigma, t)=\frac{1}{2}[{\bf a}(\sigma -t) +{\bf b}(\sigma+t)]
\ee
where 
\be
{\bf a'}^2 ={\bf b'}^2=1
\ee
Here $\sigma$ is the length parameter along the string coordinates ${\bf a}(\sigma)$ and ${\bf b}(\sigma)$ so the curves ${\bf a'}$ and $ {\bf b'}$ move on the surface of an unit sphere. A cusp is formed when they intersect while a gap in either curve indicates a kink.
Physically, a cusp appears periodically, while a kink moves around the closed string loop continuously.
The gravitational burst from a cusp has a distinct wave form $|t|^{1/3}$ while that from a kink has the wave form $|t|^{2/3}$ \cite{Damour:2004kw}.

\section{String network with gravitational radiation and
  axion emission} 

We outline a model for the string loop network which starts from
Kibble's original network model \cite{Kibble:1984hp},
supplemented with improved
understanding of the spectrum of loops created and allowing for
evaporation by both
gravitational radiation and axion emission.
Our goal is to describe the size
spectrum of loops in cosmology and the clustering enhancement of loops
within bound cosmological objects like our own Galaxy.

Let $\zeta^i=(\zeta^0,\zeta^1)$ be time-like and spatial coordinates
of the string world sheet. The string tension is $\mu$ with dimensionless form $G \mu$.  Choose the
gauge with $\zeta^0=\tau$ where $\tau$ is the conformal time in
cosmology ($d\tau = dt/a(t)$ for physical time $t$, scale factor $a(t)$ in
the preferred FRW frame) and
$\zeta^1=\sigma$. The energy $E$ is
\ba
E & = & \mu a \int d\sigma \epsilon \\
\epsilon & \equiv & \sqrt{\frac{ {{\bf x}'}^2}{1 - {\dot {\bf x}}^2}}
\ea
and ${\bf x}$ is the comoving coordinate along the string.
Dot means derivative with respect to conformal time $\tau$ and
prime means with respect to $\sigma$. The equations of motion are
\ba
\frac{2 {\dot a}}{a} {\dot {\bf x}}^2 + \frac{\partial \log \epsilon}{\partial \tau} & = & 0 \\
-\frac{ \partial (a^2 \epsilon {\dot x}^i)}{\partial \tau} +
\frac{ \partial (a^2 {x'}^i \epsilon^{-1} )}{\partial \sigma} & = & 0 .
\ea
From these we can write the energy change as
\ba
{\dot E} & = & \frac{{\dot a}}{a} E \left( 1 - 2 \left< v^2 \right> \right) \\
\left< v^2 \right> & \equiv & \frac{\int d\sigma \epsilon {\dot x}^2}{\int d\sigma \epsilon}
\ea
Note that rate of energy change has a similar form when expressed in
terms of cosmological time (${\dot E} \to a dE/dt$
and ${\dot a} \to a da/dt$) and reads
\be
\frac{dE}{dt} = \frac{1}{a} \frac{da}{dt} E \left( 1 - 2 \left< v^2 \right> \right) .
\ee
This description applies to isolated long, horizon crossing
strings and loops in cosmology. It omits collisional
interactions (intercommutation leading to chopping
and reconnection) and radiative dissipation.

Consider physical volume $V$ and the energy in a network of long
(horizon-scale) strings of tension $\mu$. The relation between the
length and energy of a string is $L=E/\mu$. $L$ is defined such
that in a volume $V=L^3$ the string is
of length $L$. The physical energy density is $\rho_{\infty} = \mu L/V =
\mu/L^2$.  The effective number density of strings (of length L) is
$1/L^3$, the number per area is $\sim 1/L^2$. Collisions generate
loops which are not counted in length $L$.

A string segment of length $dL$ (a part of the long string network)
moving with velocity $v$ will encounter $dL v \delta t/L^2$ other
segments in time $\delta t$ to give an effective encounter rate is $dL
v/L^2$. Let $p$ be the (macroscopic)
intercommutation probability per collision. For
field theory strings $p \sim 1$ but for superstrings $p$ can be
substantially smaller, i.e. $p \sim 10^{-3}$. The rate at which the length $dL$ suffers
reconnections is $p dL v/L^2$. Length $L = \int dL $ experiences
intercommutation rate $p v/L$.

A scaling solution for the network is one in which all characteristic
lengths scale with the cosmological horizon size.
Simulations of string networks find that the scaling solution
is an attractor for cosmologies with
powerlaw expansion, $a(t) \propto t^\beta$ (with $\beta=1/2$ for radiation era and
$\beta=2/3$ for matter era). In powerlaw models the
horizon scales $\propto t$.
The convergence of macroscopic properties of the string
network to the scaling solution is rapid, generally a
few doublings of the horizon suffices.  We assume that the network will be
close to scaling in the sense that the ratio of each
network length scale to the horizon is nearly constant in time
even if the assumption that the horizon $\propto t$ is not exactly satisfied.
Length $L$ is an example: $L = \gamma t$ for slowly varying $\gamma$.

Intercommutations produce loops from the network.
Let the loop have length $\ell$ with energy $\mu \ell$.
Write $y=\ell/L$ where $0<y<1$
and let $a(y)$ be the PDF for loop size $y$
to be cut out. The PDF is a function of $y$ because
$\ell$, $L$ and the horizon scale together.
The PDF satisfies $\int dy a(y)=1$. The expected energy
transformed from network to loops with sizes $y$ to $y+dy$ in a
single intercommutation is $\mu L y a(y) dy = E y a(y) dy$.  The
energy transformation from the network to loops in a comoving volume
is
\ba
\frac{d^2 E_{\infty \to \ell}}{dt dy} & = & E \frac{p v}{L} y a(y) \\
\frac{d E_{\infty \to \ell}}{dt} & = & \int dy \frac{d^2 E_{\infty \to \ell}}{dt dy} . 
\ea
There is no energy ``flux'' into or out of the comoving volume.

Conversely, a small loop of size $\ell = y L$ and energy
$\mu \ell$ encounters a long segment of string and
reconnects with probability $p$ and rate $p \ell v/L^2$.
Assume the loop's entire energy is transferred back to the network.
The energy transformation from loops to network in a comoving volume
is
\ba
\frac{d^2 E_{\ell \to \infty}}{dt dy} & = & \frac{dE_{\ell}}{dy} \frac{p v \ell}{L^2} 
                                = \frac{dE_{\ell}}{dy} y \frac{p v}{L} \\
\frac{d E_{\ell \to \infty}}{dt}    & = & \int dy \frac{d^2 E_{\ell \to \infty}}{dt dy} .
\ea
Here $y$ appears as an explicit factor
because the size of the loop influences
the rate of reconnection. In writing $dE_{\ell}/dy$ as a
function of $y$ we are again
assuming that the distribution of loop sizes, $L$ and the horizon
scale together.

The energy in the network's long strings is increased by stretching, lost
by formation of loops and gained by reconnection. Assume that the
loops are small compared to the horizon scale and the effect of
stretching is negligible. This is a reasonable assumption
  because even the largest loops formed are $\lta 0.1$
  the size of the horizon. The total energy in loops is increased by
the formation of loops and lost by reattachement to the network. These processes
are
\ba
{\dot E}_{\infty} & = & 
\frac{{\dot a}}{a} E \left( 1 - 2 \left< v^2 \right> \right)
- \frac{d E_{\infty \to \ell}}{dt} + \frac{d E_{\ell \to \infty}}{dt} \\
{\dot E}_{\ell} & = & \frac{d E_{\infty \to \ell}}{dt} - \frac{d E_{\ell \to \infty}}{dt} .
\ea
This is essentially Kibble's original model \cite{Kibble:1984hp}
with the addition of $p$, the macroscopic probability of intercommutation,
that can differ greatly between field strings and superstrings.

Consider the fate of a loop that has been cut out of the network and
radiates in isolation. For purely gravitational wave emission $dE_{GW}/dt =
\Gamma G \mu^2 c$ with $\Gamma \sim 50$ based on studies of a variety
of simple loops having a few kinks or cusps. The characteristic
timescale for a loop of initial physical size $\ell$ to evaporate
completely is $\Delta t = \ell /(\Gamma G \mu)$, so $dE/dt = dE_{GW}/dt$.

Many loops of different sizes are continually created by network
intercommutation. If the loop formed when the universe was young
then the time available today is $\Delta t \sim t_0$, where $t_0$ is
the age of the universe. A given physical process in the scaling
solution creates loops with size proportional to the horizon scale.
We write the characteristic loop size for the process
as $\alpha t$ for constant $\alpha$; the process accounts for
a fraction of all network losses $f$ (so $0<f<1$). Loops
formed at time $t$ by the process such that $0 < t < t_0 \Gamma \chi G
\mu/\alpha$ evaporate by the current
epoch. Conversely, loops formed at $t_0 \Gamma G \mu/\alpha
< t < t_0$ are present with size diminished from that at formation.

We can estimate the total loop length contributed by the
process and still present in the universe today as
$L_\ell = (E_{\ell}/\mu)$. The largest contribution to $E_{\ell}$ comes
from the small end of the loop distribution function, i.e. the oldest
loops created by that process
that have not yet evaporated. Ultimately the small end dominates
because the universe was densest when the oldest loops were formed.
If the formation epoch is the radiation era
(quantitatively,
$\Gamma G \mu/\alpha < t_{eq}/t_0 \simeq 3.5 \times 10^{-6}$)
the total string length is
$L_\ell/t \sim (\alpha/\Gamma G \mu) f \Psi $
where $\Psi = 1$ in the radiation era and $\Psi = (t_{eq}/t_0)^{1/2}$
for matter era ($t_{eq}<t_0$). We will provide proportionality constants
later.

The ratio of energy in loops to horizon cross strings is
\be
\frac{E_{\ell}}{E} \sim \left( \frac{c^2 \alpha}{\Gamma G \mu}\right ) f \Psi .
\ee
Not-too-small $f \alpha$ and very small $ \mu$ ensure a dominant
loop component. This is typical of the case we have explored
in the past: superstrings with $G \mu << 10^{-7}$, a modest
number of loops comparable to the
horizon scale ($f=0.1-0.2$, and $\alpha \sim 0.1$).

Even though loops dominate in the sense $E_{\ell}/E > 1$
the rate of loop reattachment to the network is a
small effect.  We can estimate
\be
   \frac{d E_{\ell \to \infty}}{d E_{\infty \to \ell}}
   \sim
   \frac{\alpha f \Psi}{\int y a(y) dy} .
\ee
String network simulations give $f \alpha \lta 10^{-2}$ for
the large loops,
$\int y a(y) dy \sim 0.2$ so we have ${d E_{\ell \to \infty}}/{d
  E_{\infty \to \ell}} \lta 5 \times 10^{-2} \Psi$ where
$\Psi = 1$ during the radiation era and dropping to $\Psi \sim 10^{-3}$
in today's matter era. The remaining part
of the long string length chopped out of the network turns
into very small loops with size $\alpha \sim (G
\mu/c^2)^\eta$, exponent $\eta \sim 1.5$ (radiation era) or
$\sim 1.2$ (matter era), and contributes
practically nothing to the reattachment rate.

Ignoring reconnect terms we have
\ba
{\dot E}_{\infty} & = &
\frac{{\dot a}}{a} E \left( 1 - 2 \left< v^2 \right> \right)
- \frac{d E_{\infty \to \ell}}{dt}
\\
{\dot E}_{\ell} & = & \frac{d E_{\infty \to \ell}}{dt} .
\ea
Letting $C=\int dy y a(y)$ we have
\ba
{\dot E}_{\infty} & = & 
\frac{{\dot a}}{a} E \left( 1 - 2 \left< v^2 \right> \right)
- C E \frac{pv}{L}
\\
{\dot E}_{\ell} & = & C E \frac{pv}{L} .
\ea

Let the comoving coordinates be defined at time $t_0$ so
physical volume $V_0$ equals the comoving volume today.
At other times the physical volume defined by fixed
comoving coordinates is $V = (a/a_0)^3 V_0$.
In the scaling solution the energy in the comoving
volume $E = E_{\infty} = \rho_{\infty} V$. Here,
$\rho_{\infty}$ is the physical energy density.
When the scaling solution is achieved
$\rho_{\infty} \propto 1/t^2$ and the source term for the loops is
\ba
\frac{{\dot E}_{\ell}}{a^3} & = & -\frac{d \rho_{\infty}}{dt} - 2 \rho_{\infty} H (1 + \left< v^2 \right> ) \\
& = & \frac{2 \rho_\infty}{t} \left(1 - H t \left( 1 + 
\left< v^2 \right> \right) \right) .
\ea
To proceed, $\rho_{\infty}$ and $\left< v^2 \right>$ are inferred from
string network simulations in the scaling regime.

An alternative approach which we adopt is to evaluate
the loop creation rate directly
\ba
\frac{{\dot E}_{\ell}}{a^3} & = & C \rho_{\infty} \frac{pv}{L} \\
& = & C \mu \frac{pv}{\gamma^3 t^3}
\ea
where $\rho = \mu/L^2 = \mu/\gamma^2 t^2$. We solve for $\gamma$ as
part of a velocity one scale model.
In exact scaling $\gamma$, $Ht$, $v^2$ and $C$ are constant in time
but we will now allow all to vary slowly.
Anticipating $\gamma \propto p$ we expect the combination
\be
\frac{p^2 t^3}{\mu} \frac{{\dot E}_{\ell}}{a^3} = C \frac{p^3 v}{\gamma^3} .
\label{integrateEdot}
\ee
should be nearly constant for a range of $p$, $\mu$ and times.

The summary of the model is
\ba
\label{modelstart}
\rho_\infty & = & \frac{\mu}{ \gamma^2 t^2} \\
\frac{t}{\gamma}\frac{d\gamma}{dt} & = & -1 + Ht \left( 1 + v^2 \right) + \frac{C(t) p v}{2 \gamma}
\ea
with equation for velocity
\ba
\frac{dv}{dt} & = & \left( 1 - v^2 \right) H \left( \frac{k(v)}{Ht \gamma} - 2 v \right) \\
k(v) & = & \frac{2 \sqrt{2}}{\pi} \frac{1 - 8 v^6}{1 + 8 v^6}
\ea
and chopping
\ba
C(t) & = & \frac{c_r + \frac{g c_m}{1+z}}{1 + \frac{g}{1+z}} \\
g & = & 300 \\
c_r & = & 0.23 \\
c_m & = & 0.18
\label{modelend}
\ea
where $k(v)$ and $C(t)$ are fits\cite{Martins:1995tg,Martins:2000cs,Battye:1997hu,Pogosian:1999np}.
To solve these two coupled ODEs for $\gamma$ and $v$
we begin at large $z$ when $Ht=1/2$ setting the left hand sides to
zero ($d/dt \to 0$) and find an equilibrium point for $\gamma$ and
$v$.  As $z$ decreases the system begins to evolve because $C(t)$ and $H t$ vary.
The chopping function $C(t)$ has been fit from numerical
simulations and $H t$ follows from the multicomponent $\Lambda$-CDM
cosmology. The specifics of the $\Lambda$-CDM cosmology are given in
Appendix \ref{implementationLCDM}. We integrate Eq.(\ref{integrateEdot})
to find the loop density. Note that the subsequent evolution of the
loops that are created do not effect the solution for $\gamma$ and
$v$ in any way.

\section{Results}

\begin{figure}[ht]
  \centering
\includegraphics{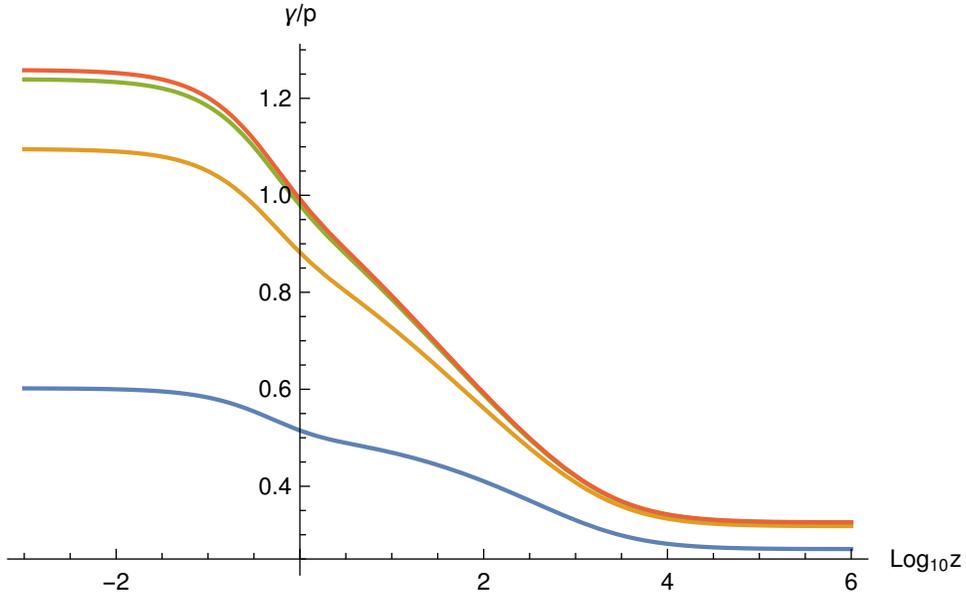}
\caption{\label{gammap} The string separation $\gamma/p$ where $L=\gamma t$
  for  $p=1$ (bottom, blue) to $p=10^{-3}$ (top, red) in powers of 10.}
\end{figure}

The result is $L = \gamma t$ where $\gamma$ would be
exactly constant for a scaling network solution with a pure power law cosmology.
Fig. \ref{gammap} displays $\gamma/p$ as a function of redshift
in the $\Lambda$CDM model. The multiple lines are 
for intercommutation probability $p=10^{-3} \to 1$ in powers of $10$.
The variation with redshift for $0 < \log_{10} z < 6$ is $\lta 3$
and with $p$ is $\lta 2$.
The detailed shape depends upon the background cosmology
and the assumed form for the chopping function. 
At late times $\Lambda$ alters the expansion away from
power law so that the chopping function (extracted from power-law string network
simulations) is probably inapplicable. Late time values for $\gamma/p$
should not to be taken too seriously for $\log_{10} z < 0$.
The part of the plot that is most relevant to the loop distribution
today is large $\log_{10} z$, because small $\mu$ means that the
loops formed long ago. The results are rather simple
$\gamma \propto p$ or $\rho_{\infty} \propto 1/p^2$.

These results are inconsistent with network simulations that show much more
modest enhancement of $\rho_\infty$ as the microscopic intercommutation
probability is diminished.
Strings in a realistic network are dense with small scale
structure. The intercommutation probability for such strings depends
not only on the microscopic intercommutation probability for long,
straight strings but also on the complexity of the collision of two
strings with small scale structure. One macroscopic string collision
may involve many repeated microscopic encounters \cite{Avgoustidis:2005nv}.

Let $p$ be the intercommutation probability for two segments of the
network and $q$ the intercommutation probability for two straight
infinite strings. These are related but distinct probabilities because
of the possible small scale structure that the network segments
possess.  Assume that small scale structure implies $N$ independent
collision attempts take place when two macroscopic network strings
meet. The probability for no collision is $e^{-N q}$ and one or more
collisions, the macroscopic probability, is $p=1-e^{-N q}$.  For $N q
<< 1 $ we have $p \sim N q$ whereas if $N q >> 1$ we have $p \sim 1$.
A network simulation for $\gamma$ in which small scale structure has
built up on the long, horizon crossing strings
will scale with $q$ in the same manner as with $p$ if $N$ is independent of
$q$ and $N q << 1$.  In that case, $\gamma \propto p \propto q$.  If
$N q >> 1$ or if $N$ depends upon $q$ then there is no reason to
expect the
macroscopic and microscopic probabilities to be linearly related. In
fact, $\gamma$ might be nearly constant.

The degree of microscopic structure, parameterized above by $N$, is
critical for understanding how $\gamma$ and loop production varies
with intercommutation probability $q$. At this point
the relevant scales have not been directly probed via simulations
for the long, horizon crossing string network.  We
will use $p$ as a parameter keeping in mind the substantial
uncertainty in linking the intercommuntation probability $q$
calculable in string theory to the macroscopic intercommutation
probability $p$ that appears in the one scale model.

\begin{figure}[ht]
\centering
  \includegraphics{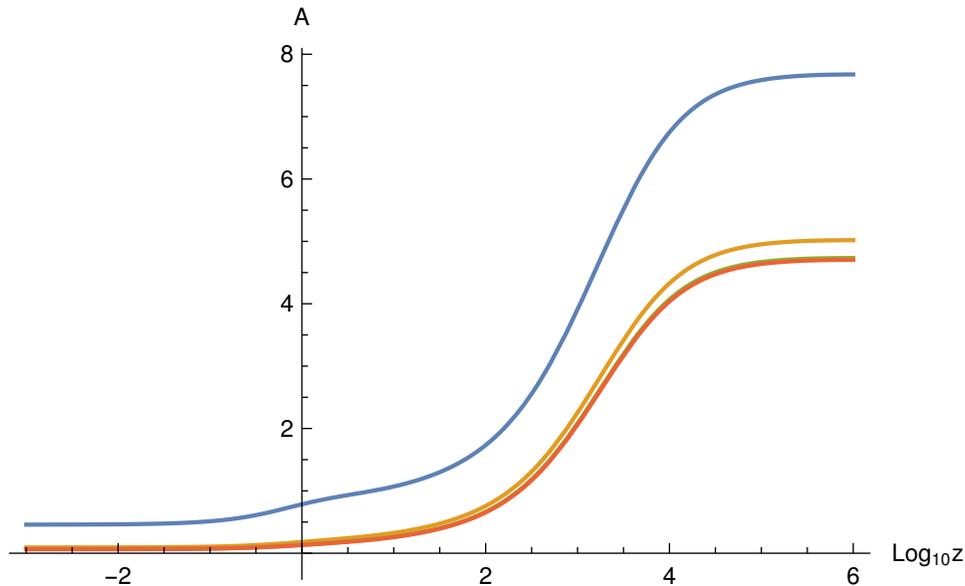}
  \caption{\label{edot1list} The scaled loop source for $p=1$ (top, blue) to $p=10^{-3}$ (bottom, red) in powers of 10.}
\end{figure}

The results for the loop energy source ${\dot E_{\ell}}$ as a function
of redshift is given in Fig. \ref{edot1list}. The quantity plotted
is ${\cal A} = (p^2 t^3/\mu) a^{-3} {\dot E_{\ell}}$ for $p=10^{-3} \to 1$.
We see that ${\cal A}$ varies modestly with redshift and with $p$.
As with $\gamma$ late time values for ${\cal A}$
should not to be taken too seriously for $\log_{10} z < 0$.
The part of the plot that is most relevant to the loop distribution
today for microlensing is large $\log_{10} z$. The results
are rather simple: ${\dot E}_\ell \propto {\cal A}/p^2$
and there is only about a factor of $2$ variation in $\cal A$ as
$p$ ranges from $10^{-3}$ to $1$ near $\log_{10} z = 5$.

An approximate fit is
\be
\frac{{\dot E_{\ell}}}{a^3} = \frac{\mu}{p^2 t^3} {\cal A}
\ee
where
\be
{\cal A} = {\cal A}_p {\cal A}_z 
\ee
is given in Appendix \ref{fittoA}.

\section{Numbers of loops today in the universe as a whole}

Loops formed at a given time $t$ will have size
$\alpha t$. As described previously, about
80\% of the energy goes into loops with $\alpha \sim (G \mu)^{1.2}$ (radiation
era) or $\sim (G \mu)^{1.5}$ (matter era) and about 20\% goes into loops
with $\alpha \sim 0.1$. Set ${\it f}=0.2$ to describe the
fraction of long string length that ends up in long loops.
The comoving energy and the loop energy density
is
\be
E_{\ell} = e_{\ell} V = e_{\ell} a^3 
\ee
where we have set $V_0/a_0^3=1$.
The velocity one scale model implies
\be
\frac{{\dot E_{\ell}}}{a^3} = \frac{\mu}{p^2 t^3} {\cal A}
\ee
and integrating in time
\be
\int dt {\dot E_{\ell}} = \int dt a^3 \frac{\mu}{p^2 t^3} {\cal A} .
\ee
For short time increments this is
\be
\Delta \left( e_{\ell} a^3 \right) = \Delta t a^3 \frac{\mu}{p^2 t^3} {\cal A}
\ee
Large loops of a given size $\alpha t$ are made only
at one instant so if we distinguish the loops of size then
$a$ is nearly constant and
\ba
\frac{d e_{\ell}}{dt d\ell} & = & \frac{{\it f} \mu}{p^2 t^3} {\cal A} \delta(\ell - \alpha t) \\
\frac{dn_{\ell}}{dt d\ell} & = & \frac{de_{\ell}}{dt d\ell} .
\frac{1}{\mu \ell} 
\ea
Restoring ``c'' the
birth rate density for loops of size $\ell_b$ born at time $t_b$ is
\be
\left( \frac{dn_{\ell}}{dt d\ell} \right)_b =
\frac{{\it f}}{\alpha p^2 t_b^4} {\cal A} \delta(\ell_b - \alpha c t_b)
\ee
and the resultant density size distribution at time $t$
\ba
\frac{dn_{\ell}}{d\ell}(t,\ell)  & = & \frac{1}{a^3} \int dt_b d\ell_b a_b^3 \left( \frac{dn_{\ell}}{dt d\ell} \right)_b \delta \left( \ell - \ell\left[ \ell_b, t_b, t \right] \right) \\
\ell\left[ \ell_b, t_b, t \right] & = & \ell_b -  \Gamma G \mu \left( t- t_b \right)  .
\ea
Integrating over $\ell_b$ and $t_b$ we have
\ba
\Phi & = & 1 + \frac{\Gamma G \mu}{\alpha} \\
t_b & = & \frac{\ell + \Gamma G \mu t}{\alpha \Phi } \\
\frac{dn}{d\ell} & = & 
\left( \frac{ {\cal A}_b {\it f} \alpha^2} {p^2} \right)
\left( \frac{a_b}{a} \right)^3  
\frac{\Phi^3}
{\left( \ell + \Gamma G \mu t \right)^4}
\ea
where we require $t_b < t$ or $\ell < \alpha t$. The quantity
$({\cal A}_b f \alpha^2/p^2)$ is a slowly varying function of time (on
account of ${\cal A}$ but we will
treat the value as constant, taking typical numerical values at $t_b$
in the radiation era because the numerous small loops are of
greatest observational interest. The form for
$dn/d\ell$ peaks at $\ell=0$ but the quantity of interest for
observational purposes is usually weighted by $\ell$ (or higher
powers). The characteristic dissipative
scale of the loop at the end of its life is $\ell_d \equiv \Gamma G \mu t$.
For a simple numerical estimate today $t=t_0$ when most loops are generated
in the radiative era $t<t_{eq}$ we write
\ba
\frac{a_b}{a} & = & \left( \frac{a_b}{a_{eq}}\right) \left( \frac{a_{eq}}{a} \right) \\
& \simeq & \left( \frac{t_b}{t_{eq}}\right)^{1/2}
\left( \frac{t_{eq}}{t} \right)^{2/3}
\ea
and using $x \equiv \ell/\ell_d$ we have
\ba
\ell \frac{dn}{d\ell} & = & \frac{x}{(1+x)^{5/2}}
\left( \frac{ {\cal A} {\it f} } {p^2} \right)
\left( {\Gamma G \mu} \right)^{-3/2}
\left( \frac{\alpha t_{eq}}{t_0} \right)^{1/2}
\left( \frac{1}{t_0} \right)^3
\ea
The approximate expression suitable for describing the
today's loops formed in the radiative era,
explicit numerical expressions for density of loops,
the characteristic length and mass of the loops just evaporating,
and the mass densities today are given in the main text.

It is also useful to have accurate descriptions of the loop
distribution at earlier
times. When the universe experiences pure power law expansion $a \propto t^\beta$
the loop distribution function at time $t$ is
\ba
\ell \frac{dn}{d\ell} & = & \frac{x}{(1+x)^{4 - 3 \beta}}
\left( \frac{ {\cal A} {\it f} } {p^2} \right)
\left( \Gamma G \mu \right)^{-3 + 3 \beta}
\alpha^{2-3\beta}
\left( \frac{1}{t} \right)^3 .
\ea
During the radiation era ($\beta=1/2$) the result is identical
to Eq.(\ref{approxdndl}) except for the factor $(t_{eq}/t)^{1/2}$
which accounts for the switch from radiation to matter expansion
at $t_{eq}$. One derives $\ell_d$, $M_{\ell_d}$,
$d\rho/d\ell$, $d\rho/d\log M_{\ell}$ and $d\Omega_\ell/d \log M_{\ell}$
for an observer at time $t<t_{eq}$.

\section{String Loop Clustering}

Network simulations show that a loop cut of the network at time $t$
with length $\ell = \alpha t$ and
$10^{-3} < \alpha < 10^{-1}$ has a moderately relativistic
center of mass motion. The rms velocity $\sim 0.5-0.9$. Smaller loops typically
move faster than larger ones.
The loop lives $\Delta t$ before shrinking to zero size.
The characteristic number of e-foldings of the universe before the
loop evaporates,
$H \Delta t \sim \alpha /(\Gamma G \mu)$,
is large if $\alpha /\Gamma G \mu >> 1$. 
Cosmic superstrings can have very small tensions since $\mu$ is
exponentially warped.

The universe's expansion steadily damps any free particle's peculiar motion.
If a slow moving loop is at the right
position when a perturbation begins to form then it can be captured. In the
normal process of structure formation, cold dark matter is
captured in precisely this manner.

Cold dark matter must be long-lived and, once accreted,
is permanently bound to the nonlinear object ultimately formed.
Superstring loops radiate energy steadily and their residence is temporary.
The pattern of a loop's emission is generally anisotropic, leading to a
net recoil on the center of mass of the loop, the so-called ``rocket effect.''  When
$\mu$ is less than a critical value the loop experiences the following
sequence of events: (1) damped
peculiar motion, (2) possible capture by growing gravitational
perturbation, (3) residence in the bound object for a
time comparable to $\sim \Delta t$, (4) ejection by the rocket effect
towards the end of its life, (5) complete evaporation
in the IGM. By contrast, when $\mu$ is too big the acceleration from the
rocket effect occurs before capture. Such a loop never
slows and evaporates in the IGM.

This process was studied in some detail in ref. \cite{Chernoff:2009tp}.
An analytic estimate of
the critical string tension for capture to radius $r$ in the Galaxy is
\be
\left( G \mu \right)_{critical}
= 4.12 \times 10^{-9} 
\left( \frac{\alpha}{0.1} \right)
\left( \frac{ 0.1}{v_i} \right)^{3/2}
\left( \frac{10 {\rm kpc}}{r} \right)^{5/16}
{\rm min} 
\left(
1, \left( \frac{r}{8.5 {\rm kpc} } \right)^{5/4}
\right)
\ee
where $v_i$ is the initial loop velocity. The criterion
$\mu < \mu_{critical}$ leads to capture. This is a conservative
  estimate for clustering. It is based on assuming that the rocket
  effect is maximally effective in two senses. First, the orientation of
  the force on the center of mass of the loop is fixed. The loop
  receives the maximum cumulative impulse.
  The loop breaks free the gravitational potential
  at the earliest possible moment. Second, the
  magnitude of the rocket effect is estimated using loops with
  cusps. Studies of various loop configurations (with a few kinks or cusps) show
  that the cusps generate the greatest anisotropy in gravitational wave emission.

To check this understanding ref. \cite{Chernoff:2009tp} carried out a
numerical calculation of how loops (born from a scaling string network)
interact with a growing, galactic scale perturbation in cosmology.
Initial conditions (loop sizes, velocities) were sampled
and loop positions calculated. All the dynamical phases above
were identified. The summary is simple:
small tension loops of all sizes simply track the dark matter that collapses to
form the bound object.

Let the enhancement of the dark matter within the Galaxy today
with respect to the universe as a whole be
\be
{\cal E} \equiv \frac{\rho_{DM}}{\left<\rho_{DM}\right>}
\ee
where ${\cal E}$ and $\rho_{DM}$ are
spatially dependent. The simulation shows that the
enhancement of the loops in the Galactic halo with respect
to the universe as a whole, ${\cal F}$, is
proportional to ${\cal E}$ with proportionality constant
that is primarily a function of tension:
\ba
   {\cal F} & = & {\cal E} \beta(\mu)
\ea
If loops followed CDM perfectly then $\beta=1$. In fact,
for the Galactic halo 
\ba
   \beta(\mu) & = & 10^{f(y)} \\
y & = & \log_{10} \mu_{-15} \\
f(y) & = &
\left\{
\begin{array}{cc}
  -0.337 - 0.064 y^2 & {\ \ \rm for \ \ } 0 \le y < 5 \\
  -0.337 & {\ \ \rm for \ \ } \  y < 0
\end{array}
\right.
\ea
For $G\mu < 10^{-15}$ the limiting enhancement is ${\cal F} \sim 0.4 {\cal E}$.
In the limit $\mu \to 0$ we do not have
${\cal F} \to {\cal E}$ because the rocket effect always removes
the loop from the Galaxy before it fully evaporates.

At the local solar position we estimate ${\cal E} \sim 10^{5.6}$ so
${\cal F}$ is a huge enhancement with respect to the
universe as a whole for low tension strings.

Assume the anisotropy
of the gravitational wave
emission. As in the original study, this assumption minimizes the
extent of clustering because it maximizes the effect of the rocket
recoil and removes loops from the halo as soon as possible.  The
capture and ejection of the loops is the same for the joint axion and
gravitational case as it is for the pure gravitational case with the
substitution $\mu \to \mu$. The degree of clustering is set by
the tension-dependent proportionality constant that becomes $y =
\log_{10} \mu_{-15}$; the spectrum $dn/d\ell$ is given by
eq. \ref{approxdndl}.

\section{Galactic fit to the String Density}
The dark matter halo of the galaxy is from Binney and Tremaine
and fully described in Appendix \ref{darkmattermodel}. One
version is for a NFW-like cusp at the center and the other
is for a core. We adopt a spherical version of the model
$\rho_{DM}({\vec r})$. The dark matter enhancement is
\ba
   {\cal E} & = & \frac{\rho_{DM}}{\rho_c} \frac{\rho_c}{\left<\rho_{DM}\right>}
   = \frac{\rho_{DM}}{\Omega_{DM} \rho_c} ,
\ea
the critical density (for $H_0=70$ km/s/Mpc) is
\be
\rho_c = \frac{3 H_0^2}{8 \pi G} \sim 1.36 \times 10^{-7} \msun {\rm pc}^{-3} 
\ee
and the dark matter fraction is
\be
\Omega_{DM} = \frac{\left<\rho_{DM}\right>}{\rho_c} \sim 0.25 .
\ee
Numerically, the spatially varying, dark matter enhancement is
\be
{\cal E}({\vec r}) = 2.94 \times 10^{7} \frac{\rho_{DM}({\vec r})}{ \msun {\rm pc}^{-3} }  .
\ee
The string enhancement is
\be
   {\cal F}({\vec r}) = {\cal E}\beta(\mu)
\ee
and the spatially varying local string density as
\be
\frac{dn}{d\ell} = \left< \frac{dn}{d\ell} \right> {\cal F}({\vec r})
\ee
and the bracket is the average for the loop density for the
universe as a whole. Note that the string loop distribution,
a function of $\ell$, is enhanced by a single factor.

\section{Implementation for $\Lambda$-CDM}
\label{implementationLCDM}
\begin{figure}[ht]
\centering
\includegraphics{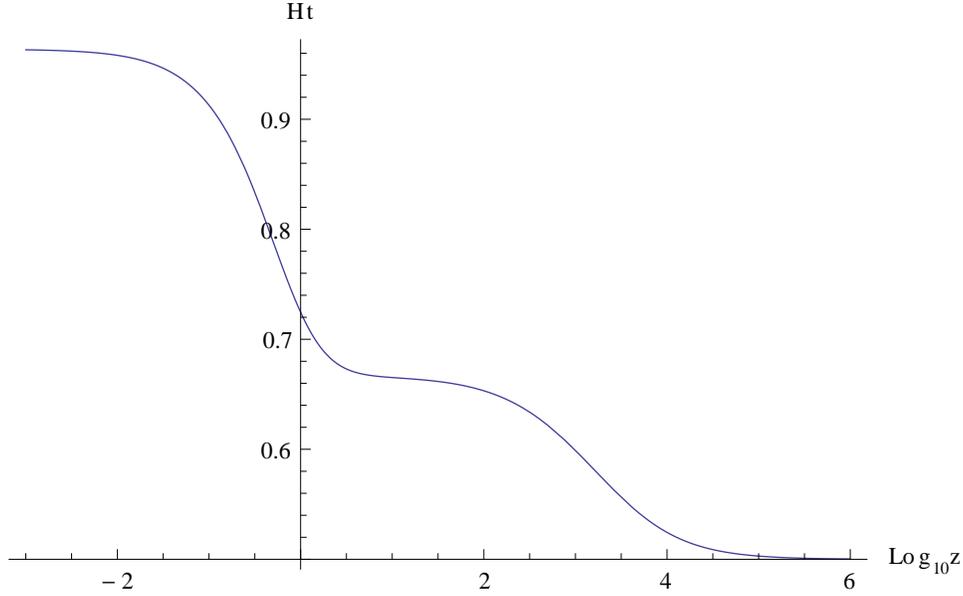}
\caption{\label{htofz} The effective powerlaw index for expansion as derived from $H(t) t$; the radiation era (1/2) and the matter era (2/3) are evident, as well as the transition  and the recent epoch of exponential expansion.
}
\end{figure}

Friedmann's equation is
\ba
H^2 & = & H_0^2 \left(
\Omega_{r,0} \left( \frac{a_0}{a} \right)^4 + 
\Omega_{m,0} \left( \frac{a_0}{a} \right)^3 + 
\Omega_{\Lambda,0} +
\Omega_K  \left( \frac{a_0}{a} \right)^2 
\right) \\
    & = & H_0^2 Q^2 \\
\Omega_0 & = & \Omega_{r,0}+\Omega_{m,0}+\Omega_{\Lambda,0} \\
\Omega_K & = & 1 - \Omega_0
\ea
with
\ba
\frac{a_0}{a} & = & 1 + z \\
\frac{da}{a} & = & - \frac{dz}{1+z} = H_0 Q dt \\
\int \frac{ da }{a Q} & = & -\int \frac{ dz }{(1+z) Q} = H_0 \int dt \\
\int_0^z \frac{dz}{(1+z)Q} & = & H_0 (t_0 - t) \\
\int_z^\infty \frac{dz}{(1+z)Q} & = & H_0 t = \frac{H t}{Q} \\
\ea
And rewrite the $d/dt$ terms in K's equations by $d/dt \to -H (1+z) d/dz$
to give 
\ba
-\left(1+z\right) \frac{d \log \gamma}{dz} & = & \frac{-1}{Ht} + 1 + v^2 + \frac{C p v}{2 \gamma Ht} \\
-\left(1+z\right) \frac{d v}{dz} & = & \left( 1 - v^2 \right) 
\left( \frac{k}{Ht \gamma} - 2 v \right) \\
Ht & = & Q \int_z^\infty \frac{dz}{(1+z)Q}
\ea
The numerical results were pre-calculated for flat $\Lambda$-CDM 
($\Omega_{r,0}= 8.4 \times 10^{-5}$, $\Omega_{m,0}=0.3$ and 
$\Omega_{\Lambda,0}=1-\Omega_{r,0}-\Omega_{m,0}$) and then used in the
string network calculation.  The quantity $Ht$ is shown in Fig. \ref{htofz}.

\section{Fit for ${\cal A}$ the loop creation rate }
\label{fittoA}
The loop creation rate is written
\be
\frac{{\dot E_{\ell}}}{a^3} = \frac{\mu}{p^2 t^3} {\cal A}
\ee
where
\be
{\cal A} = {\cal A}_p {\cal A}_z .
\ee
The factor ${\cal A}$ varies modestly.
We use a simple Pade approximant for each factor and find
\ba
{\cal A}_z & = & \frac{1 + b q + c q^2 + d q^4}{1 + f q^4}\\
q & = & z^{0.217} \\
b & = & 1.97386 \\
c & = & 0.130798 \\
d & = & 0.068323 \\
f & = & 0.00105563
\ea
and
\ba
{\cal A}_p & = &  \frac{p + g }{p + h} \\
g & = & 1.28751 \\
h & = & 17.9383 .
\ea
This fit reproduces the numerical results to relative error $0.25$ and
absolute error $0.3$ for $z> 10^2$.
\begin{figure}
  \centering
\includegraphics{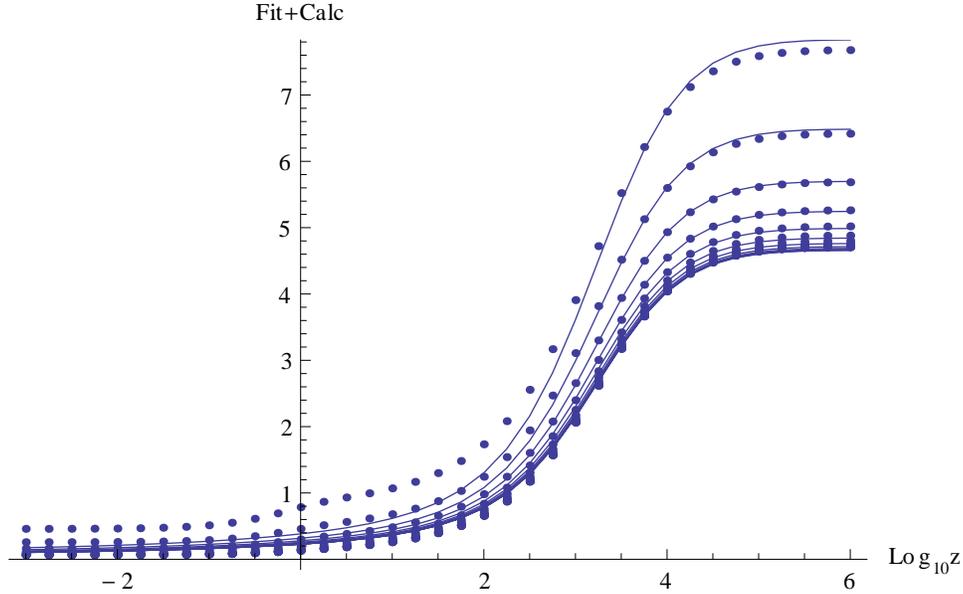}
\caption{\label{fitplot} The fit lines to the data.}
\end{figure}

\section{Dark matter model}
\label{darkmattermodel}
The dark matter halo of the galaxy is empirically described
(Binney and Tremaine) by an ellipsoidal distribution; two
models are given with parameters as follows:
\ba
\rho_{DM}({\vec r}) & = & \rho_{h0} \left( \frac{m}{a_h} \right)^{-\alpha_h}
\left( 1 + \frac{m}{a_h} \right)^{\alpha_h-\beta_h} \\
m & = & \sqrt{R^2 + \frac{z^2}{q_h^2}} \\
q_h & = & 0.8\\
\rho_{h0} & = & (0.711,0.266) \ \ \msun {\rm pc}^{-3} \\
a_h & = & (3.83,1.90)\ \  \rm kpc \\
\alpha_h & = & (-2,1.63) \\
\beta_h & = & (2.96, 2.17)
\ea
where, $q_h$ has been set arbitrarily.
We take $q_h=1$ so that the distribution is spherical.
One model is for an NFW-like cusp and other is for a core.

\section{Lensing without light physically circling a string}
\label{stringlensing}

A straight string in 3+1 dimensions generates a deficit angle in a
plane perpendicular to it. There exist two straight line paths for
light from a point source to an observer on the opposite side of the
string in the same plane if they are all aligned, via circumnavigating
the string in opposite senses, thus producing a double image.

In the braneworld scenario, although both the source (e.g., a star)
and the observer are inside the standard model branes in ``the
standard model'' throat, there are cosmic superstrings in other warped
throats, so photons can never physically circle around them. Here we
analyze lensing in that situation. Lensing is a gravitational effect
which is transmitted through the bulk spacetime so the effect is not
unexpected.

\subsection{Lensing in Minkowski-like spacetime}

Let the background metric be Minkowski -- in other words,
choose scaled coordinates so that the warp factors are
absorbed. 
Einstein's equations in N spacetime dimensions are
\be
R_{\mu\nu}-\frac{1}{2}g_{\mu\nu}R = \kappa^{(N)} T_{\mu\nu}
\ee
where Greek indices range from $0$ to $n \equiv N-1$. For ordinary
spacetime $N=4$, $n=3$ and $\kappa^{(4)}=8 \pi G$ and we will retain $G$
as the 4-dimensional Newton's constant.
For $N \ne 4$ we write $\kappa^{(N)}=\kappa^{(4)} \lambda^{N-4}$ where
$\lambda$ is a length scale and retain units for tension $\mu \sim M/L$.
Taking the trace, solving for $R$ in terms of $T$, we
recast the Einstein equations as
\be
R_{\mu\nu}=\kappa^{(N)} \left( T_{\mu\nu}- \frac{g_{\mu\nu}}{N-2} T \right) .
\ee
Next we linearize the metric about N-dimensional Minkowski
\be
g_{\mu\nu} = \eta_{\mu\nu} + h_{\mu\nu}
\ee
where $h$ is the small quantity and find
\be
R_{\mu\nu}=\frac{1}{2} \left( (h^\sigma_\nu)_{,\mu \sigma} + (h^\sigma_\mu)_{,\sigma \nu} - h_{,\mu\nu} - {(h_{\mu\nu})_{,\sigma}^{\,\sigma}} \right)
\ee
(with summation over $\sigma$ from $0$ to $n$)
and adopt the gauge ${(h_{\mu}^{\sigma})_{,\sigma}} = \frac{1}{2} (h^\sigma_\sigma)_{,\mu} = \frac{1}{2} h_{,\mu}$ to find
\be
-{(h_{\mu\nu})_{,\sigma}^{\,\sigma}} = 2 \kappa^{(N)} \left( T_{\mu\nu}- \frac{\eta_{\mu\nu}}{N-2} T \right) .
\ee
where we have taken $T$ to be first order.

Let the string lie along axis $n=N-1$. Since the background metric
is flat and has been scaled to Minkowski we can be agnostic as to
whether $n$ is in the Standard model brane or not.

The stress energy tensor is
\be
T^{\mu\nu} = \mu \delta_1 \delta_2 \cdots \delta_{n-1}
\left(
\begin{array}{ccccc}
  1 &   &         &    & \\
    & 0 &         &    & \\
    &   & \ddots  &    & \\ 
    &   &         & 0  & \\ 
    &   &         &    & -1 \\
\end{array}
\right)
\ee
where $\delta_i = \delta^1(x^i)$ is the delta function in coordinate direction
i. The first and last elements correspond to $i=0$ and $i=n$.
The stress energy combination
\be
\left( T_{\mu\nu}- \frac{\eta_{\mu\nu}}{N-2} T \right) = 
\mu \frac{\delta_1 \cdots \delta_{n-1}}{n-1}
\left(
\begin{array}{ccccc}
  n-3 &   &         &    & \\
    & 2 &         &    & \\
    &   & \ddots  &    & \\ 
    &   &         & 2  & \\ 
    &   &         &    & 3-n \\
\end{array}
\right)
\ee
and the first order set of equations is
\be
-{h_{\mu\nu,}^\sigma}_\sigma = 2 \kappa \mu \frac{\delta_1 \cdots \delta_{n-1}}{n-1}
\left(
\begin{array}{ccccc}
  n-3 &   &         &    & \\
    & 2 &         &    & \\
    &   & \ddots  &    & \\ 
    &   &         & 2  & \\ 
    &   &         &    & 3-n \\
\end{array}
\right)
\ee

We restrict to time-independent solutions with $\sigma \to a$ for $a=1$ to $n$.
The equations for the off-diagonal elements of $h_{\mu\nu}$ vanish, so
if the off-diagonal elements vanish at infinity they vanish everywhere.
We restrict to solutions that are independent of the coordinate
along the string (labeled by $a=n$), the perturbed metric elements
are functions of the form $f(x^1,x^2, \cdots, x^{n-1})$ and we
drop the derivatives with respect to $x^n$. Define the $m=n-1$
dimensional Laplacian:
$\Delta^2 = \sum_{i=1}^{m} \partial_i^2$. We have
\be
-\Delta^2 h_{00} = \Delta^2 h_{nn} = 2 \kappa \mu \frac{m-2}{m} \delta_1 \cdots \delta_{m}
\ee
and for $i=1$ to $i=m$
\be
-\Delta^2 h_{ii} = 4 \kappa \mu \frac{1}{m} \delta_1 \cdots \delta_{m} .
\ee
All these equations are of the form
\be
\Delta^2 \phi = A \delta_1 \cdots \delta_{m} .
\ee
The equation describes the potential 
of a point particle of mass $A/(G S_{m-1})$ in $m$ dimensions
where $S_k$ is the area of k-sphere ($S_1=2 \pi$, $S_2=4 \pi$; in general,
$S_{m-1} = 2 \pi^{m/2}/\Gamma(m/2)$). If $m=2$ ($m=3$) then the 
mass is $A/(2 \pi G)$ ($A/(4 \pi G)$).

Explicitly, the constants are
\be
A_{0} = -A_{n} = -2 \kappa \mu \frac{m-2}{m}
\ee
and
\be
A_i = -4 \kappa \mu \frac{1}{m} = \frac{2}{m-2} A_0
\ee

For normal spacetime, we have
$m=2$, $A_0 = A_n = 0$ so that $h_{00}=h_{nn}=0$ and for the
two dimensions perpendicular to axis $n=3$ we have
\be
h_{ii} = \frac{A_i}{2\pi} \log \frac{r}{r_0} = - {8 G \mu } \log \frac{r}{r_0}
\ee
where $r_0$ represents the constant of integration since it's not
possible to set $h_{ii}=0$ at infinity.

For integer $m \ge 3$ we have (restoring an explicit c)
\ba
h_{00}(r) & = & \frac{A_0}{(2-m) S_{m-1} r^{m-2}}  \\
& = & \left( \frac{G \mu}{c^2} \right ) \left(\frac{16 \pi }{m S_{m-1}}\right) \left(\frac{\lambda}{r}\right)^{N-4}
\ea
and $h_{nn}=-h_{00}$ and $h_{ii} = \frac{2}{m-2} h_{00}$.

For example, for $N=5$, $n=4$ and $m=3$ we have
\be
h_{00} = -h_{44} = \left( \frac{4 G \mu}{3 c^2} \right ) \frac{\lambda}{r}
\ee
and
\be
h_{ii}=2 h_{00} = \left( \frac{8 G \mu}{3 c^2} \right ) \frac{\lambda}{r}
\ee

Now we consider the geodesic equation for the photon.  The string lies
in direction $n$. Let the photon move in direction $1$ for the
unperturbed Minkowski metric, $x^0 = x^1 = t$. The coordinates $x^j$
for $j=2 \cdots m$ give the transverse separation from the string
which lies along direction $n=m+1$.

We seek the effect of the gravitational perturbation on the photon's
path. For the transverse directions $j=2 \cdots m$ the geodesic
equation is
\be
\frac{d^2 x^j}{dt^2} + \Gamma^i_{\mu\nu} \frac{dx^\mu}{dt} \frac{dx^\nu}{dt} = 0
\ee
and to lowest order (for the unperturbed motion)
\be
\frac{d^2 x^j}{dt^2} + \Gamma^j_{00} + 2 \Gamma^j_{01} + \Gamma^j_{11}  = 0
\ee
and we have
\be
\Gamma^j_{00} = -\frac{1}{2} h_{00,j}
\ee
\be
\Gamma^j_{11} = -\frac{1}{2} h_{11,j}
\ee
with other terms vanishing. So
\be
\frac{d^2 x^j}{dt^2} = \frac{1}{2} \left( h_{00,j} + h_{11,j} \right) .
\ee
With $h_{00}+h_{ii} = \frac{m}{2} h_{11}$, we have
\be
\frac{d^2 x^j}{dt^2} = \frac{m}{2} h_{11,j} .
\ee

\subsection{3+1 spacetime}

Now we will be explicit for the case of
non-compact spacetime with $N=4$: the photon moves in the direction $x
\equiv x^1$, the string lies along $z \equiv x^3$ and there is one
transverse dimensions $y \equiv x^2$. The string passes through
$x=y=0$. The photon separation from the string is $r=\sqrt{x^2 + y^2}$
and the unperturbed photon path is $x=t$, so
$r=\sqrt{t^2 + y^2}$. The transverse change in velocity of
the photon is
\be
\Delta v^y = \frac{m}{2} \int dt h_{11,y} .
\ee
We need the results for $m=2$
\be
h_{11} = - 8 G \mu \log \frac{r}{r_0}
\ee
which gives
\be
h_{11,y} = - 8 G \mu \frac{y}{r^2} .
\ee
The integral over all $t$ gives 
\ba
\Delta v^y & = & - 8 G \mu \int dt \frac{y}{t^2 + y^2} \\
           & = & - 8 G \mu \pi \sigma(y) \label{3dbending}
\ea
where $\sigma(y)$ is the sign of $y$. To lowest order, this
is the angle of bending of the massless particle
\be
\Delta \Theta = \Delta v^y = -8 \pi G \mu \sigma(y).
\ee
This illustrates the usual bending angle in 4 spacetime dimensions
where $y$ is the macroscopic dimension perpendicular to the line of
sight and the string. The alignment of string, source and observer
varies in the macroscopic dimension because all three elements are
typically moving in the macroscopic dimensions. In string lensing the
sign of $y$ switches, i.e.
the photon circumnavigates the string in two distinct paths
because of the variation of the macroscopic geometry. The two paths have different bending angles. When both paths are accessible the observer sees two images of the source. 

\subsection{$N>4$ non-compact spacetime}

For $N>4$ we have $m \ge 3$. Here we carry out the calculation
for the non-compact $N$ dimensional space. 

We will be explicit for $N>4$: the photon moves in the direction $x
\equiv x^1$, the string lies along $z \equiv x^n$ and there are $n-2$
transverse dimensions, $y \equiv x^2$ is macroscopic and the rest
are microscopic. Label $w \equiv x^3$; for $N>5$ we will set
$x^4 \dots x^{n-1}$ to zero. The string passes through
$x=y=0$ and $w=w_s$ (and the remaining microscopic dimensions are
zero). The Standard model brane has $w=0$ (and the remaining microscopic
dimensions are zero) and the photon is confined to the brane.
The photon separation from the string is $r=\sqrt{x^2 + y^2 + w_s^2}$
and the unperturbed photon path is $x=t$, so
$r=\sqrt{t^2 + y^2 + w_s^2}$.

From the previous calculations, we have
\ba
h_{11} & = & \frac{2}{m-2} 
\left( G \mu \right ) \left(\frac{16 \pi }{m S_{m-1}}\right)
\left(\frac{\lambda}{r}\right)^{N-4} \\
& = & K r^{4-N}
\ea
which gives
\be
h_{11,y} = (4-N) K r^{2-N} y .
\ee
The velocity impulse in the y direction is
\be
\Delta v^y  = (4-N) K \int dt \frac{y}{(t^2 + y^2 + w_s^2)^{(N-2)/2}} 
\ee
where the integral
\be
\int dt \frac{y}{(t^2+y^2+w_s^2)^{(N-2)/2}} =
\frac{\sqrt{\pi} \Gamma((N-3)/2) }{\Gamma((N-2)/2)} y (y^2+w_s^2)^{(3-N)/2}
\ee
(for $N>3$). The net result is
\be
\Delta v^y  = -\left(G \mu \right) \sigma(y)
\left(\frac{\lambda}{|y|}\right)^{N-4}
\left( 1 + \frac{w_s^2}{y^2} \right)^{(3-N)/2}
\frac{16 \pi^{(5-N)/2} \Gamma((N-3)/2) }{N-2}
\ee
The bending angle $\Delta \Theta = \Delta v^j$
therefore depends upon the impact parameter measured
with respect to the physical scale $\lambda$ that was
introduced at the beginning and that relates 
$\kappa^{(4)}$ to $\kappa^{(N)}$.

Set $w_s \sim \lambda$, the scale of the
maximum separation in the small dimensions transverse to the brane of the standard model. As the motion in the macroscopic dimensions sends $y \to 0$,
changing the sign of $\sigma(y)$ we have, asymptotically,
\be
\Delta v^y  = \pm \left(G \mu \right)
\left(\frac{|y|}{\lambda}\right)
\frac{16 \pi^{(5-N)/2} \Gamma((N-3)/2) }{N-2}
\ee
If one assumes that the minimal scale for $|y| \sim \lambda$ then
one infers an angular bending at that scale is
qualitatively similar but quantitatively
different than the case $N=4$. Note however that these
results depend, in general, upon $y$ unlike those in
eqn. \ref{3dbending}.

\subsection{$N=5$ compact spacetime}
Here, we consider normal spacetime having one extra
microscopic toroidal coordinate: $N=5$, $n=4$, $m=3$. The photon moves in
the direction $x \equiv x^1$, the string lies along $z \equiv x^4$,
the macroscopic transverse dimension is $y \equiv x^2$ and the
microscopic, periodic transverse dimension is $w \equiv x^3$ with
length $d$. The string passes through $x=y=0$ and $w=w_s$
where $0 \le w_s < d$ and the Standard model brane has $w=w_f$
(``s'' for source and ``f'' for field point).

The equations of motion for the non-compact spacetime were
\be
(\partial_x^2+\partial_y^2+\partial_w^2) \phi(x,y,w) =
A \delta(x) \delta(y) \delta(w - w_s) .
\ee
We now impose $\phi$ is periodic in $w$ but there is
a subtlety. If we require that $\phi$ vanishes at large
macroscopic distances then the $w$-integral of the source on the
right hand side must vanish just as the volume integral of
stress-energy sources in general relativity
must vanish for a closed manifold. In string theory constructions
with orientifolds there is a source of {\it stable} negative energy density
in the macroscopic dimensions, namely, the rigid negative tension orientifold planes. Because the volume of the macroscopic
dimensions is so large compared to that of the compact dimensions only
a very small negative energy density is actually needed. We
expect the orientifold contribution cancels that of the strings
in the microscopic dimension.

We will first construct a ``jellium'' like solution in which a local
source of negative tension in the compact dimensions exactly balances
that of the positive string contribution. We will show that the metric
perturbations are finite and that the individual lensing contributions
from positive and negative terms exactly cancel. Then we argue that
the orientifold shifts the boundary conditions at infinity so that we
can simultaneously turn on the orientifold contribution and turn off
the smooth jellium term.  For photons moving in the Standard model
brane we recover the exact 3+1 result for lensing by a string in a
hidden throat.

The periodic solution $\phi$ may be constructed
from the non-compact solution with two steps. First, adding image strings at
positions $w_s^{(i)} = w_s + i d$ for integer $i$ with $-\infty < i < \infty$.
Second, by including a local homogeneous contribution (the jellium) that
cancels the string component for each image: $\delta(w-w_s^{(i)}) \to \delta(w-w_s^{(i)}) - 1/d$

The schematic form for one string ($i=0$) in the non-compact solution is
\ba
h_{11} & = & {\cal A} \frac{1}{r} \\
r & = & \frac{1}{\sqrt{\perp^2 + (w_f - w_s)^2}} \\
\perp^2 & = & x^2 + y^2 \\
{\cal A} & = & \left(\frac{8 G \mu \lambda}{3 c^2}\right)
\ea
with $h_{11}=h_{22}=h_{33}=2h_{00}=-2h_{44}$. Here,
$\perp$ is the macroscopic distance from the field
point to the string.

The schematic form for the background contribution for one string ($i=0$)
\ba
h_{11} & = & - {\cal A} \frac{1}{d} \int_0^d \frac{dw_s}{\sqrt{\perp^2 + (w_f-w_s)^2}} \\
& = & - {\cal A} \frac{1}{d} \log\left(
\frac{ q_{1} + \sqrt{\perp^2 + q_{1}^2}}
     { q_{0} + \sqrt{\perp^2 + q_{0}^2}} \right) \\
q_i & = & i d - w_f 
\ea

Since the equations are linear we add the solutions for each
discrete image and also the contribution of
strings of negative tension uniformly distributed
in the compact dimension
(tension per compact length $-\mu/d$). The individual
sums are not absolutely convergent but the
series with paired positive and negative contributions
of each image grouped together is. 
Large $i$ contributions are small and convergent.

The total result summed over all images is
\ba
h_{11} & = & {\cal A} \sum_{i} \left(
\frac{1}{\sqrt{\perp^2 + (w_f-w_s^{(i)})^2}} -
\frac{\delta_i}{d} \log\left(
\frac{ \delta_i q_{i+1} + \sqrt{\perp^2 + q_{i+1}^2}}
     { \delta_i q_{i} + \sqrt{\perp^2 + q_{i}^2}} \right) \right) \\
\delta_i & = & \left\{
\begin{array}{cc}
  1 & i \ge 0 \\
  -1 & i \le -1
\end{array}
\right.
\ea
     
Taking the derivative with respect $y$, integrating over the photon
path gives 
\ba
\int h_{11,y} dt & = & -2 {\cal A} {\cal R}
-2 {\cal A} \frac{\sigma(y)}{d} {\cal S}(\infty) \\
{\cal R} & = & \sum_{i} \left(
\frac{y}{y^2 + (w_f-w_s^{(i)})^2} \right) \\
& = & \frac{\pi \sinh\left( \frac{2 \pi y}{d} \right) }
     {d \left( \cosh\left( \frac{2 \pi y}{d} \right) - \cos\left(
       \frac{2 \pi (w_f-w_s)}{d} \right) \right)}  \\
{\cal S}(I) & = & \sum_{i=-I}^I \delta_i
\left( \arctan\left( \frac{q_{i} \delta_i}{|y|} \right) -
\arctan\left( \frac{q_{i+1} \delta_i}{|y|} \right) \right)
\ea
For $|y|>>d $ the ${\cal R}$ term simplifies to
\be
   {\cal R} = \sigma(y) \frac{\pi}{d}
\ee
and the sum in ${\cal S}$ telescopes and successive terms cancel.
Examining the representation
for large but finite $I$ with $q_{-I}< 0$ and $q_{I+1}>0$ gives
\ba
{\cal S}(I) & = & \arctan\frac{q_{-I}}{|y|} - \arctan\frac{q_{I+1}}{|y|} \\
& = & -\pi .
\ea
Since this result is independent of $|y|$ and $I$, the final result is
\ba
\int h_{11,y} dt & = & -2 {\cal A} \sigma(y)\frac{\pi}{d}
+2 \pi {\cal A} \frac{\sigma(y)}{d} = 0
\ea
This cancellation is not
surprising since the net energy density is zero by construction.

The angle change for $|y| >> d$ for the two terms are
\ba
\Delta \Theta_{string} + \Delta \Theta_{homog} & = & \frac{3}{2} \int h_{11,y} dt = 0\ea
The size of the jellium term is
\ba
\Delta \Theta_{homog} & = & 3 \pi {\cal A} \frac{\sigma(y)}{d} \\
& = & \frac{8 \pi G \mu}{c^2} \left( \frac{\lambda}{d} \right ) \sigma(y) \\
& = & -\Delta \Theta_{string}
\ea

Now, we argue that we can alter the boundary condition at large
macroscopic distances replacing the jellium contribution with
that of a properly chosen orientifold contribution. We are
left with only the string lensing.
If we take $\lambda = d$ then
this is identical to the result in normal
spacetime.

\end{document}

%% file: muQ-LIGO-CuspShort.tex
$ -15.7 $ & $   $ & $   $ & $   $ & $   $ & $   $ & $ 0.01 $ \\ 
$ -15.6 $ & $   $ & $   $ & $   $ & $   $ & $   $ & $ 0.24 $ \\ 
$ -15.5 $ & $   $ & $   $ & $   $ & $   $ & $   $ & $ 0.55 $ \\ 
$ -15.4 $ & $   $ & $   $ & $   $ & $ 0.12 $ & $   $ & $ 0.93 $ \\ 
$ -15.3 $ & $   $ & $   $ & $   $ & $ 0.39 $ & $   $ & $ 1.41 $ \\ 
$ -15.2 $ & $   $ & $   $ & $   $ & $ 0.72 $ & $   $ & $ 1.99 $ \\ 
$ -15.1 $ & $   $ & $   $ & $   $ & $ 1.13 $ & $   $ & $ 2.69 $ \\ 
$ -15.0 $ & $   $ & $   $ & $   $ & $ 1.63 $ & $   $ & $ 3.55 $ \\ 
$ -14.9 $ & $   $ & $   $ & $   $ & $ 2.21 $ & $   $ & $ 4.57 $ \\ 
$ -14.8 $ & $   $ & $ 0.03 $ & $   $ & $ 2.91 $ & $   $ & $ 5.77 $ \\ 
$ -14.7 $ & $   $ & $ 0.19 $ & $   $ & $ 3.72 $ & $   $ & $ 7.18 $ \\ 
$ -14.6 $ & $   $ & $ 0.37 $ & $   $ & $ 4.65 $ & $   $ & $ 8.79 $ \\ 
$ -14.5 $ & $   $ & $ 0.58 $ & $   $ & $ 5.72 $ & $   $ & $ 10.64 $ \\ 
$ -14.4 $ & $   $ & $ 0.81 $ & $   $ & $ 6.90 $ & $   $ & $ 12.69 $ \\ 
$ -14.3 $ & $   $ & $ 1.07 $ & $   $ & $ 8.25 $ & $   $ & $ 15.02 $ \\ 
$ -14.2 $ & $   $ & $ 1.36 $ & $   $ & $ 9.73 $ & $   $ & $ 17.59 $ \\ 
$ -14.1 $ & $   $ & $ 1.68 $ & $   $ & $ 11.37 $ & $   $ & $ 20.43 $ \\ 
$ -14.0 $ & $   $ & $ 2.04 $ & $   $ & $ 13.21 $ & $   $ & $ 23.61 $ \\ 
$ -13.9 $ & $   $ & $ 2.42 $ & $   $ & $ 15.26 $ & $   $ & $ 27.16 $ \\ 
$ -13.8 $ & $   $ & $ 2.82 $ & $   $ & $ 17.54 $ & $   $ & $ 31.12 $ \\ 
$ -13.7 $ & $   $ & $ 3.24 $ & $   $ & $ 20.08 $ & $   $ & $ 35.51 $ \\ 
$ -13.6 $ & $   $ & $ 3.62 $ & $   $ & $ 22.89 $ & $   $ & $ 40.39 $ \\ 
$ -13.5 $ & $   $ & $ 3.95 $ & $   $ & $ 25.99 $ & $   $ & $ 45.76 $ \\ 
$ -13.4 $ & $   $ & $ 4.16 $ & $   $ & $ 29.37 $ & $ 0.01 $ & $ 51.61 $ \\ 
$ -13.3 $ & $   $ & $ 4.19 $ & $   $ & $ 33.00 $ & $ 0.14 $ & $ 57.89 $ \\ 
$ -13.2 $ & $   $ & $ 3.99 $ & $   $ & $ 36.78 $ & $ 0.32 $ & $ 64.43 $ \\ 
$ -13.1 $ & $   $ & $ 3.57 $ & $   $ & $ 40.53 $ & $ 0.52 $ & $ 70.93 $ \\ 
$ -13.0 $ & $   $ & $ 2.95 $ & $ 0.01 $ & $ 43.94 $ & $ 0.74 $ & $ 76.84 $ \\ 
$ -12.9 $ & $   $ & $ 2.26 $ & $ 0.16 $ & $ 46.70 $ & $ 1.01 $ & $ 81.61 $ \\ 
$ -12.8 $ & $   $ & $ 1.56 $ & $ 0.34 $ & $ 48.12 $ & $ 1.31 $ & $ 84.07 $ \\ 
$ -12.7 $ & $   $ & $ 0.94 $ & $ 0.54 $ & $ 47.73 $ & $ 1.66 $ & $ 83.40 $ \\ 
$ -12.6 $ & $   $ & $ 0.43 $ & $ 0.77 $ & $ 45.09 $ & $ 2.06 $ & $ 78.83 $ \\ 
$ -12.5 $ & $   $ & $ 0.06 $ & $ 1.04 $ & $ 40.30 $ & $ 2.53 $ & $ 70.53 $ \\ 
$ -12.4 $ & $   $ & $   $ & $ 1.35 $ & $ 33.92 $ & $ 3.06 $ & $ 59.48 $ \\ 
$ -12.3 $ & $   $ & $   $ & $ 1.70 $ & $ 27.05 $ & $ 3.68 $ & $ 47.59 $ \\ 
$ -12.2 $ & $   $ & $   $ & $ 2.11 $ & $ 20.55 $ & $ 4.38 $ & $ 36.33 $ \\ 
$ -12.1 $ & $   $ & $   $ & $ 2.57 $ & $ 14.96 $ & $ 5.19 $ & $ 26.65 $ \\ 
$ -12.0 $ & $   $ & $   $ & $ 3.12 $ & $ 10.63 $ & $ 6.13 $ & $ 19.14 $ \\ 
$ -11.9 $ & $ 0.02 $ & $ 0.04 $ & $ 3.74 $ & $ 7.84 $ & $ 7.20 $ & $ 14.32 $ \\ 
$ -11.8 $ & $ 0.17 $ & $ 0.18 $ & $ 4.45 $ & $ 6.67 $ & $ 8.43 $ & $ 12.28 $ \\ 
$ -11.7 $ & $ 0.34 $ & $ 0.35 $ & $ 5.26 $ & $ 6.53 $ & $ 9.84 $ & $ 12.04 $ \\ 
$ -11.6 $ & $ 0.54 $ & $ 0.54 $ & $ 6.21 $ & $ 6.95 $ & $ 11.49 $ & $ 12.77 $ \\ 
$ -11.5 $ & $ 0.76 $ & $ 0.77 $ & $ 7.30 $ & $ 7.75 $ & $ 13.37 $ & $ 14.15 $ \\ 
$ -11.4 $ & $ 1.02 $ & $ 1.02 $ & $ 8.54 $ & $ 8.82 $ & $ 15.53 $ & $ 16.00 $ \\ 
$ -11.3 $ & $ 1.31 $ & $ 1.32 $ & $ 9.97 $ & $ 10.13 $ & $ 17.99 $ & $ 18.29 $ \\ 
$ -11.2 $ & $ 1.65 $ & $ 1.65 $ & $ 11.60 $ & $ 11.72 $ & $ 20.82 $ & $ 21.03 $ \\ 
$ -11.1 $ & $ 2.03 $ & $ 2.03 $ & $ 13.49 $ & $ 13.56 $ & $ 24.10 $ & $ 24.23 $ \\ 
$ -11.0 $ & $ 2.45 $ & $ 2.45 $ & $ 15.65 $ & $ 15.69 $ & $ 27.84 $ & $ 27.92 $ \\ 
$ -10.9 $ & $ 2.93 $ & $ 2.93 $ & $ 18.11 $ & $ 18.14 $ & $ 32.10 $ & $ 32.14 $ \\ 
$ -10.8 $ & $ 3.46 $ & $ 3.46 $ & $ 20.90 $ & $ 20.92 $ & $ 36.94 $ & $ 36.96 $ \\ 
$ -10.7 $ & $ 4.06 $ & $ 4.06 $ & $ 24.09 $ & $ 24.10 $ & $ 42.45 $ & $ 42.47 $ \\ 
$ -10.6 $ & $ 4.72 $ & $ 4.72 $ & $ 27.74 $ & $ 27.75 $ & $ 48.79 $ & $ 48.80 $ \\ 
$ -10.5 $ & $ 5.44 $ & $ 5.44 $ & $ 31.88 $ & $ 31.88 $ & $ 55.95 $ & $ 55.96 $ \\ 
$ -10.0 $ & $ 10.06 $ & $ 10.06 $ & $ 61.45 $ & $ 61.45 $ & $ 107.17 $ & $ 107.17 $ \\ 
$ -9.5 $ & $ 15.92 $ & $ 15.92 $ & $ 109.48 $ & $ 109.48 $ & $ 190.35 $ & $ 190.35 $ \\ 
$ -9.0 $ & $ 21.80 $ & $ 21.80 $ & $ 174.21 $ & $ 174.21 $ & $ 302.47 $ & $ 302.47 $ \\ 
$ -8.5 $ & $ 26.56 $ & $ 26.56 $ & $ 243.94 $ & $ 243.94 $ & $ 423.25 $ & $ 423.25 $ \\ 
$ -8.0 $ & $ 30.00 $ & $ 30.00 $ & $ 303.32 $ & $ 303.32 $ & $ 526.09 $ & $ 526.09 $ \\ 
$ -7.5 $ & $ 32.84 $ & $ 32.84 $ & $ 345.62 $ & $ 345.62 $ & $ 599.36 $ & $ 599.36 $ \\ 
$ -7.1 $ & $ 35.37 $ & $ 35.37 $ & $ 371.58 $ & $ 371.58 $ & $ 644.33 $ & $ 644.33 $ \\ 

%% file: muQ-LIGO-KinkShort.tex
$ -13.6 $ & $   $ & $   $ & $   $ & $   $ & $   $ & $ 0.11 $ \\ 
$ -13.5 $ & $   $ & $   $ & $   $ & $   $ & $   $ & $ 0.26 $ \\ 
$ -13.4 $ & $   $ & $   $ & $   $ & $   $ & $   $ & $ 0.41 $ \\ 
$ -13.3 $ & $   $ & $   $ & $   $ & $   $ & $   $ & $ 0.56 $ \\ 
$ -13.2 $ & $   $ & $   $ & $   $ & $ 0.01 $ & $   $ & $ 0.71 $ \\ 
$ -13.1 $ & $   $ & $   $ & $   $ & $ 0.07 $ & $   $ & $ 0.85 $ \\ 
$ -13.0 $ & $   $ & $   $ & $   $ & $ 0.16 $ & $   $ & $ 1.00 $ \\ 
$ -12.9 $ & $   $ & $   $ & $   $ & $ 0.25 $ & $   $ & $ 1.16 $ \\ 
$ -12.8 $ & $   $ & $   $ & $   $ & $ 0.34 $ & $   $ & $ 1.32 $ \\ 
$ -12.7 $ & $   $ & $   $ & $   $ & $ 0.43 $ & $   $ & $ 1.48 $ \\ 
$ -12.6 $ & $   $ & $   $ & $   $ & $ 0.53 $ & $   $ & $ 1.66 $ \\ 
$ -12.5 $ & $   $ & $   $ & $   $ & $ 0.64 $ & $   $ & $ 1.84 $ \\ 
$ -12.4 $ & $   $ & $   $ & $   $ & $ 0.75 $ & $   $ & $ 2.03 $ \\ 
$ -12.3 $ & $   $ & $   $ & $   $ & $ 0.86 $ & $   $ & $ 2.23 $ \\ 
$ -12.2 $ & $   $ & $   $ & $   $ & $ 0.98 $ & $   $ & $ 2.44 $ \\ 
$ -12.1 $ & $   $ & $   $ & $   $ & $ 1.10 $ & $   $ & $ 2.64 $ \\ 
$ -12.0 $ & $   $ & $   $ & $   $ & $ 1.22 $ & $   $ & $ 2.85 $ \\ 
$ -11.9 $ & $   $ & $   $ & $   $ & $ 1.32 $ & $   $ & $ 3.02 $ \\ 
$ -11.8 $ & $   $ & $   $ & $   $ & $ 1.37 $ & $   $ & $ 3.11 $ \\ 
$ -11.7 $ & $   $ & $   $ & $   $ & $ 1.34 $ & $   $ & $ 3.05 $ \\ 
$ -11.6 $ & $   $ & $   $ & $   $ & $ 1.18 $ & $   $ & $ 2.77 $ \\ 
$ -11.5 $ & $   $ & $   $ & $   $ & $ 0.90 $ & $   $ & $ 2.29 $ \\ 
$ -11.4 $ & $   $ & $   $ & $   $ & $ 0.55 $ & $   $ & $ 1.69 $ \\ 
$ -11.3 $ & $   $ & $   $ & $   $ & $ 0.20 $ & $   $ & $ 1.08 $ \\ 
$ -11.2 $ & $   $ & $   $ & $   $ & $   $ & $   $ & $ 0.54 $ \\ 
$ -11.1 $ & $   $ & $   $ & $   $ & $   $ & $   $ & $ 0.10 $ \\ 
$ -9.4 $ & $   $ & $   $ & $   $ & $   $ & $ 0.02 $ & $ 0.02 $ \\ 
$ -9.3 $ & $   $ & $   $ & $   $ & $   $ & $ 0.11 $ & $ 0.11 $ \\ 
$ -9.2 $ & $   $ & $   $ & $   $ & $   $ & $ 0.22 $ & $ 0.22 $ \\ 
$ -9.1 $ & $   $ & $   $ & $   $ & $   $ & $ 0.33 $ & $ 0.33 $ \\ 
$ -9.0 $ & $   $ & $   $ & $   $ & $   $ & $ 0.45 $ & $ 0.45 $ \\ 
$ -8.9 $ & $   $ & $   $ & $   $ & $   $ & $ 0.59 $ & $ 0.59 $ \\ 
$ -8.8 $ & $   $ & $   $ & $ 0.02 $ & $ 0.02 $ & $ 0.73 $ & $ 0.73 $ \\ 
$ -8.7 $ & $   $ & $   $ & $ 0.09 $ & $ 0.09 $ & $ 0.89 $ & $ 0.89 $ \\ 
$ -8.6 $ & $   $ & $   $ & $ 0.19 $ & $ 0.19 $ & $ 1.06 $ & $ 1.06 $ \\ 
$ -8.5 $ & $   $ & $   $ & $ 0.29 $ & $ 0.29 $ & $ 1.24 $ & $ 1.24 $ \\ 
$ -8.0 $ & $   $ & $   $ & $ 0.91 $ & $ 0.91 $ & $ 2.31 $ & $ 2.31 $ \\ 
$ -7.5 $ & $   $ & $   $ & $ 1.65 $ & $ 1.65 $ & $ 3.59 $ & $ 3.59 $ \\ 
$ -7.1 $ & $   $ & $   $ & $ 2.25 $ & $ 2.25 $ & $ 4.62 $ & $ 4.62 $ \\ 

%% file: muQ-LISA-CuspShort.tex
$ -16.3 $ & $   $ & $   $ & $   $ & $   $ & $   $ & $ 0.05 $ \\ 
$ -16.2 $ & $   $ & $   $ & $   $ & $   $ & $   $ & $ 0.32 $ \\ 
$ -16.1 $ & $   $ & $   $ & $   $ & $   $ & $   $ & $ 0.66 $ \\ 
$ -16.0 $ & $   $ & $   $ & $   $ & $ 0.21 $ & $   $ & $ 1.09 $ \\ 
$ -15.9 $ & $   $ & $   $ & $   $ & $ 0.52 $ & $   $ & $ 1.63 $ \\ 
$ -15.8 $ & $   $ & $   $ & $   $ & $ 0.91 $ & $   $ & $ 2.31 $ \\ 
$ -15.7 $ & $   $ & $   $ & $   $ & $ 1.41 $ & $   $ & $ 3.17 $ \\ 
$ -15.6 $ & $   $ & $ 0.13 $ & $   $ & $ 2.03 $ & $   $ & $ 4.24 $ \\ 
$ -15.5 $ & $   $ & $ 0.42 $ & $   $ & $ 2.81 $ & $   $ & $ 5.59 $ \\ 
$ -15.4 $ & $   $ & $ 0.78 $ & $   $ & $ 3.79 $ & $   $ & $ 7.29 $ \\ 
$ -15.3 $ & $   $ & $ 1.24 $ & $   $ & $ 5.02 $ & $   $ & $ 9.42 $ \\ 
$ -15.2 $ & $   $ & $ 1.80 $ & $   $ & $ 6.56 $ & $   $ & $ 12.10 $ \\ 
$ -15.1 $ & $   $ & $ 2.49 $ & $   $ & $ 8.50 $ & $   $ & $ 15.46 $ \\ 
$ -15.0 $ & $   $ & $ 3.35 $ & $   $ & $ 10.93 $ & $   $ & $ 19.66 $ \\ 
$ -14.9 $ & $   $ & $ 4.40 $ & $   $ & $ 13.97 $ & $ 0.15 $ & $ 24.92 $ \\ 
$ -14.8 $ & $   $ & $ 5.68 $ & $   $ & $ 17.74 $ & $ 0.34 $ & $ 31.46 $ \\ 
$ -14.7 $ & $   $ & $ 7.21 $ & $   $ & $ 22.43 $ & $ 0.55 $ & $ 39.59 $ \\ 
$ -14.6 $ & $   $ & $ 9.06 $ & $ 0.03 $ & $ 28.23 $ & $ 0.79 $ & $ 49.64 $ \\ 
$ -14.5 $ & $   $ & $ 11.25 $ & $ 0.20 $ & $ 35.39 $ & $ 1.07 $ & $ 62.03 $ \\ 
$ -14.4 $ & $   $ & $ 13.79 $ & $ 0.38 $ & $ 44.21 $ & $ 1.39 $ & $ 77.31 $ \\ 
$ -14.3 $ & $   $ & $ 16.73 $ & $ 0.59 $ & $ 54.99 $ & $ 1.75 $ & $ 95.97 $ \\ 
$ -14.2 $ & $   $ & $ 20.07 $ & $ 0.83 $ & $ 68.04 $ & $ 2.17 $ & $ 118.59 $ \\ 
$ -14.1 $ & $   $ & $ 23.83 $ & $ 1.11 $ & $ 83.71 $ & $ 2.65 $ & $ 145.72 $ \\ 
$ -14.0 $ & $   $ & $ 27.97 $ & $ 1.43 $ & $ 102.27 $ & $ 3.21 $ & $ 177.88 $ \\ 
$ -13.9 $ & $   $ & $ 32.63 $ & $ 1.80 $ & $ 124.32 $ & $ 3.84 $ & $ 216.06 $ \\ 
$ -13.8 $ & $   $ & $ 37.76 $ & $ 2.22 $ & $ 149.84 $ & $ 4.57 $ & $ 260.26 $ \\ 
$ -13.7 $ & $   $ & $ 43.41 $ & $ 2.70 $ & $ 178.92 $ & $ 5.42 $ & $ 310.63 $ \\ 
$ -13.6 $ & $   $ & $ 49.67 $ & $ 3.27 $ & $ 211.95 $ & $ 6.39 $ & $ 367.84 $ \\ 
$ -13.5 $ & $ 0.06 $ & $ 56.60 $ & $ 3.91 $ & $ 248.63 $ & $ 7.51 $ & $ 431.36 $ \\ 
$ -13.4 $ & $ 0.22 $ & $ 64.34 $ & $ 4.65 $ & $ 288.78 $ & $ 8.79 $ & $ 500.91 $ \\ 
$ -13.3 $ & $ 0.40 $ & $ 72.91 $ & $ 5.50 $ & $ 333.50 $ & $ 10.26 $ & $ 578.38 $ \\ 
$ -13.2 $ & $ 0.61 $ & $ 82.32 $ & $ 6.49 $ & $ 382.34 $ & $ 11.98 $ & $ 662.96 $ \\ 
$ -13.1 $ & $ 0.86 $ & $ 92.56 $ & $ 7.63 $ & $ 435.79 $ & $ 13.94 $ & $ 755.54 $ \\ 
$ -13.0 $ & $ 1.14 $ & $ 103.50 $ & $ 8.93 $ & $ 494.51 $ & $ 16.20 $ & $ 857.25 $ \\ 
$ -12.9 $ & $ 1.46 $ & $ 114.86 $ & $ 10.42 $ & $ 559.15 $ & $ 18.78 $ & $ 969.21 $ \\ 
$ -12.8 $ & $ 1.83 $ & $ 126.29 $ & $ 12.15 $ & $ 630.30 $ & $ 21.78 $ & $ 1092.40 $ \\ 
$ -12.7 $ & $ 2.26 $ & $ 136.91 $ & $ 14.15 $ & $ 708.43 $ & $ 25.24 $ & $ 1227.80 $ \\ 
$ -12.6 $ & $ 2.76 $ & $ 145.18 $ & $ 16.44 $ & $ 793.60 $ & $ 29.21 $ & $ 1375.30 $ \\ 
$ -12.5 $ & $ 3.33 $ & $ 149.96 $ & $ 19.06 $ & $ 885.39 $ & $ 33.75 $ & $ 1534.30 $ \\ 
$ -12.4 $ & $ 3.98 $ & $ 149.00 $ & $ 22.08 $ & $ 982.24 $ & $ 38.98 $ & $ 1702.00 $ \\ 
$ -12.3 $ & $ 4.73 $ & $ 141.15 $ & $ 25.60 $ & $ 1080.90 $ & $ 45.07 $ & $ 1873.00 $ \\ 
$ -12.2 $ & $ 5.59 $ & $ 126.47 $ & $ 29.63 $ & $ 1175.30 $ & $ 52.06 $ & $ 2036.40 $ \\ 
$ -12.1 $ & $ 6.60 $ & $ 107.10 $ & $ 34.26 $ & $ 1258.30 $ & $ 60.07 $ & $ 2180.20 $ \\ 
$ -12.0 $ & $ 7.75 $ & $ 85.68 $ & $ 39.55 $ & $ 1315.80 $ & $ 69.24 $ & $ 2279.80 $ \\ 
$ -11.9 $ & $ 9.06 $ & $ 65.34 $ & $ 45.71 $ & $ 1333.40 $ & $ 79.90 $ & $ 2310.20 $ \\ 
$ -11.8 $ & $ 10.57 $ & $ 47.92 $ & $ 52.81 $ & $ 1293.30 $ & $ 92.21 $ & $ 2240.80 $ \\ 
$ -11.7 $ & $ 12.33 $ & $ 34.55 $ & $ 60.96 $ & $ 1192.90 $ & $ 106.32 $ & $ 2066.80 $ \\ 
$ -11.6 $ & $ 14.35 $ & $ 26.31 $ & $ 70.29 $ & $ 1034.90 $ & $ 122.49 $ & $ 1793.30 $ \\ 
$ -11.5 $ & $ 16.66 $ & $ 23.08 $ & $ 81.04 $ & $ 848.44 $ & $ 141.09 $ & $ 1470.30 $ \\ 
$ -11.4 $ & $ 19.30 $ & $ 22.94 $ & $ 93.56 $ & $ 657.50 $ & $ 162.77 $ & $ 1139.50 $ \\ 
$ -11.3 $ & $ 22.36 $ & $ 24.50 $ & $ 107.91 $ & $ 488.73 $ & $ 187.63 $ & $ 847.25 $ \\ 
$ -11.2 $ & $ 25.90 $ & $ 27.19 $ & $ 124.34 $ & $ 353.62 $ & $ 216.09 $ & $ 613.23 $ \\ 
$ -11.1 $ & $ 29.95 $ & $ 30.73 $ & $ 143.15 $ & $ 266.66 $ & $ 248.67 $ & $ 462.61 $ \\ 
$ -11.0 $ & $ 34.58 $ & $ 35.06 $ & $ 165.02 $ & $ 229.39 $ & $ 286.55 $ & $ 398.05 $ \\ 
$ -10.9 $ & $ 39.87 $ & $ 40.19 $ & $ 190.21 $ & $ 225.88 $ & $ 330.19 $ & $ 391.97 $ \\ 
$ -10.8 $ & $ 46.04 $ & $ 46.23 $ & $ 219.08 $ & $ 239.99 $ & $ 380.19 $ & $ 416.41 $ \\ 
$ -10.7 $ & $ 53.07 $ & $ 53.19 $ & $ 252.16 $ & $ 264.73 $ & $ 437.49 $ & $ 459.27 $ \\ 
$ -10.6 $ & $ 61.09 $ & $ 61.17 $ & $ 290.23 $ & $ 298.09 $ & $ 503.42 $ & $ 517.04 $ \\ 
$ -10.5 $ & $ 70.23 $ & $ 70.27 $ & $ 334.55 $ & $ 339.25 $ & $ 580.19 $ & $ 588.34 $ \\ 
$ -10.0 $ & $ 138.74 $ & $ 138.75 $ & $ 674.76 $ & $ 675.16 $ & $ 1169.50 $ & $ 1170.10 $ \\ 
$ -9.5 $ & $ 260.44 $ & $ 260.44 $ & $ 1340.00 $ & $ 1340.00 $ & $ 2321.70 $ & $ 2321.70 $ \\ 
$ -9.0 $ & $ 449.18 $ & $ 449.18 $ & $ 2563.90 $ & $ 2563.90 $ & $ 4441.50 $ & $ 4441.50 $ \\ 
$ -8.5 $ & $ 689.27 $ & $ 689.27 $ & $ 4573.50 $ & $ 4573.50 $ & $ 7922.20 $ & $ 7922.20 $ \\ 
$ -8.0 $ & $ 928.96 $ & $ 928.96 $ & $ 7325.80 $ & $ 7325.80 $ & $ 12689.00 $ & $ 12689.00 $ \\ 
$ -7.5 $ & $ 1119.40 $ & $ 1119.40 $ & $ 10307.00 $ & $ 10307.00 $ & $ 17853.00 $ & $ 17853.00 $ \\ 
$ -7.1 $ & $ 1228.10 $ & $ 1228.10 $ & $ 12373.00 $ & $ 12373.00 $ & $ 21431.00 $ & $ 21431.00 $ \\ 

%% file: muQ-LISA-KinkShort.tex
$ -15.4 $ & $   $ & $   $ & $   $ & $   $ & $   $ & $ 0.02 $ \\ 
$ -15.3 $ & $   $ & $   $ & $   $ & $   $ & $   $ & $ 0.28 $ \\ 
$ -15.2 $ & $   $ & $   $ & $   $ & $   $ & $   $ & $ 0.61 $ \\ 
$ -15.1 $ & $   $ & $   $ & $   $ & $ 0.17 $ & $   $ & $ 1.03 $ \\ 
$ -15.0 $ & $   $ & $   $ & $   $ & $ 0.48 $ & $   $ & $ 1.56 $ \\ 
$ -14.9 $ & $   $ & $ 0.06 $ & $   $ & $ 0.86 $ & $   $ & $ 2.21 $ \\ 
$ -14.8 $ & $   $ & $ 0.32 $ & $   $ & $ 1.33 $ & $   $ & $ 3.04 $ \\ 
$ -14.7 $ & $   $ & $ 0.65 $ & $   $ & $ 1.93 $ & $   $ & $ 4.07 $ \\ 
$ -14.6 $ & $   $ & $ 1.05 $ & $   $ & $ 2.67 $ & $   $ & $ 5.36 $ \\ 
$ -14.5 $ & $   $ & $ 1.54 $ & $   $ & $ 3.61 $ & $   $ & $ 6.98 $ \\ 
$ -14.4 $ & $   $ & $ 2.13 $ & $   $ & $ 4.77 $ & $   $ & $ 8.99 $ \\ 
$ -14.3 $ & $   $ & $ 2.84 $ & $   $ & $ 6.22 $ & $ 0.01 $ & $ 11.50 $ \\ 
$ -14.2 $ & $   $ & $ 3.68 $ & $   $ & $ 8.01 $ & $ 0.09 $ & $ 14.61 $ \\ 
$ -14.1 $ & $   $ & $ 4.65 $ & $   $ & $ 10.23 $ & $ 0.20 $ & $ 18.45 $ \\ 
$ -14.0 $ & $   $ & $ 5.76 $ & $   $ & $ 12.96 $ & $ 0.31 $ & $ 23.18 $ \\ 
$ -13.9 $ & $   $ & $ 6.97 $ & $   $ & $ 16.30 $ & $ 0.43 $ & $ 28.96 $ \\ 
$ -13.8 $ & $   $ & $ 8.27 $ & $   $ & $ 20.35 $ & $ 0.57 $ & $ 35.97 $ \\ 
$ -13.7 $ & $   $ & $ 9.61 $ & $ 0.01 $ & $ 25.22 $ & $ 0.72 $ & $ 44.41 $ \\ 
$ -13.6 $ & $   $ & $ 10.96 $ & $ 0.09 $ & $ 31.04 $ & $ 0.88 $ & $ 54.49 $ \\ 
$ -13.5 $ & $   $ & $ 12.31 $ & $ 0.19 $ & $ 37.82 $ & $ 1.06 $ & $ 66.24 $ \\ 
$ -13.4 $ & $   $ & $ 13.64 $ & $ 0.30 $ & $ 45.57 $ & $ 1.25 $ & $ 79.66 $ \\ 
$ -13.3 $ & $   $ & $ 14.99 $ & $ 0.42 $ & $ 54.25 $ & $ 1.47 $ & $ 94.69 $ \\ 
$ -13.2 $ & $   $ & $ 16.35 $ & $ 0.56 $ & $ 63.57 $ & $ 1.70 $ & $ 110.84 $ \\ 
$ -13.1 $ & $   $ & $ 17.72 $ & $ 0.70 $ & $ 73.24 $ & $ 1.95 $ & $ 127.59 $ \\ 
$ -13.0 $ & $   $ & $ 19.18 $ & $ 0.86 $ & $ 83.02 $ & $ 2.22 $ & $ 144.53 $ \\ 
$ -12.9 $ & $   $ & $ 20.69 $ & $ 1.04 $ & $ 92.61 $ & $ 2.53 $ & $ 161.13 $ \\ 
$ -12.8 $ & $   $ & $ 22.23 $ & $ 1.23 $ & $ 102.00 $ & $ 2.87 $ & $ 177.39 $ \\ 
$ -12.7 $ & $   $ & $ 23.91 $ & $ 1.44 $ & $ 111.26 $ & $ 3.23 $ & $ 193.44 $ \\ 
$ -12.6 $ & $   $ & $ 25.60 $ & $ 1.67 $ & $ 120.34 $ & $ 3.62 $ & $ 209.16 $ \\ 
$ -12.5 $ & $   $ & $ 27.43 $ & $ 1.92 $ & $ 129.41 $ & $ 4.06 $ & $ 224.88 $ \\ 
$ -12.4 $ & $   $ & $ 29.32 $ & $ 2.20 $ & $ 138.89 $ & $ 4.54 $ & $ 241.30 $ \\ 
$ -12.3 $ & $   $ & $ 31.28 $ & $ 2.50 $ & $ 148.35 $ & $ 5.06 $ & $ 257.69 $ \\ 
$ -12.2 $ & $   $ & $ 33.25 $ & $ 2.82 $ & $ 158.41 $ & $ 5.62 $ & $ 275.11 $ \\ 
$ -12.1 $ & $   $ & $ 35.09 $ & $ 3.18 $ & $ 168.68 $ & $ 6.24 $ & $ 292.89 $ \\ 
$ -12.0 $ & $ 0.01 $ & $ 36.45 $ & $ 3.57 $ & $ 179.43 $ & $ 6.92 $ & $ 311.52 $ \\ 
$ -11.9 $ & $ 0.08 $ & $ 36.87 $ & $ 4.01 $ & $ 190.61 $ & $ 7.67 $ & $ 330.88 $ \\ 
$ -11.8 $ & $ 0.18 $ & $ 35.60 $ & $ 4.47 $ & $ 202.17 $ & $ 8.48 $ & $ 350.90 $ \\ 
$ -11.7 $ & $ 0.29 $ & $ 32.32 $ & $ 4.98 $ & $ 214.15 $ & $ 9.37 $ & $ 371.65 $ \\ 
$ -11.6 $ & $ 0.41 $ & $ 27.36 $ & $ 5.54 $ & $ 226.09 $ & $ 10.33 $ & $ 392.32 $ \\ 
$ -11.5 $ & $ 0.54 $ & $ 21.69 $ & $ 6.17 $ & $ 237.15 $ & $ 11.41 $ & $ 411.49 $ \\ 
$ -11.4 $ & $ 0.69 $ & $ 16.14 $ & $ 6.84 $ & $ 245.57 $ & $ 12.58 $ & $ 426.07 $ \\ 
$ -11.3 $ & $ 0.85 $ & $ 11.33 $ & $ 7.57 $ & $ 247.61 $ & $ 13.85 $ & $ 429.60 $ \\ 
$ -11.2 $ & $ 1.02 $ & $ 6.93 $ & $ 8.37 $ & $ 238.73 $ & $ 15.23 $ & $ 414.23 $ \\ 
$ -11.1 $ & $ 1.21 $ & $ 3.40 $ & $ 9.26 $ & $ 216.62 $ & $ 16.77 $ & $ 375.93 $ \\ 
$ -11.0 $ & $ 1.42 $ & $ 2.23 $ & $ 10.23 $ & $ 182.94 $ & $ 18.45 $ & $ 317.60 $ \\ 
$ -10.9 $ & $ 1.65 $ & $ 2.03 $ & $ 11.29 $ & $ 144.87 $ & $ 20.28 $ & $ 251.65 $ \\ 
$ -10.8 $ & $ 1.90 $ & $ 2.10 $ & $ 12.43 $ & $ 108.50 $ & $ 22.27 $ & $ 188.66 $ \\ 
$ -10.7 $ & $ 2.17 $ & $ 2.27 $ & $ 13.69 $ & $ 77.29 $ & $ 24.44 $ & $ 134.60 $ \\ 
$ -10.6 $ & $ 2.46 $ & $ 2.52 $ & $ 15.09 $ & $ 52.28 $ & $ 26.87 $ & $ 91.28 $ \\ 
$ -10.5 $ & $ 2.78 $ & $ 2.81 $ & $ 16.62 $ & $ 32.59 $ & $ 29.51 $ & $ 57.18 $ \\ 
$ -10.0 $ & $ 4.93 $ & $ 4.93 $ & $ 26.66 $ & $ 27.09 $ & $ 46.91 $ & $ 47.65 $ \\ 
$ -9.5 $ & $ 8.29 $ & $ 8.29 $ & $ 42.50 $ & $ 42.52 $ & $ 74.34 $ & $ 74.37 $ \\ 
$ -9.0 $ & $ 13.44 $ & $ 13.44 $ & $ 67.63 $ & $ 67.63 $ & $ 117.87 $ & $ 117.87 $ \\ 
$ -8.5 $ & $ 20.95 $ & $ 20.95 $ & $ 107.24 $ & $ 107.24 $ & $ 186.48 $ & $ 186.48 $ \\ 
$ -8.0 $ & $ 30.74 $ & $ 30.74 $ & $ 168.08 $ & $ 168.08 $ & $ 291.86 $ & $ 291.86 $ \\ 
$ -7.5 $ & $ 41.63 $ & $ 41.63 $ & $ 255.46 $ & $ 255.46 $ & $ 443.20 $ & $ 443.20 $ \\ 
$ -7.1 $ & $ 49.76 $ & $ 49.76 $ & $ 342.55 $ & $ 342.55 $ & $ 594.05 $ & $ 594.05 $ \\ 